\documentclass[12pt]{iopart}
\pdfoutput=1
\usepackage{iopams}
\usepackage{latexsym}
\usepackage{graphicx}
\usepackage{subfigure}
\usepackage{float}

\newtheorem{lem}{Lemma}[section]

\newtheorem{thm}{Theorem}[section]

\newtheorem{algorithm}{Algorithm}[section]

\def\debproof{\noindent {\bf Proof.} }
\def\finproof{\hfill {\small $\Box$} \\}
\newcommand{\bn}{\nu}

\newcommand{\pa}{\partial}
\newcommand{\la}{\langle}
\newcommand{\ra}{\rangle}

\newcommand{\na}{\nabla}

\newcommand{\Ga}{\Gamma}
\newcommand{\Om}{\Omega}
\newcommand{\de}{\delta}

\newcommand{\lam}{\lambda}

\renewcommand{\i}{\mathbf{i}}
\newcommand{\R}{{\mathbb{R}}}

\newcommand{\C}{{\mathbb{C}}}

\renewcommand{\Im}{\mathrm{Im}\,}
\renewcommand{\div}{\mathrm{div}}
\newcommand{\curl}{\mathrm{curl}\,}

\newcommand{\be}{\begin{eqnarray}}
\newcommand{\ee}{\end{eqnarray}}
\newcommand{\ben}{\begin{eqnarray*}}
\newcommand{\een}{\end{eqnarray*}}
\newcommand{\nn}{\nonumber}
\renewcommand{\theequation}
{\arabic{section}.\arabic{equation}}
\begin{document}
\title[RTM for Inverse Electromagnetic Scattering Problems]{\bf  Reverse Time Migration for Extended Obstacles: Electromagnetic Waves}
\author{Junqing Chen$^1$, Zhiming Chen$^2$,
Guanghui Huang$^2$ }
\address{$^1$ Department of Mathematical
Sciences, Tsinghua University,
Beijing 100084, China}
\address{$^2$ LSEC, Institute of Computational Mathematics, Academy of
Mathematics and System Sciences, Chinese Academy of Sciences,
Beijing 100190, China}
%\author{Junqing Chen\thanks{Department of Mathematical
%Sciences, Tsinghua University, Beijing 100084, China. The research of this author was supported in part by China NSF under the grant 11001150 and 11171040.
%({\tt jqchen@math.tsinghua.edu.cn})}
%\and Zhiming Chen \thanks{ LSEC, Institute of Computational Mathematics, Academy of
%Mathematics and System Sciences, Chinese Academy of Sciences,
%Beijing 100190, China. This author was supported in part by
%National Basic Research Project under the grant
%2011CB309700 and China NSF under the grant
%11021101. ({\tt zmchen@lsec.cc.ac.cn})}
%\and Guanghui Huang\thanks{LSEC, Institute of Computational Mathematics, Academy of
%Mathematics and System Sciences, Chinese Academy of Sciences,
%Beijing 100190, China
%({\tt ghhuang@lsec.cc.ac.cn})}
%}

%\date{}
\begin{abstract}
We propose a new single frequency reverse time migration (RTM) algorithm for imaging extended targets using electromagnetic waves. The imaging functional is defined as the imaginary part of the cross-correlation of the Green function for Helmholtz equation and the back-propagated electromagnetic field. The resolution of our RTM method for both penetrable and non-penetrable extended targets is studied by virtue of Helmholtz-Kirchhoff identity for the time-harmonic Maxwell equation. The analysis implies that our imaging functional is always positive and thus may have better stability properties. Numerical examples are provided to demonstrate the powerful imaging quality and confirm our theoretical results.
\end{abstract}
\maketitle

%% {\bf keywords:}
%% ODCIGs, RTM, MVA  \\
%% Mathematics Subject Classification (2000).

\section{Introduction}

In this paper we propose a reverse time migration algorithm for inverse electromagnetic scattering problems. Let $D$ be a bounded Lipschitz domain in $\R^3$ with $\nu$ being the unit outer normal to its boundary $\Ga_D$.  We assume the incident wave is generated by a point source at $x_s$ on a surface $\Ga_s$ far away from the obstacle and we measure the electric field $E$ on a surface $\Ga_r$ which need not to be identical to $\Ga_s$. For penetrable obstacles $D$, the measured field $E$ is the solution of the following problem:
\be
& &\curl\curl E-k^2n(x) E=\delta_{x_s}(x)p\ \ \ \ \mbox{in }\R^3,\label{p1}\\
& &r\left(\curl E\times\hat x-\i kE\right)\to 0\ \ \ \ \mbox{as }r=|x|\to\infty,\label{p2}
\ee
where $k>0$ is the wave number, $n\in L^\infty(D)$ is a positive scalar function and $n(x)-1$ is compactly supported in $D$, $\de_{x_s}$ is the Dirac source located at $x_s$, $p\in\R^3$,  $|p|=1$, is the polarization direction of the source, and $\hat x=x/|x|$. The condition (\ref{p2}) is the well-known Silver-M\"uller radiation condition. For non-penetrable obstacles $D$, the measured field $E$ is the solution of the following problem:
\be
& &\curl\curl E-k^2 E=\delta_{x_s}(x)p\ \ \ \ \mbox{in }\R^3,\label{p3}\\
& &\nu\times E=0\ \ \mbox{or }\ \ \nu\times\curl E-\i k\eta(x)(\nu\times E\times \nu)=0\ \ \ \ \mbox{on }\Ga_D,\label{p4}\\
& &r\left(\curl E\times\hat x-\i kE\right)\to 0\ \ \ \ \mbox{as }r=|x|\to\infty,\label{p5}
\ee
where $\eta(x)\ge 0$ is a bounded function on $\Ga_D$. The Dirichlet condition $\nu\times E=0$ on $\Ga_D$ corresponds to the perfectly conducting
obstacle. The second condition in (\ref{p4}) is the impedance condition. The existence and uniqueness of the problem (\ref{p1})-(\ref{p2}) such that $E^s=E-E^i$ in $H_{\rm loc}(\curl;\R^3)$ and the problem (\ref{p3})-(\ref{p5})  such that $E^s=E-E^i$ in $H_{\rm loc}(\curl;\R^3\backslash\bar D)$ is a well studied subject in the literature \cite{colton-kress, monk}, where $E^i(x,x_s)=\mathbb{G}(x,x_s)p$ and $\mathbb{G}(x,x_s)\in \R^{3\times 3}$ is the dyadic Green function for the time-harmonic Maxwell equation (see section 2 below).

The direct methods for solving inverse scattering problems have drawn considerable interest in the literature in recent years. One example is the
MUltiple SIgnal Classification (MUSIC) method \cite{music, Devaney, BHV, anomaly} which are particularly useful in identifying well-separated small inclusions.  The other class of direct method includes the linear sampling method \cite{LSM}, the factorization method \cite{kirsch_1998, fm_book}, and the point source method \cite{p96, potthast}. The third class of the method is the reverse time migration (RTM) or the closely related prestack depth migration methods \cite{ber84, cla85, bcs} that are widely used in the geophysical community.

In this paper we propose a new RTM algorithm for imaging extended targets using electromagnetic waves by extending our previous study
in \cite{cch} where we consider the single frequency RTM method for extended targets using acoustic waves. The resolution analysis in \cite{cch}, which applies in both penetrable and non-penetrable obstacles with any type of boundary conditions including sound soft, sound hard, or impedance condition on the obstacle, implies that the imaginary part of the two point correlation imaging functional is always positive and thus may have better stability properties. We also refer to \cite{fink1}, \cite{GPR} for using RTM methods to find small electromagnetic inclusions.

Let $E^s(x,x_s)$ be the scattered electric field which is measured on some surface $\Ga_r$. The first step of the RTM method is to back-propagate the complex conjugated (time reversed) of the recorded data on $\Ga_r$ into the computational domain by solving a Maxwell source problem to obtain the back-propagated field $F_b$. A direct extension of the imaging functional from acoustic waves would be to compute the cross-correlation of $E^i$ and $F_b$ which is indeed used in \cite{fink1}, \cite{GPR}. We propose to use a novel imaging
functional which computes the correlation of $g(x,x_s)p$ and $F_b$, where $g(x,x_s)$ is the fundamental solution of the Helmholtz equation. This
new imaging functional is simpler in the computation and allows to provide a resolution analysis for extended targets for both penetrable and non-penetrable targets.

The rest of this paper is outlined as follows. In section 2 we introduce the RTM algorithm. In section 3 we study the resolution of the imaging algorithm in section 2 for both penetrable and non-penetrable obstacles. In section 4 we report extensive numerical experiments to
show the competitive performance of our RTM algorithm.

\section{The reverse time migration algorithm}\label{section2}

In this section we introduce the RTM imaging method for inverse electromagnetic scattering problems. We assume that there are $N_s$ transducers on $\Ga_s=\pa B_s$ and $N_r$ transducers on $\Ga_r=\pa B_r$, where $B_s$ and $B_r$ are the balls of radius $R_s$ and $R_r$, respectively. The distribution of the transducers and receivers
are uniform in polar and azimuthal angular coordinates on the sphere. Let $(R_s, \theta_s,\phi_s)$ and $(R_r,\theta_r,\phi_r)$ be the spherical coordinates of the source $x_s$ and the receiver $x_r$, respectively. We denote by $\Om$ the sampling domain in which the obstacle is sought. We assume the obstacle $D\subset\Om$ and $\Om$ is inside in $B_s$, $B_r$. We assume that $\Om$ is far away from $\Ga_s,\Ga_r$, that is, ${\rm dist}(\Om,\Ga_s)\ge CR_s$, ${\rm dist}(\Om,\Ga_r)\ge CR_r$ for some fixed constant $C>0$.

The dyadic Green function $\mathbb{G}(x,y)$ is a $\C^{3\times 3}$ matrix defined by
\be\label{gl}
\mathbb{G}(x,y)=g(x,y)\mathbb{I}+\frac{\na_x\na_x}{k^2}g(x,y),
\ee
where $\mathbb{I}$ is the $\R^{3\times 3}$ identity matrix and $g(x,y)$ is the fundamental solution of the Helmholtz equation in 3D: $g(x,y)=\frac{e^{\i k|x-y|}}{4\pi |x-y|}$. Clearly $\mathbb{G}(x,y)$ is a symmetric matrix. We denote its column vectors by $g_1(x,y),g_2(x,y),g_3(x,y)$, which satisfy
\ben
\curl\curl g_l(x,y)-k^2g_l(x,y)=\de_y(x)e_l\ \ \ \ \mbox{in }\R^3,\ \ \ \ l=1,2,3,
\een
where $e_l$ is the unit vector of the $x_l$ axis. Let $E^i(x,x_s)=\mathbb{G}(x,x_s)p$, where $p$ is a unit polarization vector, be the incident field and $E^s(x_r,x_s)=E(x_r,x_s)-E^i(x_r,x_s)$ be the scattered electric field measured at $x_r$, where $E(x,x_s)$ is the solution of the problem either (\ref{p1})-(\ref{p2}) or (\ref{p3})-(\ref{p5}).

Our reverse time imaging algorithm consists of two steps. The first step is the back-propagation in which we back-propagate the complex conjugated data $\overline{E^s(x_r,x_s)}$ into the domain. The second step is the correlation in which we compute the cross-correlation of the modified incident field and the back-propagated field.

\begin{algorithm} {\sc (Reverse time migration algorithm)} \\
Given the data $E^s(x_r,x_s)$ which is the measurement of the scattered electric field at $x_r$ when the source is emitted at $x_s$, $s=1,\dots, N_s$ and $r=1,\dots,N_r$. \\
$1^\circ$ Back-propagation: For $s=1,\dots,N_s$, compute the solution $F_b$ of the following problem:
\be
& &\fl\qquad \curl\curl F_b(x,x_s)-k^2F_b(x,x_s)=-\frac{1}{N_r}\sum^{N_r}_{r=1}|\Delta(x_r)|\,\overline{E^s(x_r,x_s)}\de_{x_r}(x)\ \ \mbox{in }\R^3,\label{b1}\\
& &\fl\qquad r\left(\curl F_b\times\hat x-\i kF_b\right)\to 0\ \ \ \ \mbox{as }r\to\infty,\label{b2}
\ee
where $|\Delta(x_r)|=2\pi^2R_r^2\sin(\theta_r)$ is the surface element at $x_r$. \\
$2^\circ$ Cross-correlation: For $z\in\Om$, compute
\be\label{cor1}
I(z)=k^2\cdot\Im\left\{\frac{1}{N_s}\sum^{N_s}_{s=1}|\Delta(x_s)|\,g(z,x_s)p\cdot F_b(z,x_s)\right\},
\ee
where $|\Delta(x_s)|=2\pi^2R_s^2\sin(\theta_s)$ is the surface element at $x_s$.
\end{algorithm}

We remark that we use the modified incident wave $g(z,x_s)p$ instead of the incident wave $\mathbb{G}(z,x_s)p$ in the imaging functional which is simpler and cheaper in the computation. We take the imaginary part of the correlation of the modified incident field and the back-propagated field is motivated by the resolution analysis in the next section where we show that $I(z)$ is a positive function and thus is more stable than the real part of the correlation functional. By using the dyadic  Green function we can represent the solution $F_b$ of (\ref{b1})-(\ref{b2}) as
\ben
F_b(z,x_s)=-\frac{1}{N_r}\sum^{N_r}_{r=1}|\Delta(x_r)|\mathbb{G}(z,x_r)^T\overline{E^s(x_r,x_s)},
\een
which implies for $z\in\Om$,
\be\label{cor2}
\fl\ \  I(z)=-k^2\cdot\Im\left\{\frac{1}{N_sN_r}\sum^{N_s}_{s=1}\sum^{N_r}_{r=1}|\Delta(x_r)|\,|\Delta(x_s)|\,g(z,x_s)p\cdot\mathbb{G}(z,x_r)^T\overline{E^s(x_r,x_s)}\right\}.
\ee
This formula is used in our numerical experiments in section 4.

Noticing that for $z\in\Om$ which is a subdomain of $\Om_s$, $g(z,x_s)$ is a smooth function in $x_s\in\Ga_s$. Similarly, $\mathbb{G}(z,x_r)$ is smooth in $x_r\in\Ga_r$. We also know that since $E^s=E-E^i$ is the scattering solution of (\ref{p1})-(\ref{p2}) or (\ref{p3})-(\ref{p4}),
$E^s(x_r,x_s)$ is also smooth in
$x_r,x_s$. Therefore, the imaging functional $I(z)$ in (\ref{cor2}) is a good quadrature approximation of the following continuous functional:
\be\label{cor3}
\fl\qquad\hat I(z)=-k^2\cdot\Im\int_{\Ga_r}\int_{\Ga_s}g(z,x_s)p\cdot\mathbb{G}(z,x_r)^T\overline{E^s(x_r,x_s)}ds(x_s)ds(x_r)\ \ \ \ \forall z\in\Om.
\ee
This formula is the starting point of our resolution analysis in the next section.

\section{The resolution analysis}\label{section3}

In this section we consider the resolution of the imaging functional in (\ref{cor3}). We start by recalling the Helmholtz-Kirchhoff identity (see \cite{boj82}).

\begin{lem}\label{lem:3.1}
Let $\mathcal{D}$ be a bounded Lipschitz domain in $\R^3$ with $\bn$ being the unit outer normal to the boundary. For any $p,q\in\R^3$, we have
\ben
    & &\int_{\partial\mathcal{D}} \left (\overline{\mathbb{G}(x,\xi)}p\cdot\nu\times\curl(\mathbb{G}(\xi,y)q)-\nu\times\curl(\overline{\mathbb{G}(x,\xi)}p)\cdot\mathbb{G}(\xi,y)q\right)ds(\xi)  \nn\\
    &&\qquad\qquad\qquad=-2\i\,p\cdot{\rm{Im}}\mathbb{G}(x,y)q  \ \ \ \  \forall x,y \in\mathcal{D}.
\een
\end{lem}

\debproof For the sake of completeness we sketch a proof here. For any fixed $x\in\mathcal D$, since $\Im\mathbb{G}(x,\cdot)q$ satisfies the Maxwell equation, we use
the integral representation formula to get, for any $y\in\mathcal D$,
\ben
\Im\mathbb{G}(x,y)q\cdot e_l&=&\int_{\pa\mathcal D}\Big(\nu\times\curl g_l(x,\xi)\cdot\Im\mathbb{G}(\xi,y)q\\
& &\ \ \ \ -g_l(x,\xi)\cdot\nu\times\curl(\Im\mathbb{G}(\xi,y)q)\Big)ds(\xi),\ \ l=1,2,3.
\een
Thus
\ben
p\cdot\Im\mathbb{G}(x,y)q&=&\int_{\pa\mathcal D}\Big(\nu\times\curl (\mathbb{G}(x,\xi)p)\cdot\Im\mathbb{G}(\xi,y)q\\
& &\ \ \ \ -\mathbb{G}(x,\xi)p\cdot\nu\times\curl(\Im\mathbb{G}(\xi,y)q)\Big)ds(\xi).
\een
Since $\Im\mathbb{G}(\xi,y)q=\frac 1{2\i}(\mathbb{G}(\xi,y)q-\overline{\mathbb{G}(\xi,y)}q)$, we know the lemma follows if we can prove, for any $x,y\in\pa\mathcal D$,
\be\label{ccc}
\fl\ \ \int_{\pa D}\Big(\nu\times\curl(\mathbb{G}(x,\xi)p)\cdot\mathbb{G}(\xi,y)q-\mathbb{G}(x,\xi)p\cdot\nu\times\curl(\mathbb{G}(\xi,y)q)\Big)ds(\xi)=0.
\ee
Let $B_R$ be a ball of radius $R>0$ such that $\bar{\mathcal D}\subset B_R$. Since $x,y\in\mathcal D$, $\mathbb{G}(x,\cdot)p$ and $\mathbb{G}(\cdot,y)q$ satisfy
the Maxwell equation in $B_R\backslash\bar{D}$. By integration by parts we have
\ben
\fl& &\int_{\pa D}\Big(\nu\times\curl(\mathbb{G}(x,\xi)p)\cdot\mathbb{G}(\xi,y)q-\mathbb{G}(x,\xi)p\cdot\nu\times\curl(\mathbb{G}(\xi,y)q)\Big)ds(\xi)\\
\fl&=&\int_{\pa B_R}\Big(\nu\times\curl(\mathbb{G}(x,\xi)p)\cdot\mathbb{G}(\xi,y)q-\mathbb{G}(x,\xi)p\cdot\nu\times\curl(\mathbb{G}(\xi,y)q)\Big)ds(\xi)\\
\fl&=&\int_{\pa B_R}\Big(\mathbb{G}(x,\xi)p\cdot(\curl(\mathbb{G}(\xi,y)q)\times\hat x-\i k\,\mathbb{G}(\xi,y)q)\\
\fl& &\ \ \ \ -(\curl(\mathbb{G}(x,\xi)p)\times\hat x-\i k\,\mathbb{G}(x,\xi)p)\cdot\mathbb{G}(\xi,y)q\Big)ds(\xi).
\een
This show the desired identity (\ref{ccc}) by letting $R\to\infty$ and using the asymptotic relations $\mathbb{G}(x,\xi)p=O(|\xi|^{-1})$ and $\curl(\mathbb{G}(x,\xi)p)\times\hat x-\i k\,\mathbb{G}(x,\xi)p=O(|\xi|^{-2})$ as $|\xi|\to\infty$ (see e.g., \cite[Theorem 5.2.2]{nec01}). This completes the proof.
\finproof

The following corollary of the Helmholtz-Kirchhoff identity plays a key role in our analysis.

\begin{lem}\label{lem:3.2}
We have
\ben
k\int_{\Ga_r}\overline{\mathbb{G}(x,x_r)}^T\mathbb{G}(x_r,z)ds(x_r)=\Im\mathbb{G}(x,z)+\mathbb{W}_r(x,z)\ \ \ \ \forall x,z\in\Om,
\een
where $|w_r^{ij}(x,z)|+|\na_x w_r^{ij}(x,z)|\le CR^{-1}_r$ uniformly for any $x,z\in\Om$. Here $w_r^{ij}(x,z)$ is the $(i,j)$-element of the matrix $\mathbb{W}_r(x,z)$, $i,j=1,2,3$.
\end{lem}

\debproof We use the following asymptotic relations
\ben
\mathbb{G}(x,x_r)p=O(R_r^{-1}),\ \ \ \ \curl(\mathbb{G}(x,x_r)p)\times\hat x-\i k\mathbb{G}(x,x_r)p=O(R_r^{-2}),
\een
and Lemma \ref{lem:3.1} to obtain that for any $p,q\in\R^3$,
\ben
\fl\qquad k\int_{\Ga_r}\overline{\mathbb{G}(x,x_r)}p\cdot\mathbb{G}(x_r,z)qds(x_r)=p\cdot\Im\mathbb{G}(x,z)q+O(R_r^{-1})\ \ \ \ \forall x,z\in\Om.
\een
This shows the estimate for $|w^{ij}_r(x,z)|$. The estimate for $|\na_xw^{ij}_r(x,z)|$ can be proved similarly by using the following asymptotic relations:
\ben
\fl\qquad \frac{\pa}{\pa x_j}(\mathbb{G}(x,x_r)p)=O(R_r^{-1}),\ \ \frac{\pa}{\pa x_j}\Big(\curl(\mathbb{G}(x,x_r)p)\times\hat x-\i k\mathbb{G}(x,x_r)p\Big)=O(R_r^{-2}),
\een
for any $x\in\Om,x_r\in\Ga_r$, $j=1,2,3$. This completes the proof.
\finproof

Similarly we can prove the following lemma by using the Helmholtz-Kirchhoff identity for the Helmholtz equation.

\begin{lem}\label{lem:3.4}
We have
\ben
& &k\int_{\Ga_s}\overline{g(z,x_s)}{\mathbb{G}(x,x_s)}\,ds(x_s)=\Im\mathbb{G}(x,z)+\mathbb{W}_s(x,z)\ \ \ \ \forall x,z\in\Om,
\een
where $|w_s^{ij}(x,z)|+|\na_x w_s^{ij}(x,z)|\le CR^{-1}_s$ uniformly for any $x,z\in\Om$. Here $w_s^{ij}(x,z)$ is the $(i,j)$-element of the  matrix $\mathbb{W}_s(x,z)$, $i,j=1,2,3$.
\end{lem}
\debproof
By (\ref{gl}), we know that for $x,z\in\Om$,
\ben
\int_{\Ga_s}\overline{g(z,x_s)}{\mathbb{G}(x,x_s)}\,ds(x_s)=
(\mathbb{I}+\frac{\nabla_x\nabla_x}{k^2})\int_{\Ga_s}\overline{g(z,x_s)}g(x,x_s)\,ds(x_s).
%&=&\int_{\Ga_s}\overline{g(z,x_s)}g(x,x_s)\,ds(x_s)+\int_{\Ga_s}\overline{g(z,x_s)}{\frac{\nabla_x\nabla_x}{k^2}g(x,x_s)}\,ds(x_s)\\
\een
By \cite[Lemma 3.2]{cch} we have
\ben
k\int_{\Ga_s}\overline{g(z,x_s)}g(x,x_s)\,ds(x_s)=\Im g(x,z)+w_s(x,z)\ \ \ \ \forall x,z\in\Om,
\een
where $|w_s(x,z)|+|\na_x w_s(x,z)|\le CR_s^{-1}$ uniformly in $x,z\in\Om$. It is easy to show that we also have $|\pa^2 w_s(x,z)/\pa x_i\pa x_j|
+|\pa^3 w_s(x,z)/\pa x_i\pa x_j\pa x_k|\le CR_s^{-1}$ uniformly in $x,z\in\Om$, $i,j,k=1,2,3$. This completes the proof.
\finproof

Now we recall the definition of the Dirichlet-to-Neumann mapping $G_e:H^{-1/2}(\div;\Ga_D)\to H^{-1/2}(\div;\Ga_D)$ for Maxwell scattering problems (see e.g., \cite{monk}). For any $g\in H^{-1/2}(\div;\Ga_D)$, $G_e(g)=\nu\times\curl U$, where $U\in H_{\rm loc}(\curl;\R^3\backslash\bar D)$ is the solution of the following scattering problem:
\be
& &\curl\curl U-k^2U=0\ \ \ \ \mbox{in }\R^3\backslash\bar D,\label{u1}\\
& &\nu\times U=g\ \ \mbox{on }\Ga_D,\ \ \ \ r\left(\curl U\times\hat x-\i kU\right)\to 0\ \ \ \ \mbox{as }r\to\infty.\label{u2}
\ee
The far field pattern $U_\infty(\hat x)$ of the solution $U$ to the scattering problem (\ref{u1})-(\ref{u2}) is defined by the asymptotic behavior
\be
U(x)=\frac{e^{\i k|x|}}{|x|}\left\{U_\infty(\hat x)+O\left(\frac 1{|x|}\right)\right\},\ \ \ \ |x|\to\infty,\label{u3}
\ee
where $\hat x=x/|x|\in S^2:=\{x\in\R^3:|x|=1\}$.

\begin{lem}\label{lem:3.3}
Let $g\in H^{-1/2}(\div;\Ga_D)$ and $U$ be the radiation solution satisfying (\ref{u1})-(\ref{u2}); then
\ben
\Im\la g\times\nu,G_e(g)\ra_{\Ga_D}=k\int_{S^2}|U_\infty(\hat x)|^2d\hat x\ge 0,
\een
where $\la\cdot,\cdot\ra_{\Ga_D}$ is the duality pairing between $H^{-1/2}(\curl;\Ga_D)$ and $H^{-1/2}(\div;\Ga_D)$.
\end{lem}

\debproof We first remark that for the solution $U$ of the problem (\ref{u1})-(\ref{u2}), $g\times\nu=\nu\times U|_{\Ga_D}\times\nu\in H^{-1/2}(\curl;\Ga_D)$, the dual space of $H^{-1/2}(\div;\Ga_D)$ (see e.g., \cite[Theorem 5.4.2]{nec01} for smooth domains and \cite[Lemma 5.6]{buffa} for Lipschitz domains). Let $B_R$ be a ball of radius $R$ that includes $D$. By integrating by parts one easily obtains
\ben
\la g\times\nu,G_e(g)\ra_{\Ga_D}&=&\la U,\nu\times\curl U \ra_{\Ga_D}\\
&=&\int_{B_R\backslash\bar D}(|\curl U|^2-k^2|U|^2)dx+\int_{\Ga_R}U\cdot\hat x\times\curl\bar Uds(x).
\een
Thus by the Silver-M\"uller radiation condition
\ben
\fl\qquad\Im\la g\times\nu,G_e(g)\ra_{\Ga_D}=\lim_{R\to\infty}\Im\int_{\Ga_R}U\cdot\hat x\times\curl\bar Uds(x)=\lim_{R\to\infty}k\int_{\Ga_R}|U|^2ds(x).
\een
This completes the proof by (\ref{u3}).
\finproof

The following stability estimate for the forward scattering problem can be found in \cite[Theorem 4.2]{LSM-book} and \cite{kir-mon}.

\begin{lem}\label{L:3.5}
Assume that $n(x)$ is positive and piecewise smooth in $D$ and $f\in L^2(\mathbb{R}^3)$ has compact support. The the following problem
\ben
\curl\curl U-k^2n(x) U =f(x) \ \ \ \ \mbox{in } \mathbb{R}^3, \\
r\left(\curl U\times\hat x-\i kU\right)\to 0\ \ \ \ \mbox{as }r\to\infty,
\een
has a unique solution $U\in H_{\rm loc}(\curl;\R^3)$. Moreover, the solution satisfies
$\|U\|_{H({\rm curl};D)}\leq C\|f\|_{L^2(\mathbb{R}^3)}$
for some constant $C$ independent of $f$.
\end{lem}

The following theorem on the resolution of the RTM algorithm for penetrable scatterers is the first main result of this paper.

\begin{thm}\label{thm:3.1}
For any $z\in\Om$, let $\Psi(x,z)$ be the radiation solution of the Maxwell scattering problem
\be\label{ps3}
\fl\qquad\curl\curl \Psi(x,z)-k^2n(x)\Psi(x,z)=k^2(n(x)-1)\Im\mathbb{G}(x,z)p\ \ \ \ \mbox{\rm in }\R^3.
\ee
Then if the measured field $E^s=E-E^i$ and $E$ satisfies (\ref{p1})-(\ref{p2}), we have
\ben
\hat I(z)=k\int_{S^2}|\Psi_\infty(\hat x,z)|^2d\hat x+w_{\hat I}(z),
\een
where $\|w_{\hat I}\|_{L^\infty(\Om)}\le C(R^{-1}_s+R^{-1}_r)$.
\end{thm}

\debproof By (\ref{cor3}) we know that for any $z\in\Om$,
\be\label{v1}
\hat I(z)=-k^2\cdot\Im\int_{\Ga_s}g(z,x_s)p\cdot F_b(z,x_s)ds(x_s),
\ee
where $F_b(z,x_s)$ is the back-propagated field
\ben
F_b(z,x_s)=\int_{\Ga_r}\mathbb{G}(z,x_r)^T\overline{E^s(x_r,x_s)}ds(x_r).
\een
It is easy to see that $E^s(x,x_s)$ satisfies
\ben
\curl\curl E^s(x,x_s)-k^2E^s(x,x_s)=k^2(n(x)-1)E(x,x_s),
\een
which implies by using the dyadic Green function that
\be\label{v2}
E^s(x_r,x_s)=\int_{D}k^2(n(x)-1)\mathbb{G}(x_r,x)^TE(x,x_s)dx.
\ee
By Lemma \ref{lem:3.2}
\ben
F_b(z,x_s)&=&\int_D\int_{\Ga_r}k^2(n(x)-1)\mathbb{G}(z,x_r)^T\overline{\mathbb{G}(x_r,x)}\,\overline{E(x,x_s)}ds(x_r)dx\\
&=&\frac 1k\int_Dk^2(n(x)-1)\big(\Im\mathbb{G}(x,z)+\overline{\mathbb{W}_r(x,z)}\big)\overline{E(x,x_s)}dx,
\een
where we have used the fact that $\mathbb{G}(x_r,x)$ is symmetric in the first equality. From (\ref{v1}) we have then
\be\label{v3}
\hat I(z)=-k\ \Im\int_Dk^2(n(x)-1)p\cdot\big(\Im\mathbb{G}(x,z)+\overline{\mathbb{W}_r(x,z)}\big)v(x,z)dx,
\ee
where $v(x,z)=k\int_{\Ga_s}g(z,x_s)\overline{E(x,x_s)}ds(x_s)$. Since $E(x,x_s)=\mathbb{G}(x,x_s)p+E^s(x,x_s)$, we obtain by
Lemma \ref{lem:3.4} that
\ben
v(x,z)=\big(\Im\mathbb{G}(x,z)+\overline{\mathbb{W}_s(x,z)}\big)p+k\int_{\Ga_s}g(z,x_s)\overline{E^s(x,x_s)}ds(x_s).
\een
Denote $w(x,z)=k\int_{\Ga_s}g(z,x_s)\overline{E^s(x,x_s)}ds(x_s)$. Since $E^s(x,x_s)$ satisfies
\ben
\curl\curl E^s(x,x_s)-k^2n(x)E^s(x,x_s)=k^2(n(x)-1))\mathbb{G}(x,x_s)p,
\een
we know that $\overline{w(x,z)}$ satisfies
\ben
\fl\quad\curl\curl\overline{w(x,z)}-k^2n(x)\overline{w(x,z)}&=&k\int_{\Ga_s}\overline{g(z,x_s)}[k^2(n(x)-1)\mathbb{G}(x,x_s)p]ds(x_s)\\
&=&k^2(n(x)-1)\big(\Im\mathbb{G}(x,z)+{\mathbb{W}_s(x,z)}\big)p\ \ \ \ \mbox{in }\R^3,
\een
where we have used Lemma \ref{lem:3.4} again in the last equality. Now from (\ref{ps3}) we know that $\zeta(x,z):=\overline{w(x,z)}-\Psi(x,z)$ satisfies
\ben
\curl\curl\zeta(x,z)-k^2n(x)\zeta(x,z)=k^2(n(x)-1)\mathbb{W}_s(x,z)p\ \ \ \ \mbox{in }\R^3,
\een
and the Silver-M\"uller radiation condition. By Lemma \ref{L:3.5} we obtain
\ben
\|\zeta(\cdot,z)\|_{H(\curl;D)}\le C\|k^2(n(\cdot)-1){\mathbb{W}_s(\cdot,z)p}\|_{L^2(D)}\le CR_s^{-1},
\een
where we have used Lemma \ref{lem:3.4}. This implies that
\ben
v(x,z)&=&w(x,z)+\big(\Im\mathbb{G}(x,z)+\overline{{\mathbb{W}_s(x,z)}}\big)p\\
&=&\overline{\Psi(x,z)}+\overline{\zeta(x,z)}+\big(\Im\mathbb{G}(x,z)+\overline{\mathbb{W}_s(x,z)}\big)p,
\een
where $\|\zeta(\cdot,z)\|_{L^2(D)}+\|{\mathbb{W}_s(\cdot,z)}p\|_{L^2(D)}\le CR^{-1}_s$. Now by (\ref{v3}) we obtain
\ben
\fl\quad\hat I(z)&=&-\Im\int_Dk^2(n(x)-1)p\cdot\Im\mathbb{G}(x,z)(\overline{\Psi(x,z)}+\Im\mathbb{G}(x,z)p)dx+O(R^{-1}_s+R^{-1}_r)\\
\fl &=&-\Im\int_Dk^2(n(x)-1)\Im\mathbb{G}(x,z)p\cdot\overline{\Psi(x,z)}dx+O(R^{-1}_s+R^{-1}_r).
\een
Now by (\ref{ps3}) and integrating by parts we have
\ben
& &-\Im\int_Dk^2(n(x)-1)\Im\mathbb{G}(x,z)p\cdot\overline{\Psi(x,z)}dx\\
&=&-\int_D\big(\curl\curl\Psi(x,z)-k^2n(x)\Psi(x,z)\big)\overline{\Psi(x,z)}dx\\
&=&-\ \Im\int_{\Ga_D}\nu\times\curl\Psi(x,z)\cdot\overline{\Psi(x,z)}dx\\
&=&\Im\int_{\Ga_D}\Psi(x,z)\cdot\nu\times\overline{\curl\Psi(x,z)}dx.
\een
This completes the proof by Lemma \ref{lem:3.3}.
\finproof

%\begin{rem}\label{rem:3.1}
%Denote $Tr(A)$ the trace of matrix $A$, for any vectors $x,y$, it is easy to check that $x^Ty=Tr(xy^T)$. Instead $\mathbb{G}%(x,x_s)p$ of $g(x,x_s)p$  in (\ref{cor3}), (\ref{v3}) becomes
%\ben
%\hat I(z)=-k\ \ Tr( Im\int_Dk^2(n(x)-1)\big(\Im\mathbb{G}(x,z)+\overline{\mathbb{W}_r(x,z)}\big)p \tilde{v}(x,z)dx),
%\een
%where $\tilde{v}=\int_{\Gamma_s}\overline{E(z,x_s)}^T\mathbb{G}(x,x_s)ds(x_s)$. Noting that
%$\sum_{l=1}^3e_le_l^T=\mathbb{I}$. By carefully calculations, we have
%\ben
%\sum^3_{l=1}\hat{I}(z)=k\sum^3_{l=1}\int_{S^2}|\Psi_{l,\infty}(\hat x,z)|^2d\hat x+w_{\hat I}(z),
%\een
%where $\Psi_l$ is defined by (\ref{ps3}) corresponding to $e_l$ for $l=1,2,3$.
%\end{rem}

Noticing that
\ben
\hskip-1cm\curl\curl(\Im\mathbb{G}(x,z)p)-k^2n(x)(\Im\mathbb{G}(x,z)p)=k^2(1-n(x))(\Im\mathbb{G}(x,z)p),
\een
we know that $\Psi(x,z)$ is the radiation solution of the Maxwell equation with the incident wave $\Im\mathbb{G}(x,z)p$. It is known that
$\Im\mathbb{G}(x,z)p=\frac k{4\pi}[(\mathbb{I}+\frac{\na_x\na_x}{k^2})j_0(k|x-z|)]p$ which peaks when $x=z$ and decays as $|x-z|$ becomes large.
It is clear that the source in (\ref{ps3}) is supported in $D$ since $n(x)=1$ outside $D$. Thus the source becomes small when $z$ moves away from
$\pa D$ outside the scatterer. On the other hand, the source will not be small when $z$ is inside $D$.
Therefore we expect that the imaging functional will have a contrast at the boundary of the scatterer $D$ and decay away from the scatterer. This is indeed confirmed in our numerical experiments.

Now we consider the resolution of the imaging functional in the case of non-penetrable obstacles. We only prove the results for the case of impedance boundary condition. The case of Dirichlet boundary condition is similar and left to the interested readers.
We need the following result on the forward scattering problem for
non-penetrable scatterers with the impedance boundary condition. It can be proved by adapting the proof in \cite{ccm} for partially coated scatterers or by using the method of limiting absorption principle, see e.g. \cite{leis}.

\begin{lem}\label{L.3.6}
Let $\eta\ge 0$ be bounded on $\Ga_D$ and $g\in L^2(\Ga_D)$. Then the scattering problem
\ben
& &\curl\curl U-k^2 U=0\ \ \ \ \mbox{in }\R^3,\\
& &\nu\times\curl U-\i k\eta(x)(\nu\times U\times \nu)=g\ \ \ \ \mbox{on }\Ga_D,\\
& &r\left(\curl U\times\hat x-\i kU\right)\to 0\ \ \ \ \mbox{as }r=|x|\to\infty,
\een
has a unique solution $U\in H_{\rm loc}(\curl;\R^3\backslash\bar D)$ which satisfies $\|U\|_{H_{\rm loc}(\curl;\R^3\backslash\bar D)}\le C\|g\|_{L^2(\Ga_D)}$
for some constant $C$ independent of $g$.
\end{lem}

\begin{thm}\label{thm:3.2}
For any $z\in\Om$, let $\Psi(x,z)$ be the radiation solution of the Maxwell equation
\be\label{ps1}
\curl\curl \Psi(x,z)-k^2\Psi(x,z)=0\ \ \ \ \mbox{\rm in }\R^3\backslash\bar D
\ee
with the impedance boundary condition
\be\label{ps2}
\fl\quad& &\nu\times\curl\Psi(x,z)-\i k\eta(x)\nu\times\Psi(x,z)\times\nu\nn\\
\fl&=&-\big[\nu\times\curl(\Im\mathbb{G}(x,z)p)-\i k\eta(x)\nu\times(\Im \mathbb{G}(x,z)p)\times\nu\big]\ \ \ \ \mbox{\rm on }\Ga_D.
\ee
Then if the measured field $E^s=E-E^i$ and $E$ satisfies (\ref{p3})-(\ref{p5}) with the impedance condition in (\ref{p4}), we have
\ben
\fl\quad\hat I(z)=k\int_{S^2}|\Psi_\infty(\hat x,z)|^2d\hat x+\ k\int_{\Ga_D}\eta(x)\big|\nu\times(\Psi(x,z)+\Im\mathbb{G}(x,z)p)\times\nu\big|^2ds+w_{\hat I}(z),
\een
where $\|w_{\hat I}\|_{L^\infty(\Om)}\le C(R^{-1}_s+R^{-1}_r)$.
\end{thm}

\debproof By (\ref{cor3}) we know that for any $z\in\Om$,
\be\label{w1}
\hat I(z)=-k^2\cdot\Im\int_{\Ga_s}g(z,x_s)p\cdot F_b(z,x_s)ds(x_s),
\ee
where $F_b(z,x_s)$ is the back-propagated field
\be\label{w2}
F_b(z,x_s)=\int_{\Ga_r}\mathbb{G}(z,x_r)^T\overline{E^s(x_r,x_s)}ds(x_r).
\ee
Since $\curl\curl E^s(x,x_s)-k^2E^s(x,x_s)=0$ in $\R^3\backslash\bar D$, we obtain by the integral representation formula that
\ben
\fl\qquad E^s(x_r,x_s)\cdot e_l=\int_{\Ga_D}\big[g_l(x_r,x)\cdot\nu\times\curl E^s(x,x_s)-\nu\times\curl g_l(x_r,x)\cdot E^s(x,x_s)\big]ds,
\een
where $g_l$ satisfies (\ref{gl}). Now (\ref{w2}) implies that
\ben
F_b(z,x_s)\cdot e_i&=&\int_{\Ga_r}g_i(z,x_r)\cdot\overline{E^s(x_r,x_s)}ds(x_r)\\
&=&\sum_{l=1}^3\int_{\Ga_r}\int_{\Ga_D}g_i(z,x_r)\cdot e_l\Big[\overline{g_l(x_r,x)}\cdot\nu\times\curl \overline{E^s(x,x_s)}\\
& &\qquad\ -\nu\times\curl \overline{g_l(x_r,x)}\cdot\overline{E^s(x,x_s)}\Big]dsds(x_r).
\een
Denote by $g^{ij}(x,y)$ the $(i,j)$-element of the matrix $\mathbb{G}(x,y)$. By Lemma \ref{lem:3.2} we have
\ben
\fl\qquad\sum^3_{l=1}\int_{\Ga_r}(g_i(z,x_r)\cdot e_l)\overline{g_l(x_r,x)}ds(x_r)&=&\sum^3_{l=1}\sum^3_{j=1}\int_{\Ga_r}g^{il}(z,x_r)\overline{g^{jl}(x_r,x)}e_jds(x_r)\\
\fl &=&\frac 1k\left[\Im\mathbb{G}(x,z)+\mathbb{W}_r(x,z)\right]e_i.
\een
Thus
\ben
F_b(z,x_s)\cdot e_i&=&\frac 1k\int_{\Ga_D}\Big\{(\Im\mathbb{G}(x,z)+\mathbb{W}_r(x,z))e_i\cdot\nu\times\curl \overline{E^s(x,x_s)}\\
& &\qquad\ -\nu\times\curl \left[(\Im\mathbb{G}(x,z)+\mathbb{W}_r(x,z))e_i\right]\cdot\overline{E^s(x,x_s)}\Big\}ds.
\een
Substituting above identity into (\ref{w1}) we have
\be\label{w3}
\hat I(z)&=&-k\cdot\Im\int_{\Ga_D}g(z,x_s)\Big\{(\Im\mathbb{G}(x,z)+\mathbb{W}_r(x,z))p\cdot\nu\times\curl {v^s(x,z)}\nn\\
& &\quad-\nu\times\curl \left[(\Im\mathbb{G}(x,z)+\mathbb{W}_r(x,z))p\right]\cdot{v_s(x,z)}\Big\}ds,
\ee
where $v_s(x,z)=k\cdot\int_{\Ga_s}g(z,x_s)\overline{E_s(x,x_s)}ds(x_s)$. By taking the complex conjugate,
\ben
\overline{v_s(x,z)}=k\cdot\int_{\Ga_s}\overline{g(z,x_s)}E^s(x,x_s)ds(x_s).
\een
Thus $\overline{v_s(x,z)}$ is the weighted superposition of the scattered waves $E^s(x,x_s)$. Therefore, $\overline{v_s(x,z)}$ is the radiation solution of the Maxwell equation
\ben
\curl \curl \overline{v_s(x,z)}-k^2\overline{v_s(x,z)}=0\ \ \ \ \mbox{in }\R^3\backslash\bar D
\een
satisfying the impedance condition
\ben
&&\nu\times\curl \overline{v_s(x,z)}-\i k\eta(x)\nu\times\overline{v_s(x,z)}\times\nu\\
&=&-\int_{\Ga_s}\overline{g(z,x_s)}\left[\nu\times\curl (\mathbb{G}(y,x_s)p)-\i k\eta(x)\nu\times(\mathbb{G}(y,x_s)p)\times\nu\right]ds(x_s)\\
&=&-\Big\{\nu\times\curl (\Im\mathbb{G}(x,z)p+\mathbb{W}_s(x,z)p)\\
& &\quad\ -\i k\eta(x)\nu\times(\Im\mathbb{G}(x,z)p+\mathbb{W}_s(x,z)p)\times\nu\Big\}\ \ \ \ \mbox{on }\Ga_D,
\een
where we have used Lemma \ref{lem:3.4} in the last inequality. This implies by (\ref{ps1})-(\ref{ps2}) that $\overline{v_s(x,z)}= (\Psi(x,z)+\zeta(x,z))$,
where $\zeta(x,z)$ satisfies the scattering problem in Lemma \ref{L.3.6} with $g(\cdot)=-(\nu\times\curl(\mathbb{W}_s(\cdot,z)p)-\i k\eta(\cdot)\nu\times(\mathbb{W}_s(\cdot,z)p)
\times\nu)$. By Lemma \ref{lem:3.4} and Lemma \ref{L.3.6}, we know that $\zeta(x,z)$ satisfies the estimate $\|\zeta(\cdot,z)\|_{H_{\rm loc}(\curl;\R^3\backslash\bar D)}\le
C\|g\|_{L^2(\Ga_D)}\le CR^{-1}_s$ uniformly for $z\in\Om$. Substituting $v_s(x,z)= (\overline{\Psi(x,z)}+\overline{\zeta(x,z)})$ into (\ref{w3}) we obtain
\ben
\hat I(z)&=&-\ \Im\int_{\Ga_D}\Big(\Im\mathbb{G}(x,z)p\cdot\nu\times\curl \overline{\Psi(x,z)}\\
& &\quad\ -\nu\times\curl(\Im\mathbb{G}(x,z)p)\cdot\overline{\Psi(x,z)}\Big)ds+O(R_s^{-1}+R_r^{-1})\\
&=&\Im\int_{\Ga_D}\Big(\Im\mathbb{G}(x,z)p\cdot\nu\times\curl {\Psi(x,z)}\\
& &\quad\ -\nu\times\curl(\Im\mathbb{G}(x,z)p)\cdot{\Psi(x,z)}\Big)ds+O(R_s^{-1}+R_r^{-1}).
\een
By (\ref{ps2}) we have
\ben
\fl\qquad& &\Im\int_{\Ga_D}\Big(\Im\mathbb{G}(x,z)p\cdot\nu\times\curl {\Psi(x,z)}
-\nu\times\curl(\Im\mathbb{G}(x,z)p)\cdot{\Psi(x,z)}\Big)ds\\
\fl &=&\Im\int_{\Ga_D}\Big\{\Im\mathbb{G}(x,z)p\cdot\left(\nu\times\curl\Psi(x,z)-\i k\eta(x)\nu\times\Psi(x,z)\times\nu\right)\\
\fl & &\quad\ -\big[\nu\times\curl(\Im\mathbb{G}(x,z)p)+\i k\eta(x)\nu\times(\Im\mathbb{G}(x,z)p)\times\nu)\cdot\Psi(x,z)\big]\\
\fl & &\quad\ +2\i k\eta(x)\nu\times(\Im\mathbb{G}(x,z)p)\times\nu\cdot\nu\times\Psi(x,z)\times\nu\Big\}ds\\
\fl &=&\Im\int_{\Ga_D}\nu\times\curl\overline{\Psi(x,z)}\cdot\Psi(x,z)ds\\
\fl && +k\int_{\Ga_D}\eta(x)\big|\nu\times(\Psi(x,z)+\Im\mathbb{G}(x,z)p)\times\nu\big|^2ds.
\een
This completes the proof by using Lemma \ref{lem:3.3}.
\finproof

\section{Numerical results}\label{section4}

In this section we show several numerical examples to illustrate the performance of the RTM algorithm proposed in this paper.

\subsection{Numerical examples in 2D}

We first show the efficiency of our imaging algorithm in the setting of transverse electric (TE) case, that is, the electromagnetic waves are independent of $x_3$ direction. In this subsection all the vector fields are assumed to be two dimensional. Let $p=(p_1,p_2)^T$ be the polarization direction and $g(x,x_s)=\frac{\i}4H^{(1)}_0(k|x-x_s|)$ be the fundamental solution of the two-dimensional Helmholtz equation with the source at $x_s\in\R^2$. The incident electric field $E^i(x,x_s)=\mathbb{G}(x,x_s)p$, where $\mathbb{G}(x,x_s)=(\mathbb{I}_2+\frac{\na\na}{k^2})g(x,x_s)$ is the two-dimensional dyadic Green function. To obtain the synthetic data for our RTM algorithm, we use the magnetic field integral equation (MFIE) code in \cite{peterson} to obtain the equivalent surface currents then produce the scattering electric field at the receivers. The MFIE integral equations on $\Ga_D$ are solved on a uniform mesh of the boundary with ten points per probe wavelength. The boundaries of the obstacles used in our numerical experiments are parameterized as follows:
\ben
\mbox{Circle:}\ \ \ \ &&x_1=\rho\cos(\theta),\ \ x_2=\rho\sin(\theta),\ \ \theta\in (0,2\pi],\\
\mbox{Kite:}\ \ \ \ &&x_1=\cos(\theta) + 0.65\cos(2\theta) - 0.65,\ \ x_2=1.5 \sin (\theta),\ \ \theta\in (0,2\pi],\\
\mbox{$n$-leaf:}\ \ \ \ &&r(\theta)=1+0.2\cos(n\theta),\ \ \theta\in (0,2\pi].
\een

\bigskip
\textbf{Example 1}.
We first consider the imaging with single polarization $p=(1,0)^T$. The surface elements in 2D case are $|\Delta(x_s)|=|\Ga_s|$, $|\Delta(x_r)|=|\Ga_r|$, where
$|\Ga_s|=2\pi R_s$, $|\Ga_r|=2\pi R_r$.
The sources and receivers are uniformly distributed on a circle with radius 1000. The probe wavelength is $\lam=2\pi/k$.

\begin{figure}
\includegraphics[width=0.4\textwidth]{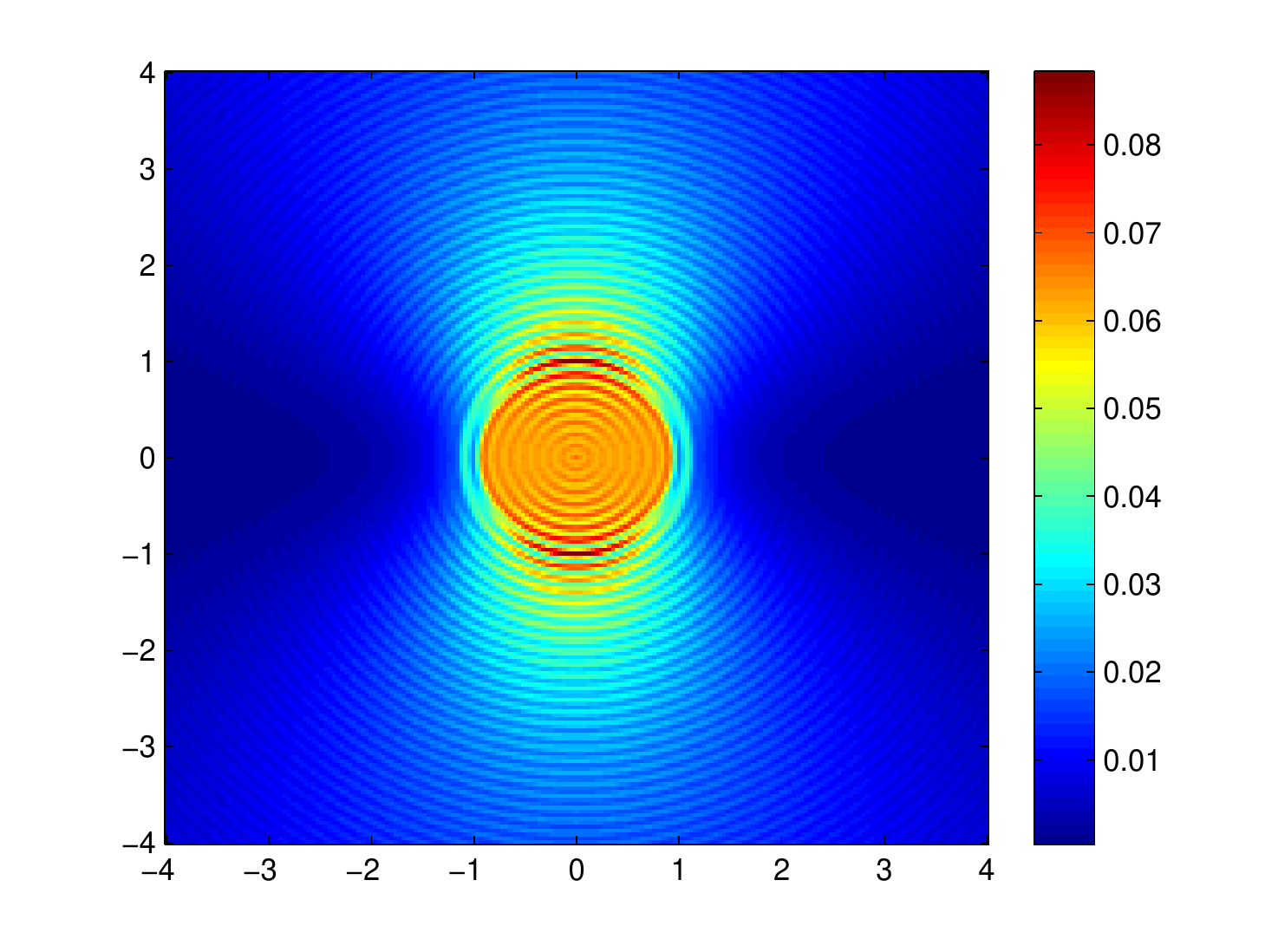} \qquad
\includegraphics[width=0.4\textwidth]{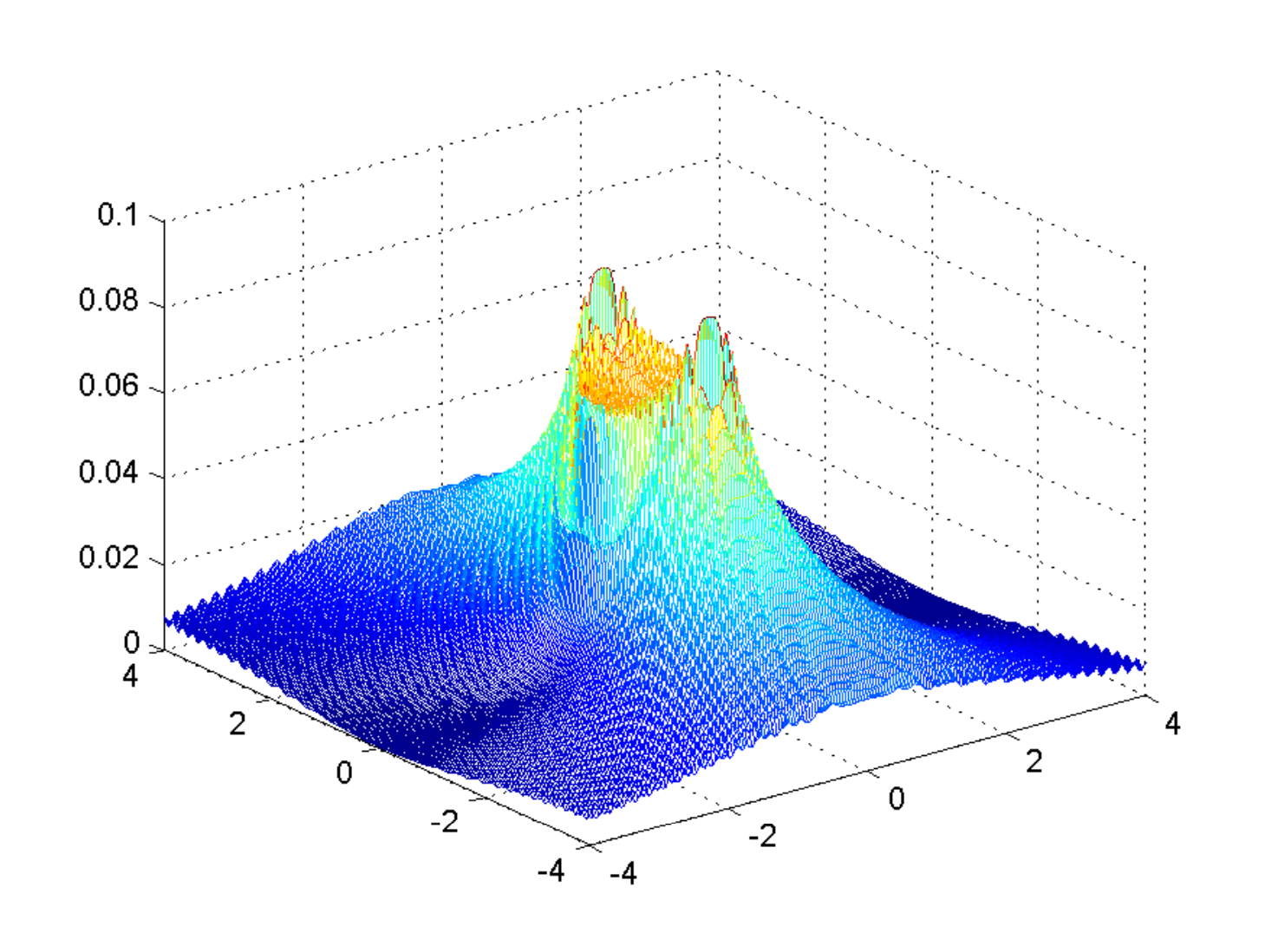}
\caption{The contour (left) and surface (right) plot of the imaging functional. The probe wavelength $\lam=1/4$. $N_s=N_r=256$ .}\label{fig:11}
\end{figure}

\begin{figure}

\includegraphics[width=0.24\textwidth]{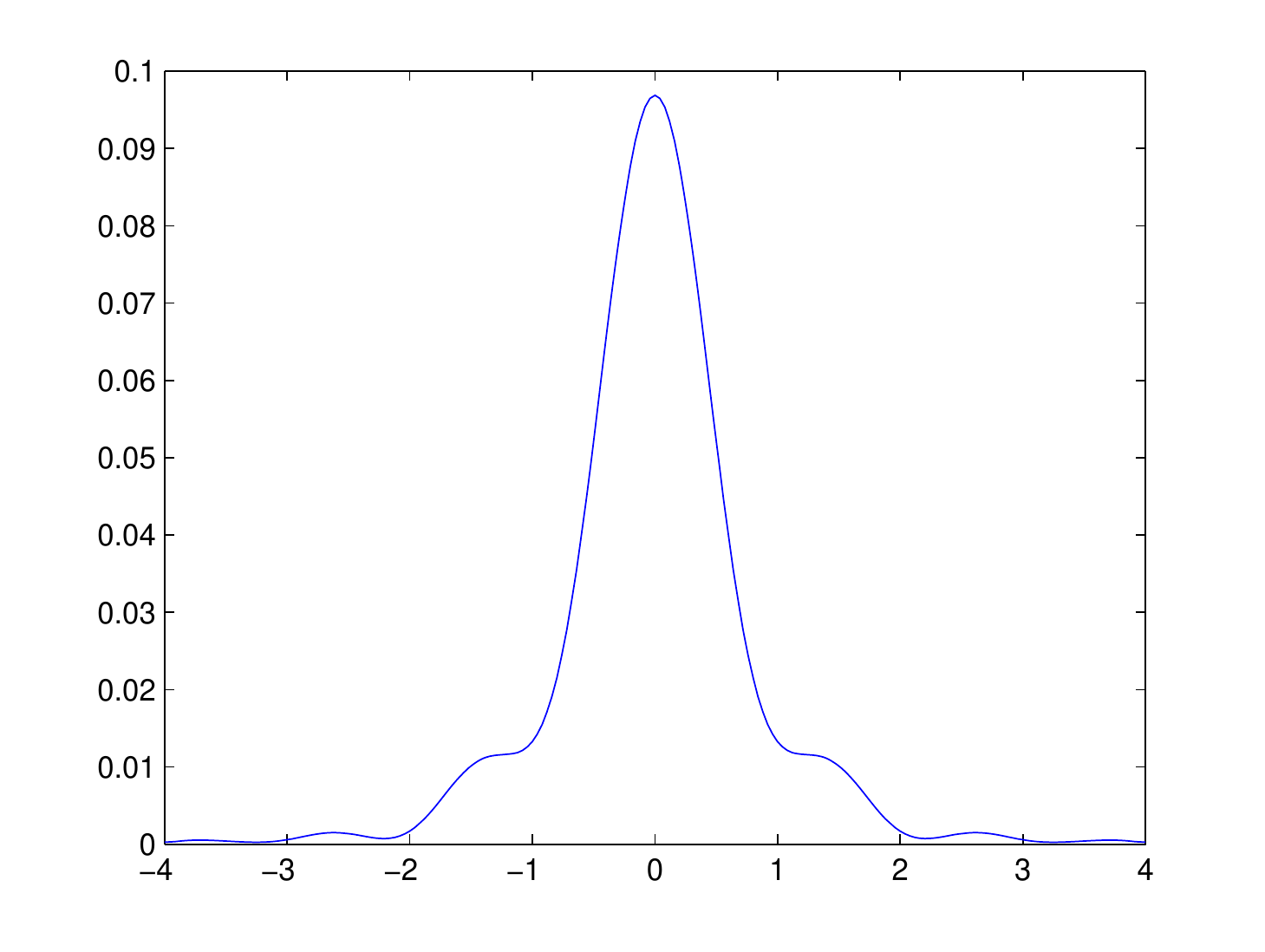}
\includegraphics[width=0.24\textwidth]{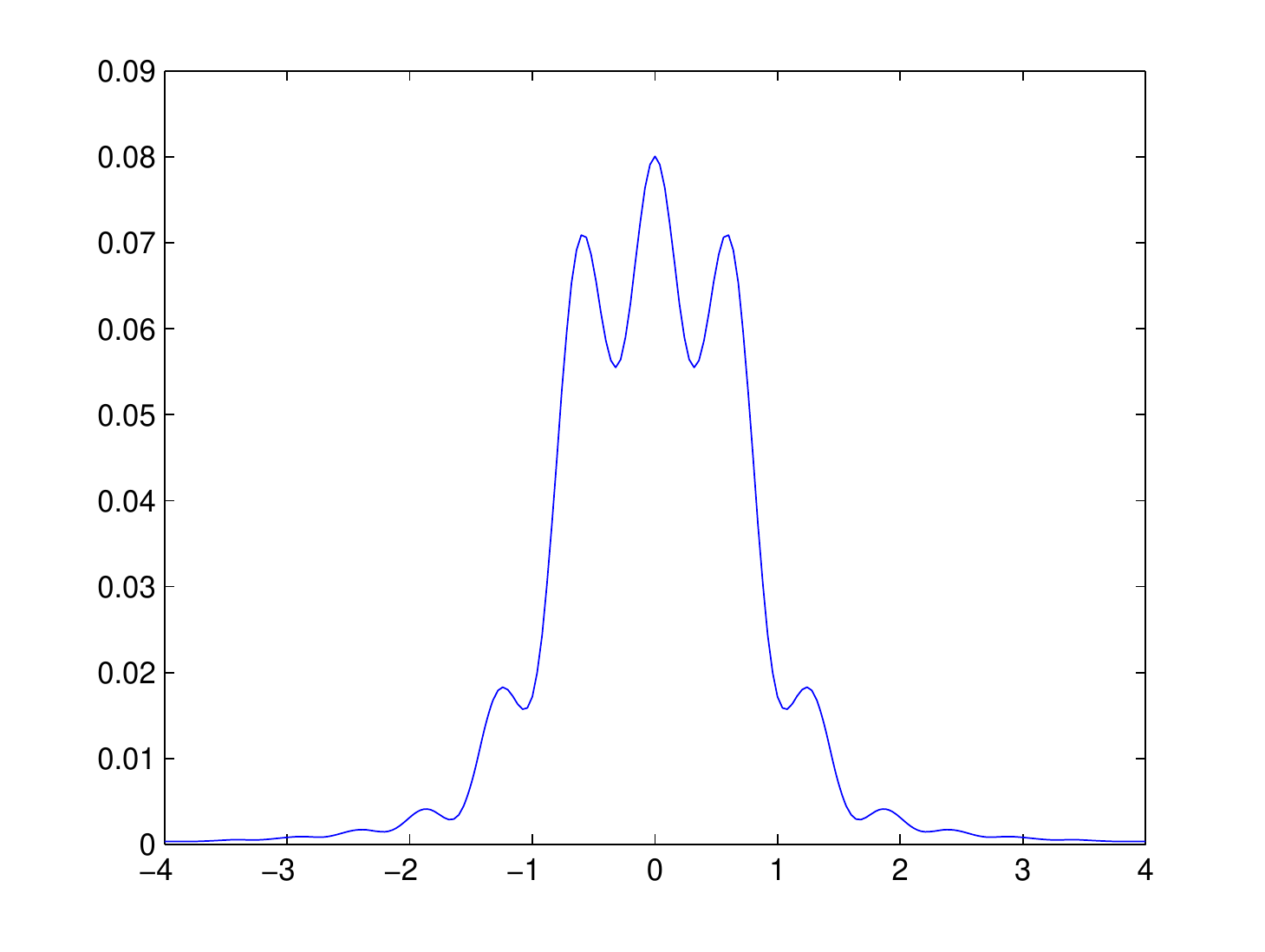}
\includegraphics[width=0.24\textwidth]{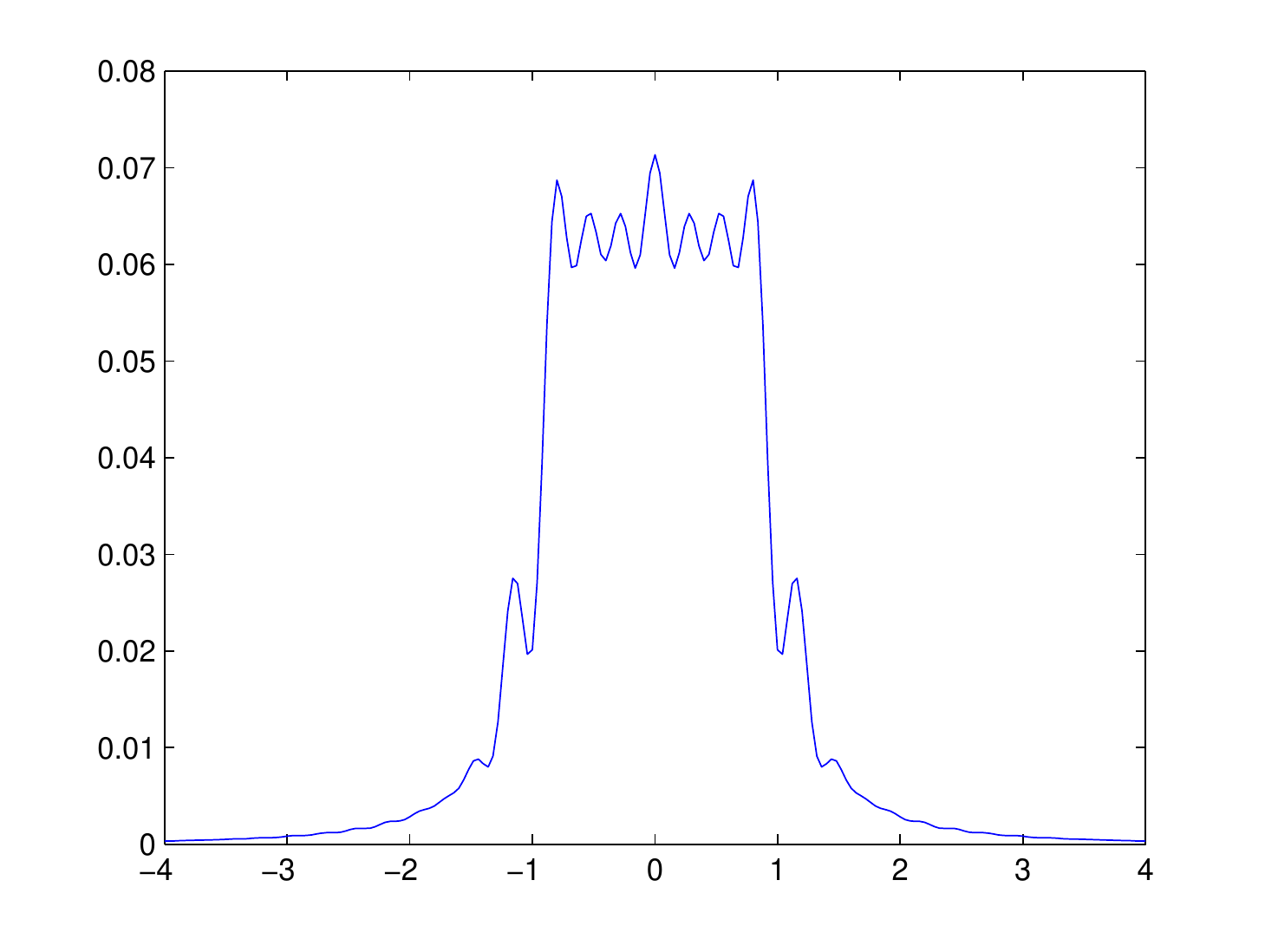}
\includegraphics[width=0.24\textwidth]{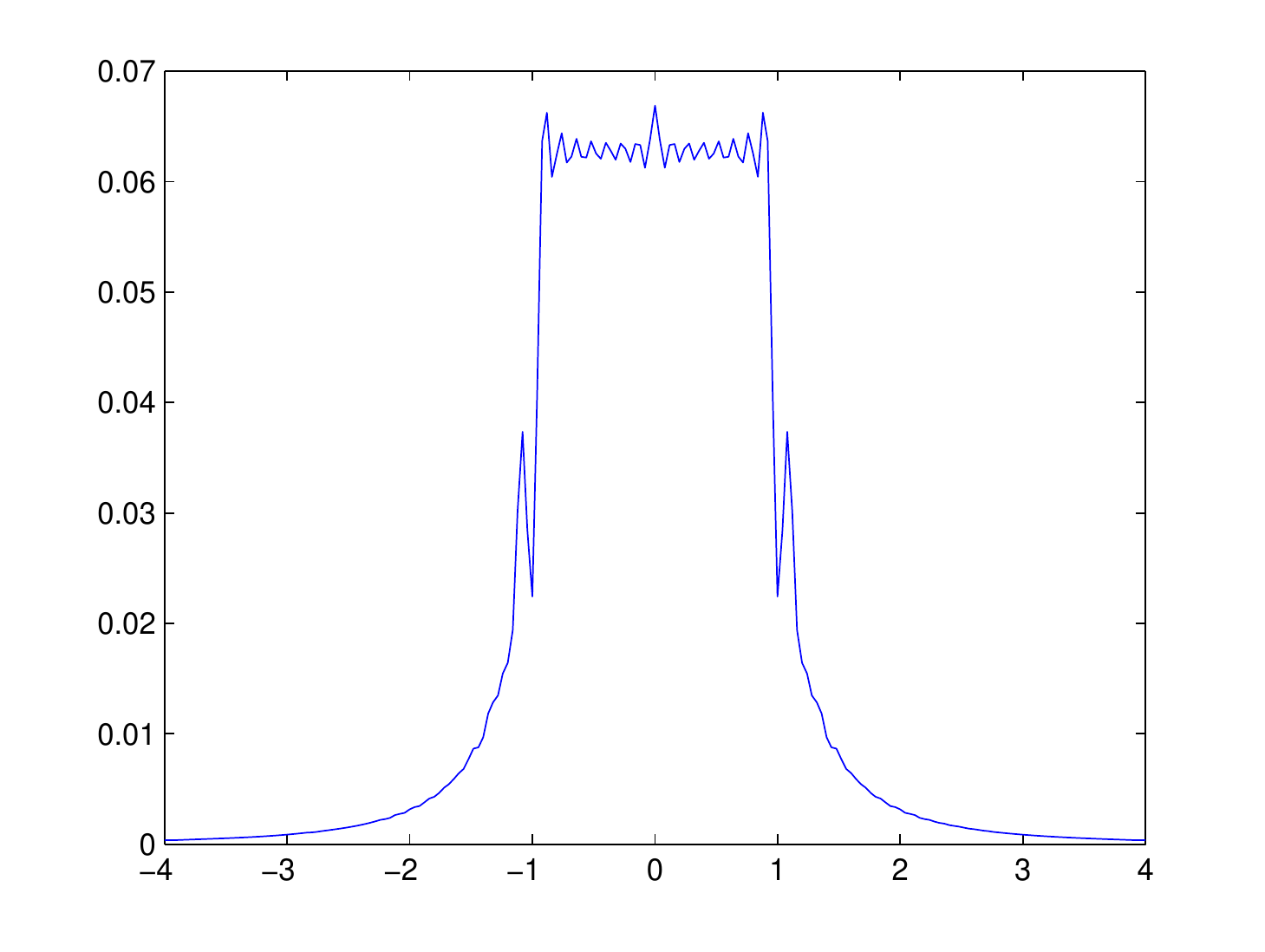}
\caption{The cross section of the imaging functional along $x_1$ axis: the probe wavelength $\lam=2,1,1/2,1/4$ (from left to right).}\label{fig:12}
\end{figure}

Figure \ref{fig:11} shows our imaging algorithm for imaging a perfectly conducting circle of radius $\rho=1$. It shows clearly that the imaging functional can capture the boundary of the scatterer.
Figure \ref{fig:12} shows the cross section of the imaging functional along the $x_1$ axis with different wave numbers. These results confirm that the imaging functional is positive and the oscillation decays with the increase of wave number. The imaging functional captures the boundary of the circle accurately.

\begin{figure}
\includegraphics[width=0.4\textwidth]{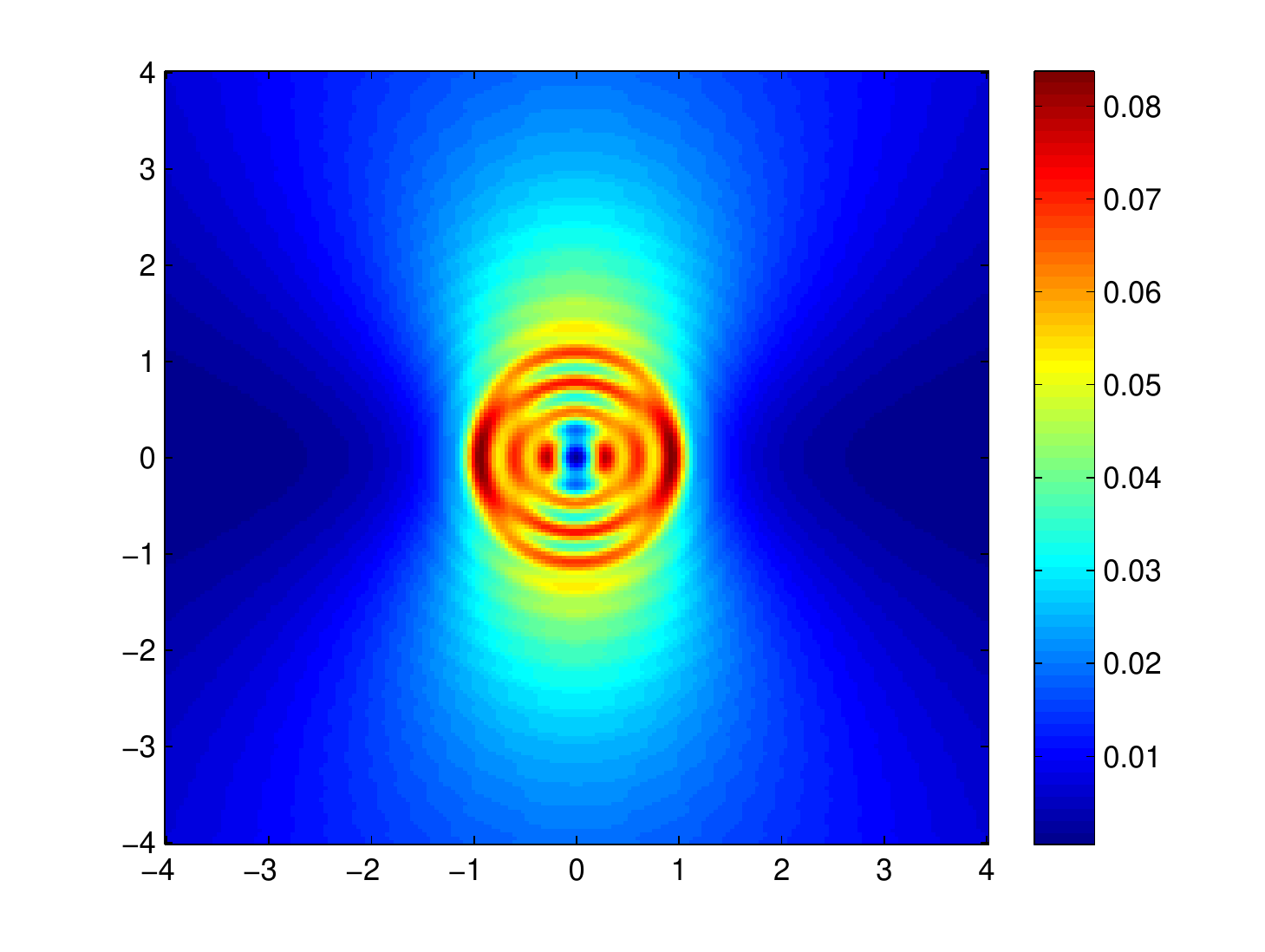} \qquad
\includegraphics[width=0.4\textwidth]{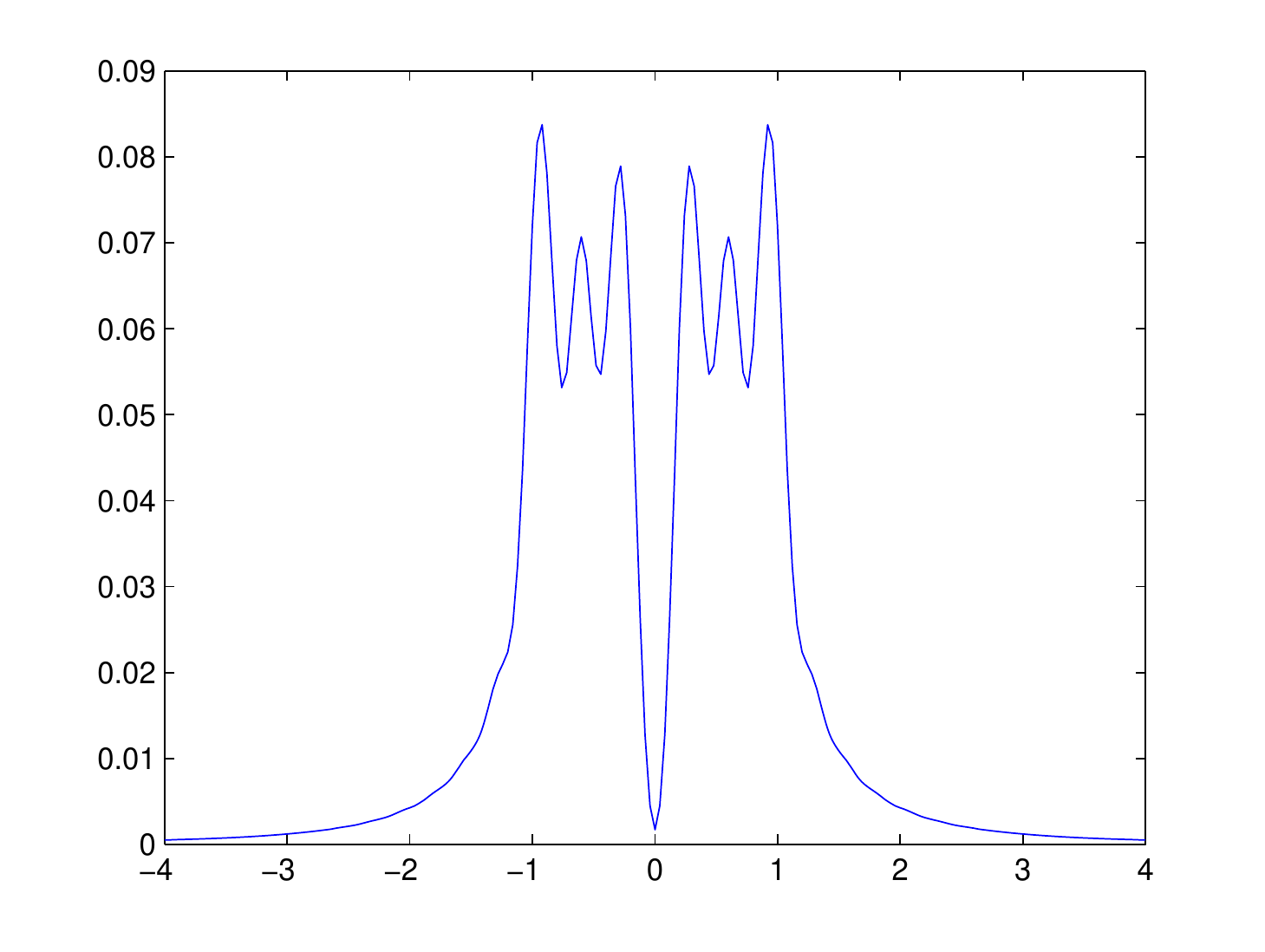}\\
\includegraphics[width=0.4\textwidth]{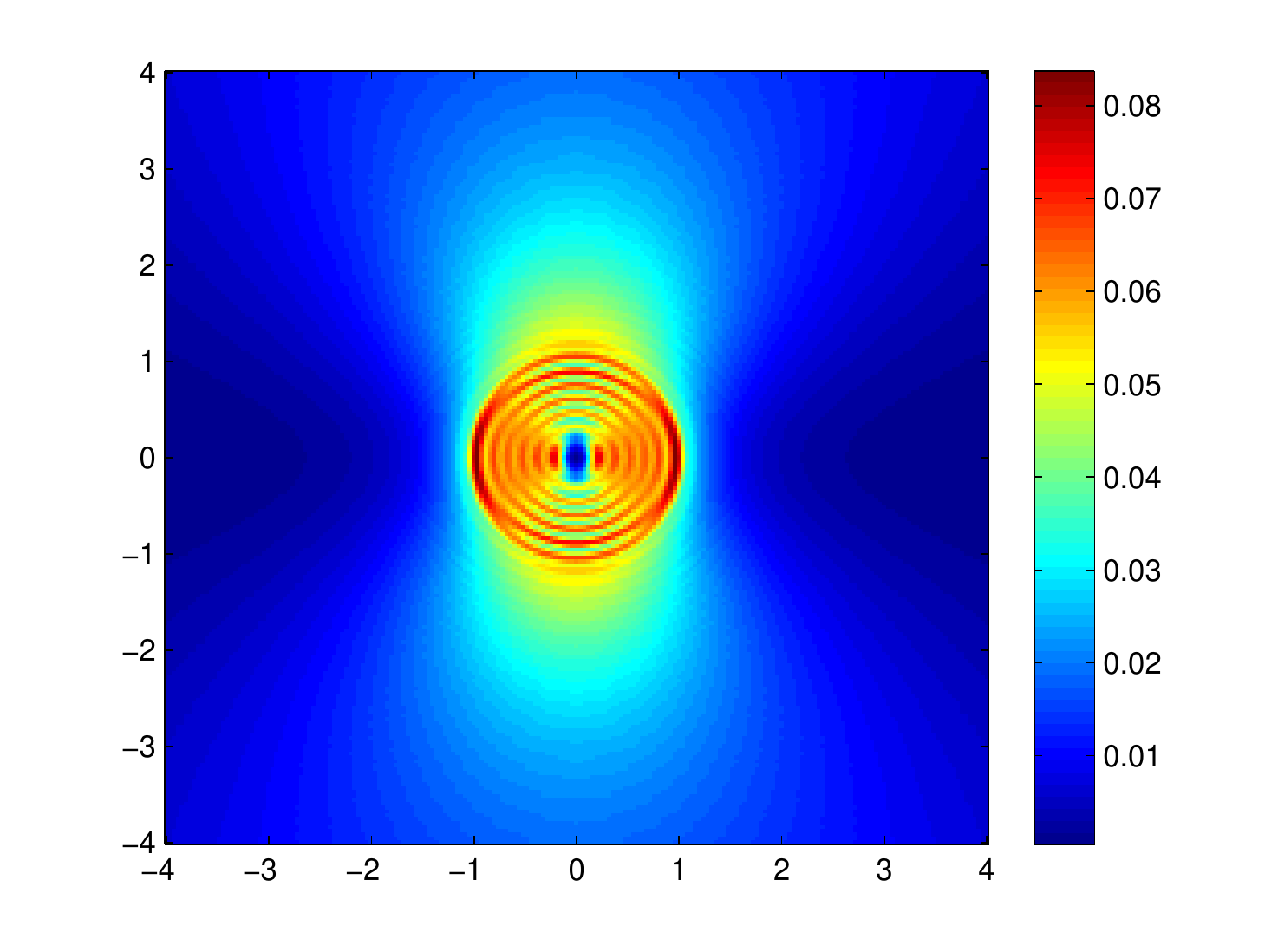} \qquad
\includegraphics[width=0.4\textwidth]{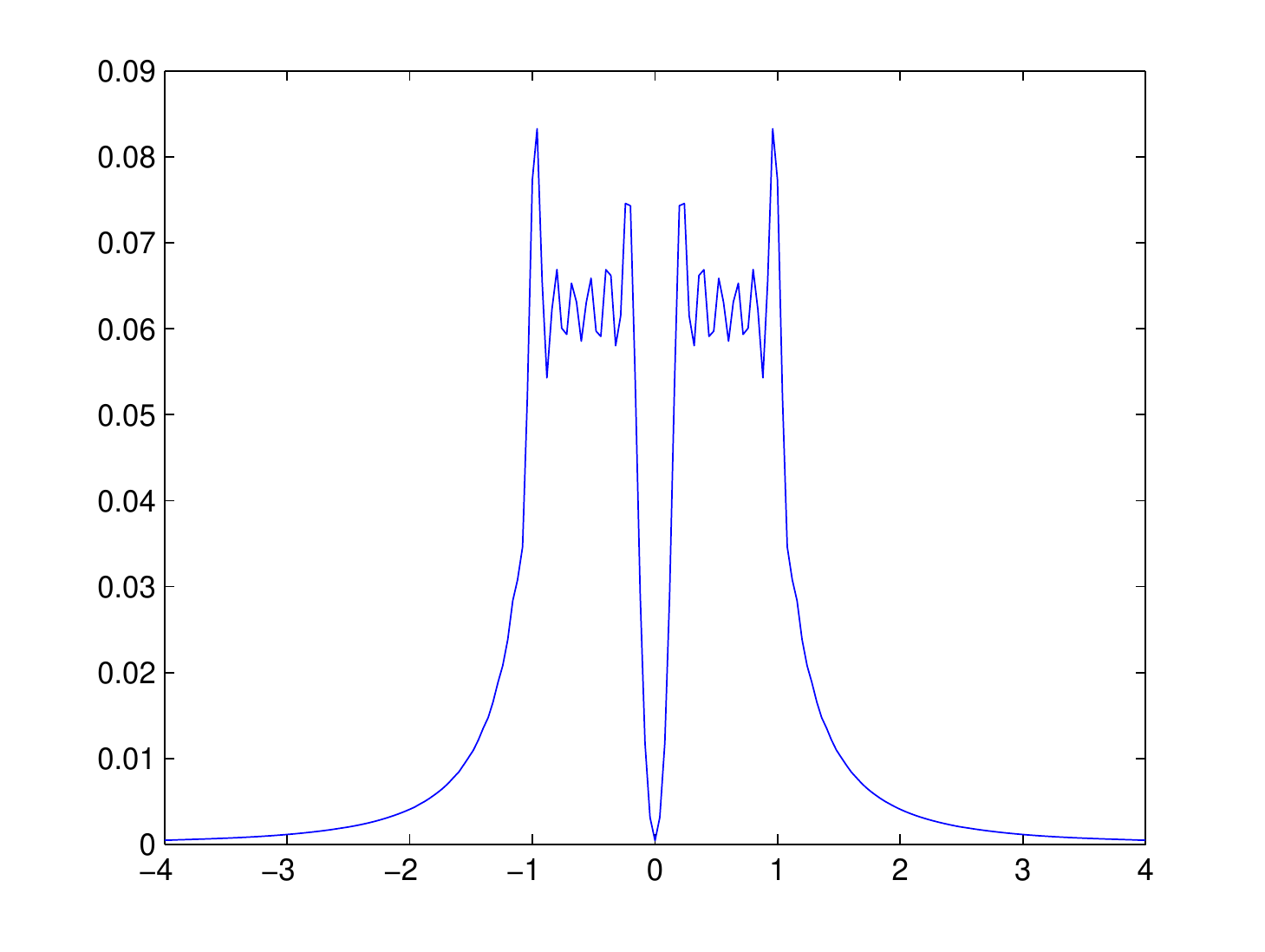}
\caption{The contour plots (left) and the cross section along $x_1=0$ of the imaging functional for the penetrable circular scatterer. The probe wavelength $\lam=1/2$ in the first row and $\lam=1/4$ in the second row. }\label{fig:13}
\end{figure}

Figure \ref{fig:13} shows the outcome of imaging a penetrable circle of radius $\rho=1$ with choice of different wave numbers. The refractive index $n(x)=0.25$. The result shows that our algorithm works well for penetrable scatterers.

\begin{figure}
\includegraphics[width=0.4\textwidth]{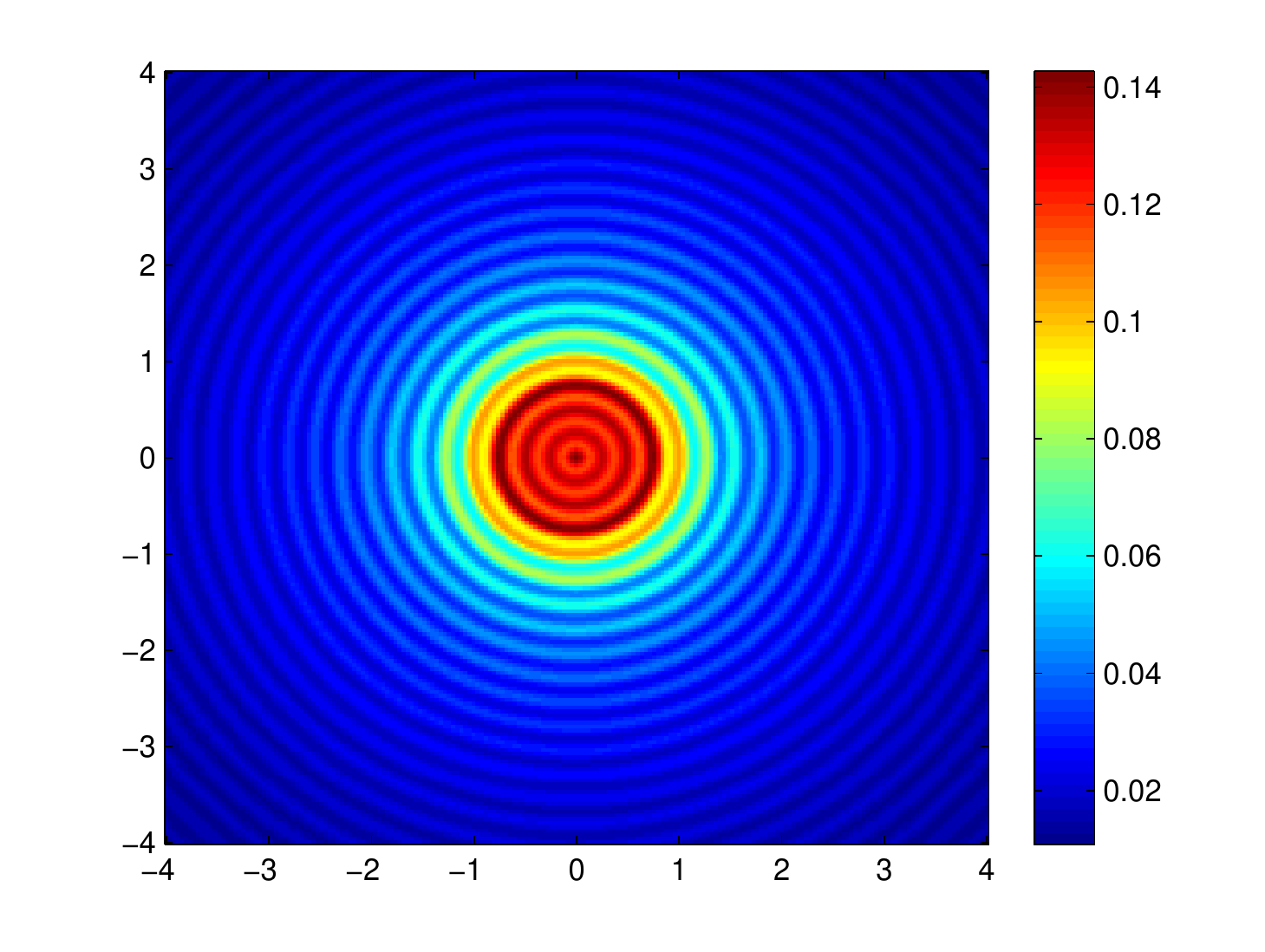} \qquad
\includegraphics[width=0.4\textwidth]{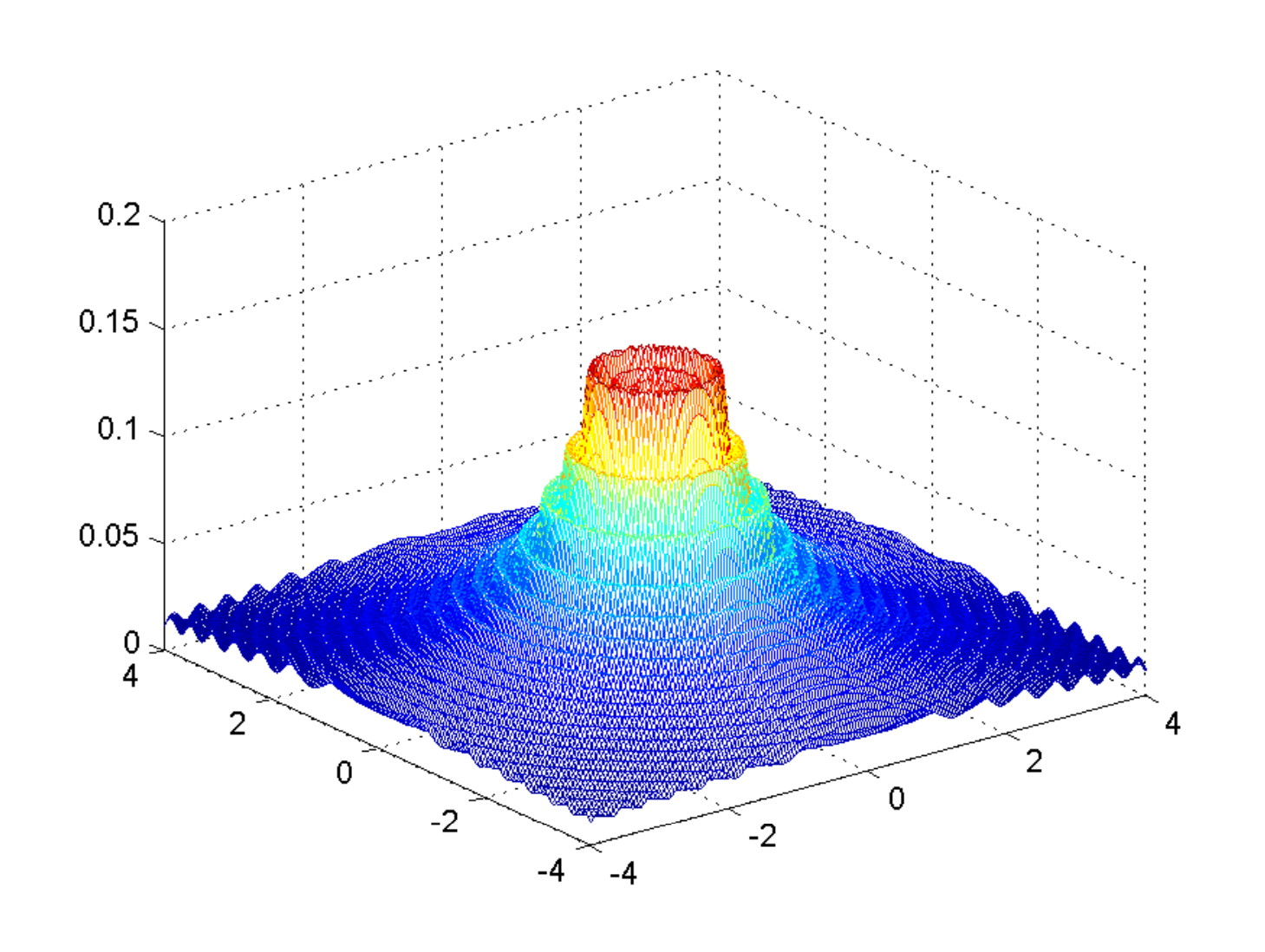}\\
\includegraphics[width=0.4\textwidth]{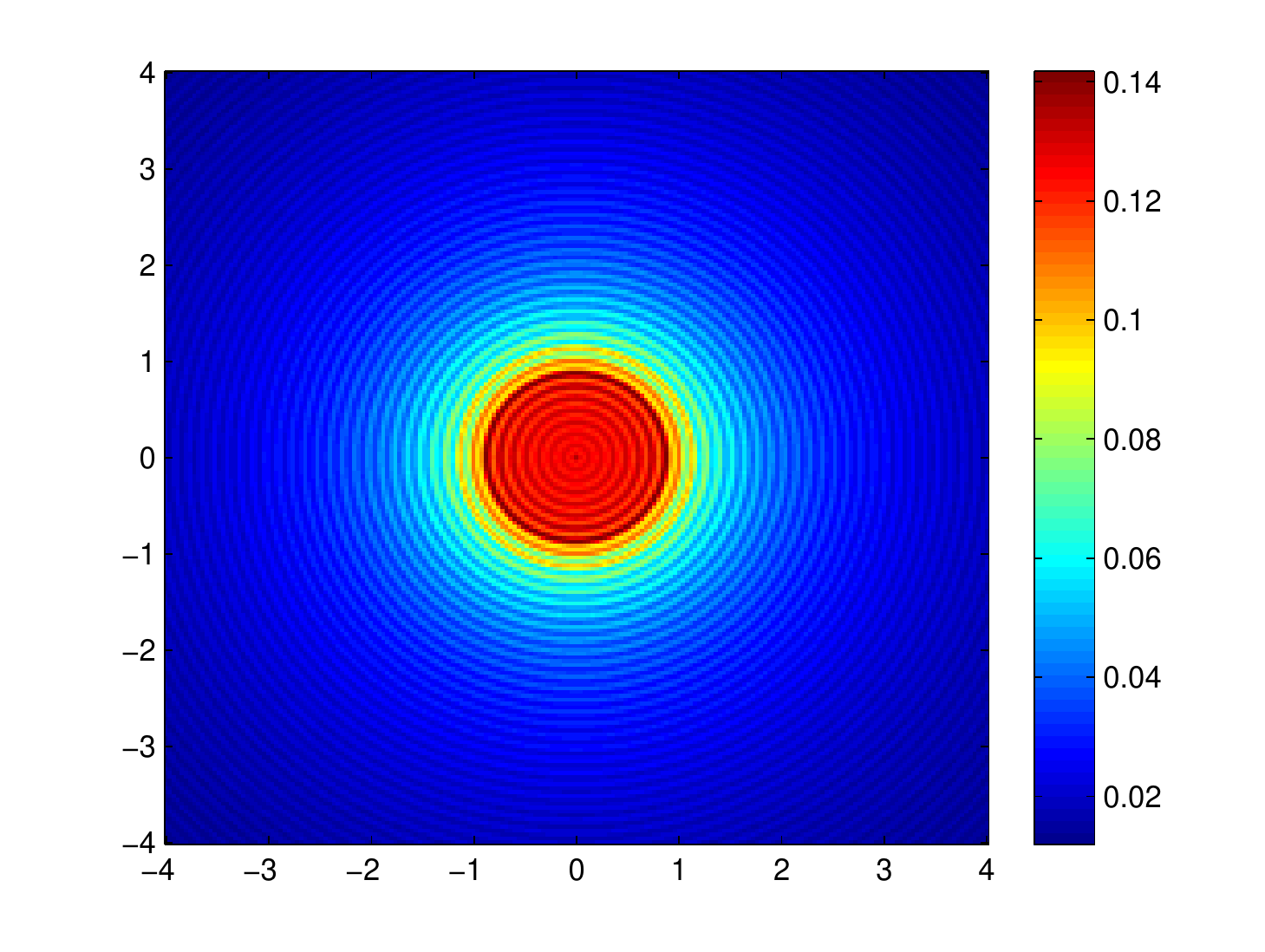} \qquad
\includegraphics[width=0.4\textwidth]{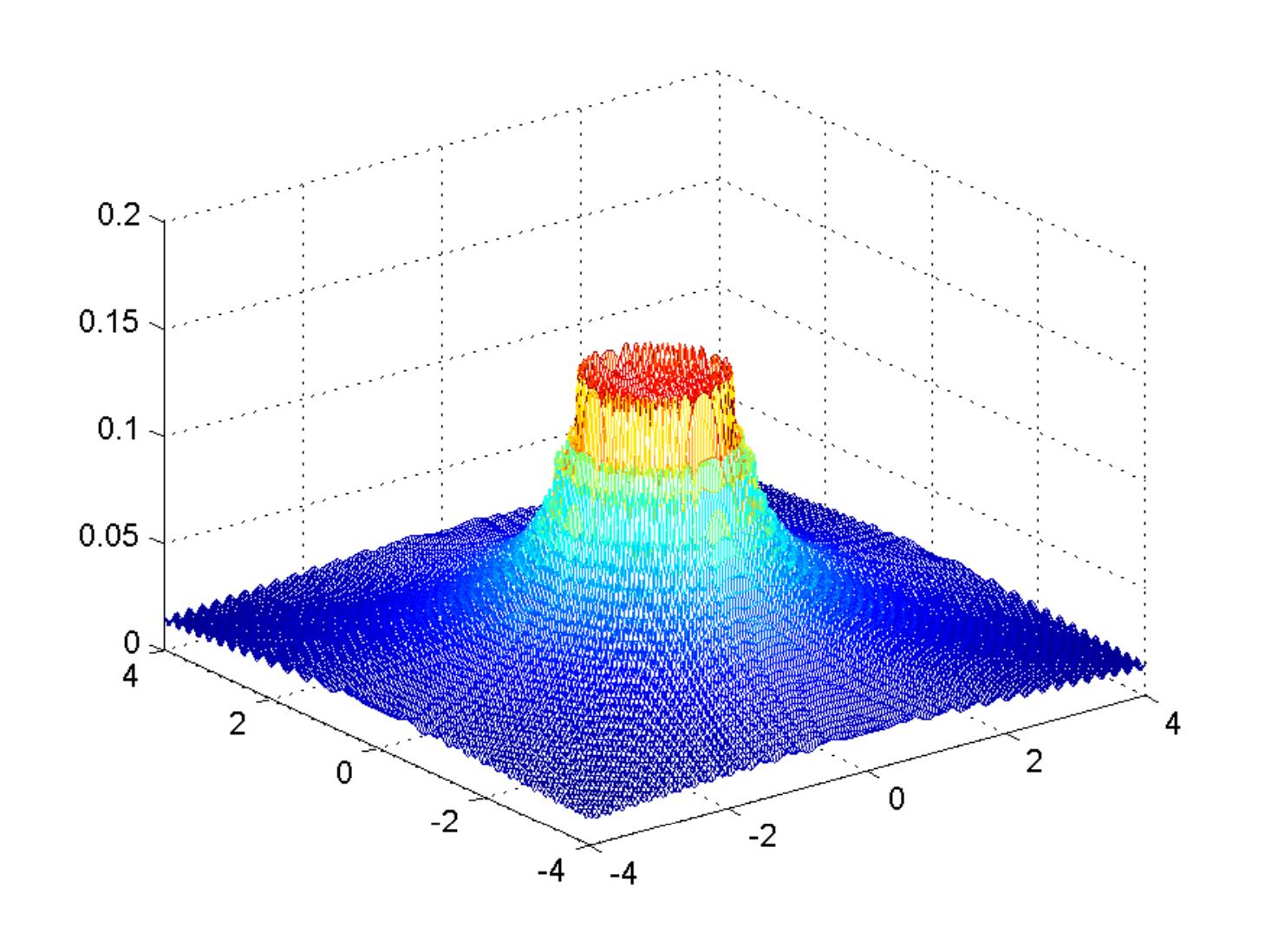}
\caption{RTM with two polarization directions for imaging a perfectly conducting circular scatterer: the wavelength in the first row is $\lam=1/2$ and in the second row $\lam=1/4$.}\label{fig:21}
\end{figure}

\begin{figure}
\includegraphics[width=0.4\textwidth]{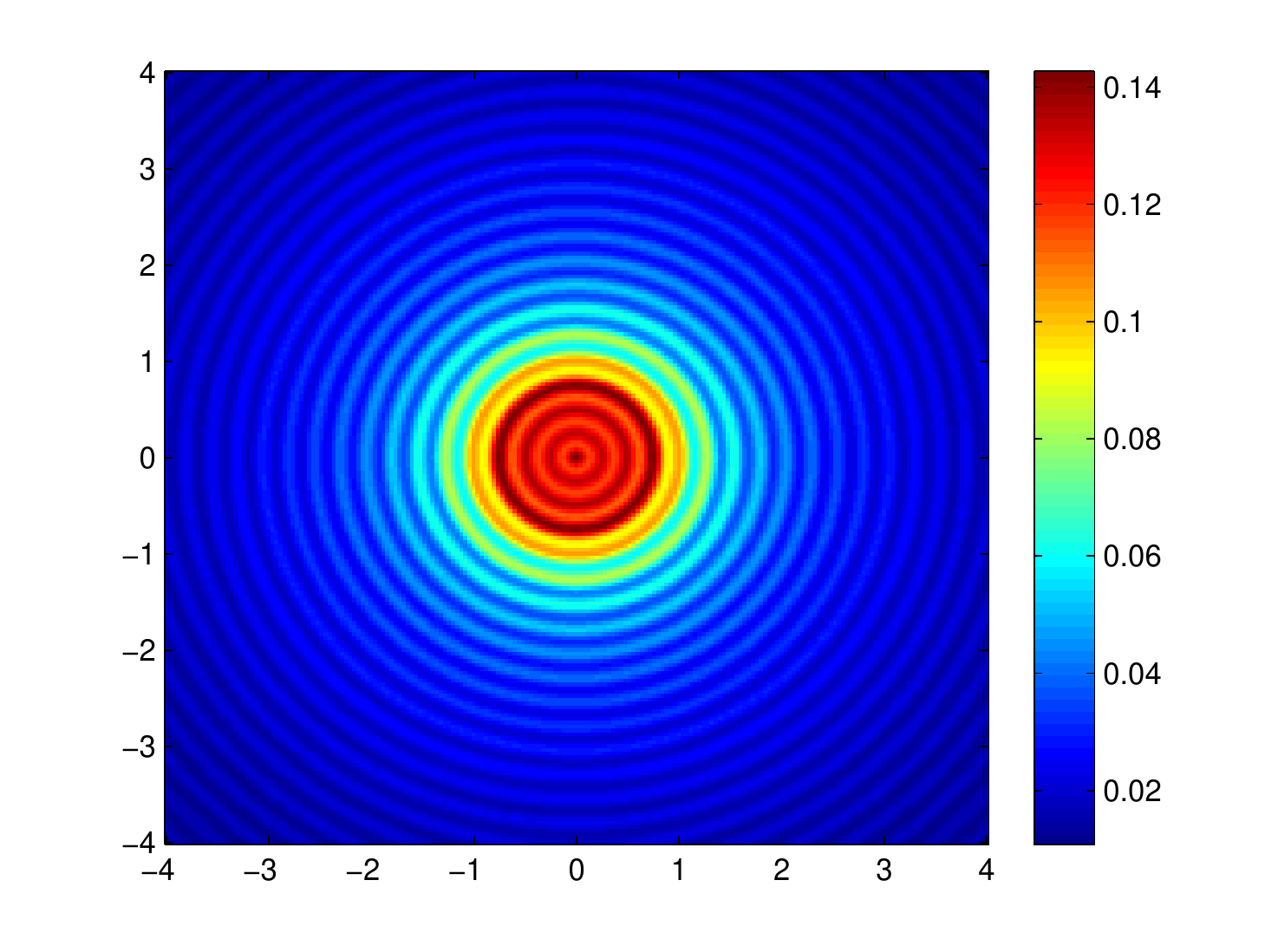} \qquad
\includegraphics[width=0.4\textwidth]{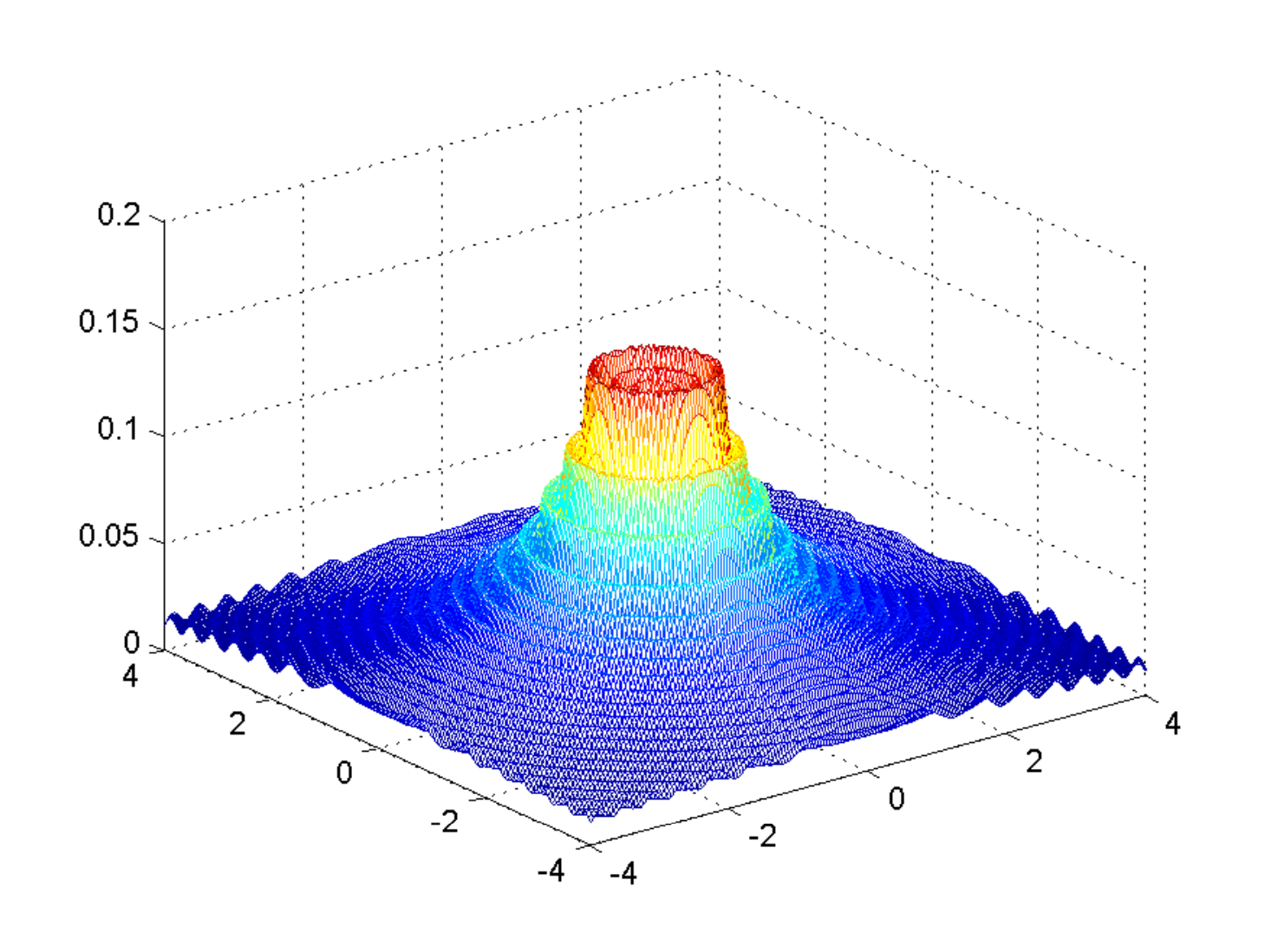}\\
\caption{RTM with two polarization directions for imaging a perfectly conducting circular scatterer, $\mathbb{G}(z,x_s)p$ instead of $g(z,x_s)p$ in (\ref{cor2}), $\lambda=1/2$. }\label{fig:22}
\end{figure}

\begin{figure}
\includegraphics[width=0.33\textwidth]{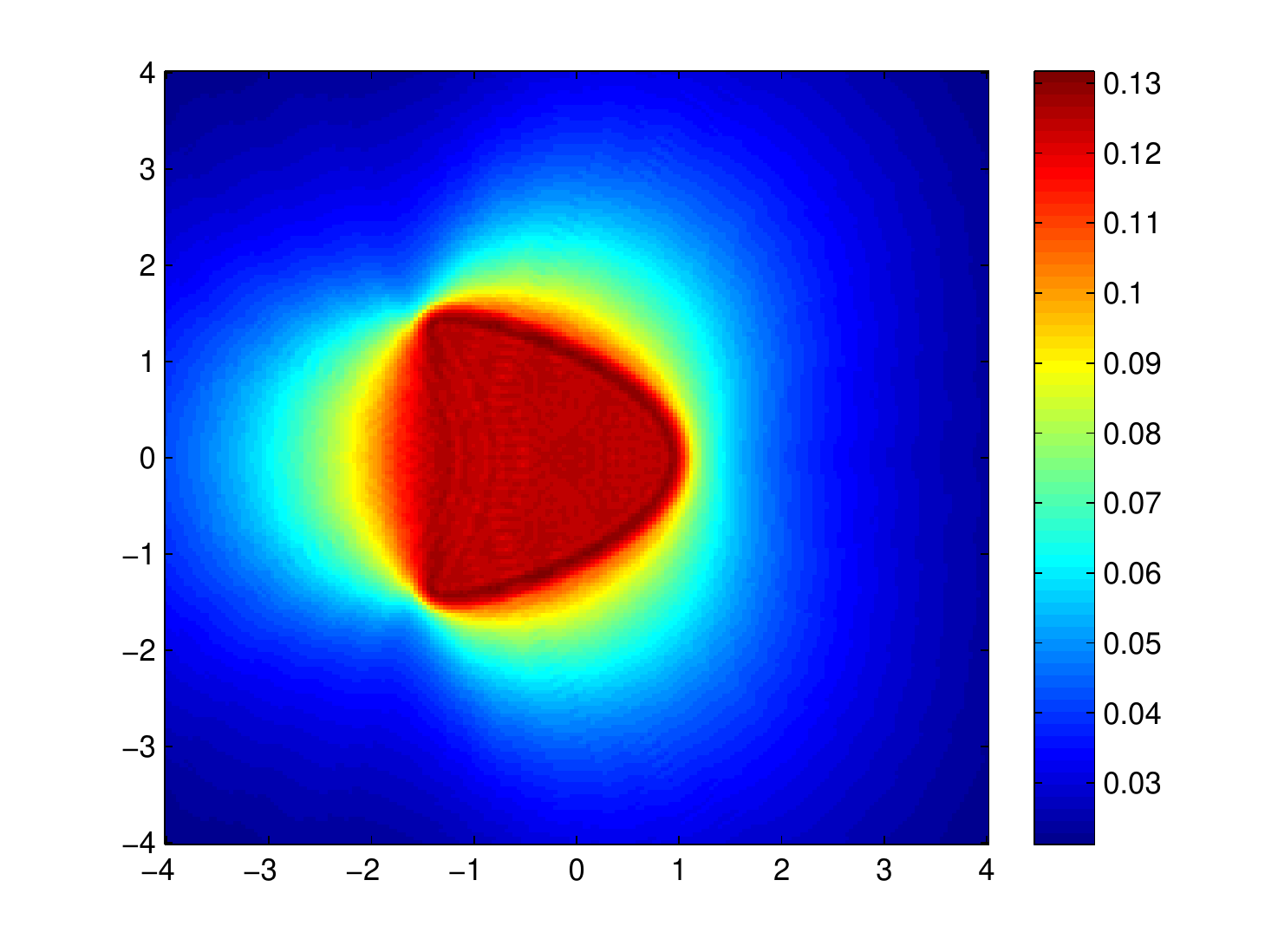}
\includegraphics[width=0.33\textwidth]{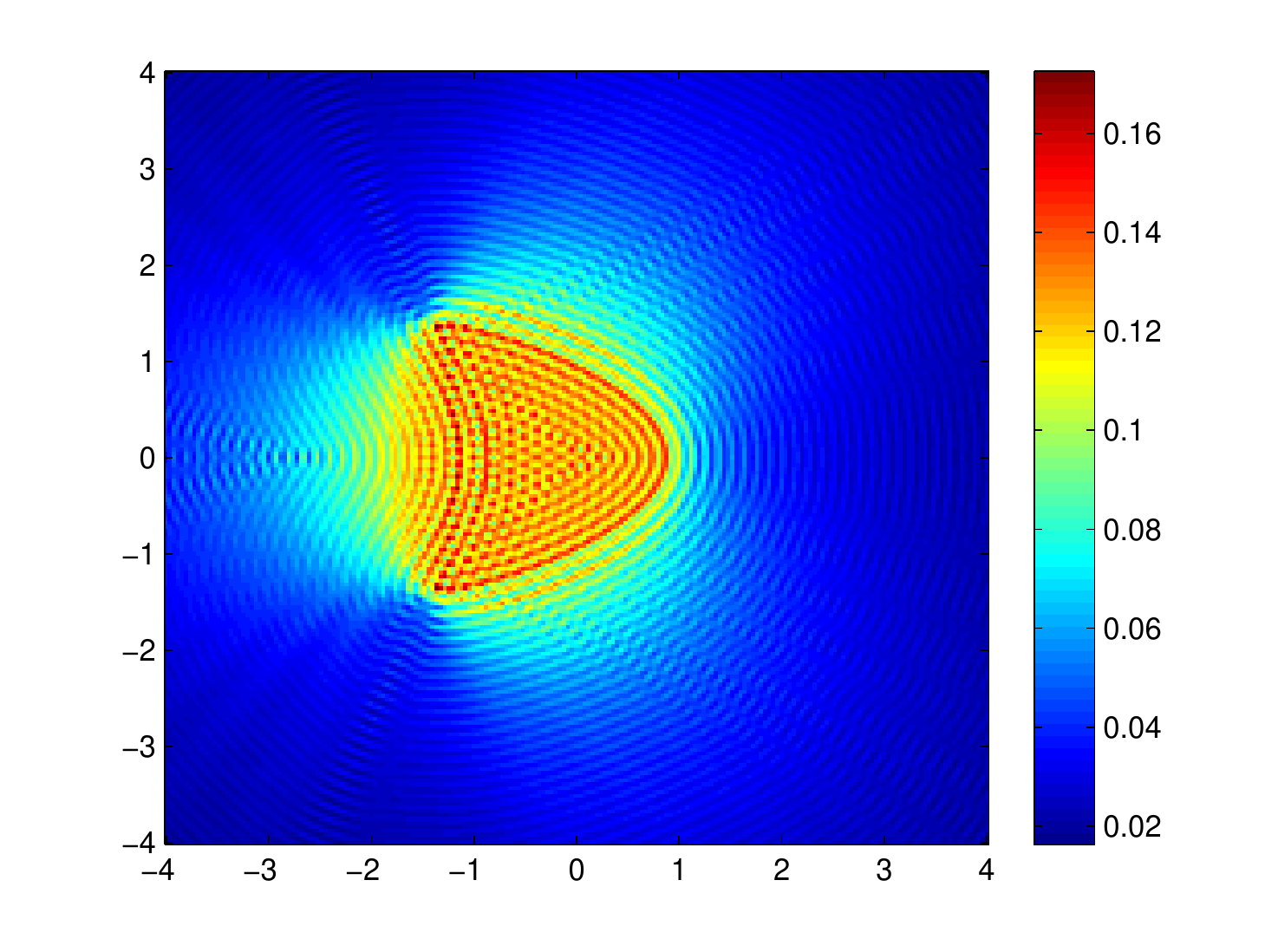}
\includegraphics[width=0.33\textwidth]{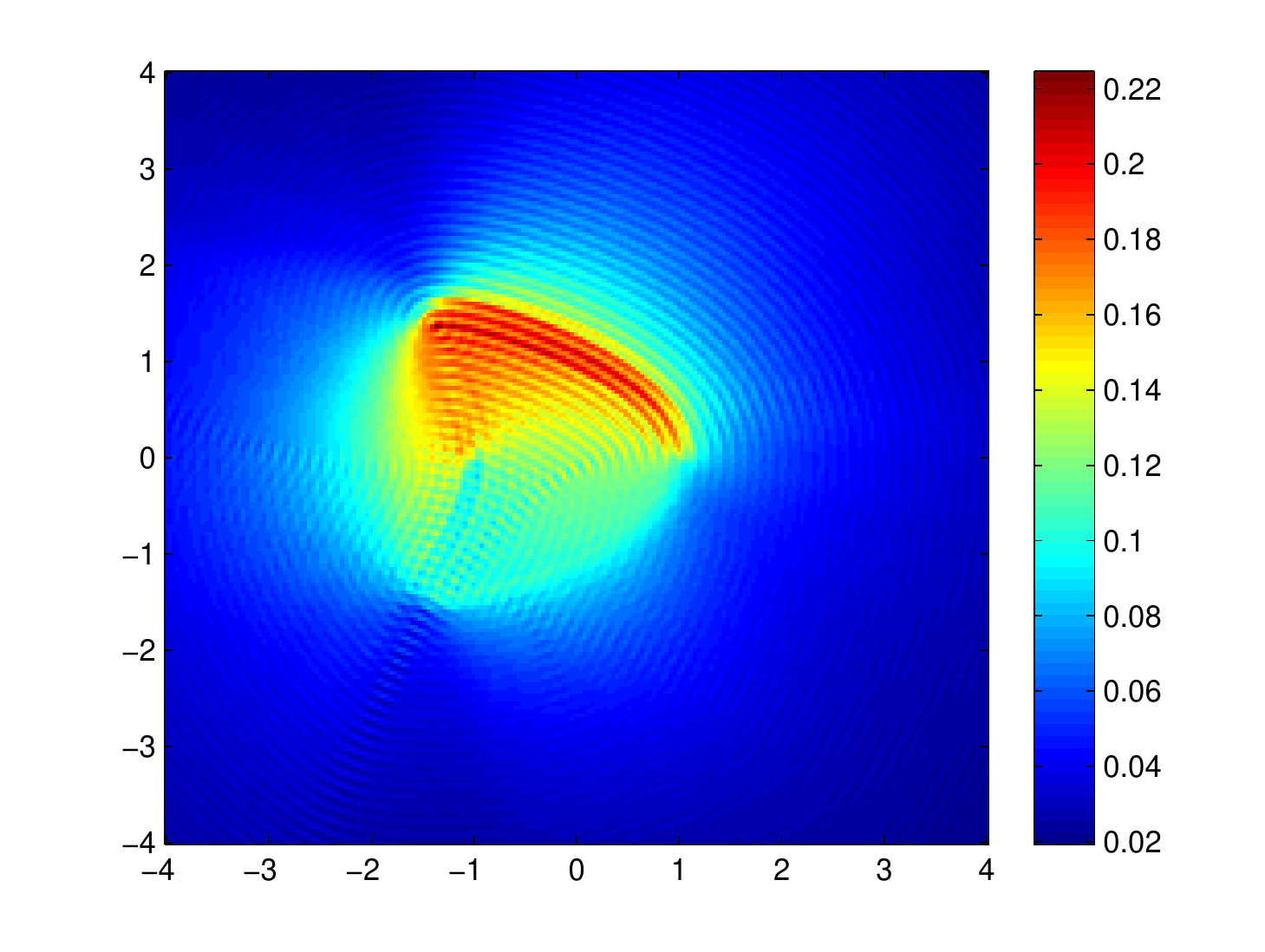}
\caption{RTM with two polarization directions for imaging with impedance boundary conditions. The impedance in the left picture is $\eta(x)=1$, in the middle picture is $\eta(x)=1000$, and the right picture is $\eta(x)=1000$ in the upper half scatterer and $\eta(x)=1$ in the lower half of the scatterer. The probe wavelength $\lam=1/4$. The sampling mesh is $201\times 201$ in the search domain.}\label{fig:23}
\end{figure}

\bigskip
\textbf{Example 2}.
We observe from Figure \ref{fig:11} that the the image has some imperfections. This can be improved by summing up the imaging functionals (\ref{cor2}) with polarization directions ${e}_1=(1,0)^T$ and ${e}_2=(0,1)^T$. In the remainder of this subsection, we will use the following imaging functional:

\ben
\fl \qquad I_1(z)=-k^2\sum_{p={e_1,e_2}}\Im\left\{\frac{|\Ga_s||\Ga_r|}{N_sN_r}\sum^{N_s}_{s=1}\sum^{N_r}_{r=1}g(z,x_s)p\cdot\mathbb{G}(z,x_r)^T\overline{E^s(x_r,x_s)}\right\}\ \ \ \ \forall z\in\Om.
\een
Figure \ref{fig:21} shows the imaging result for imaging a perfectly conducting circle. Figure \ref{fig:22} shows the imaging result when the correlational function
$g(z,x_s)p$ is changed to $\mathbb{G}(z,x_s)p$ in (\ref{cor2}). The result agrees with those shown in Figure \ref{fig:21}. We remark that the imaging functional
with $g(z,x_s)p$ as the correlation function is, however, less expansive in terms of the computational time. Figure \ref{fig:23} shows the results of imaging a kite like scatterer with impedance boundary condition. We observe that the imaging functional $I_1$ works quite well.

\begin{figure}
\includegraphics[width=0.24\textwidth]{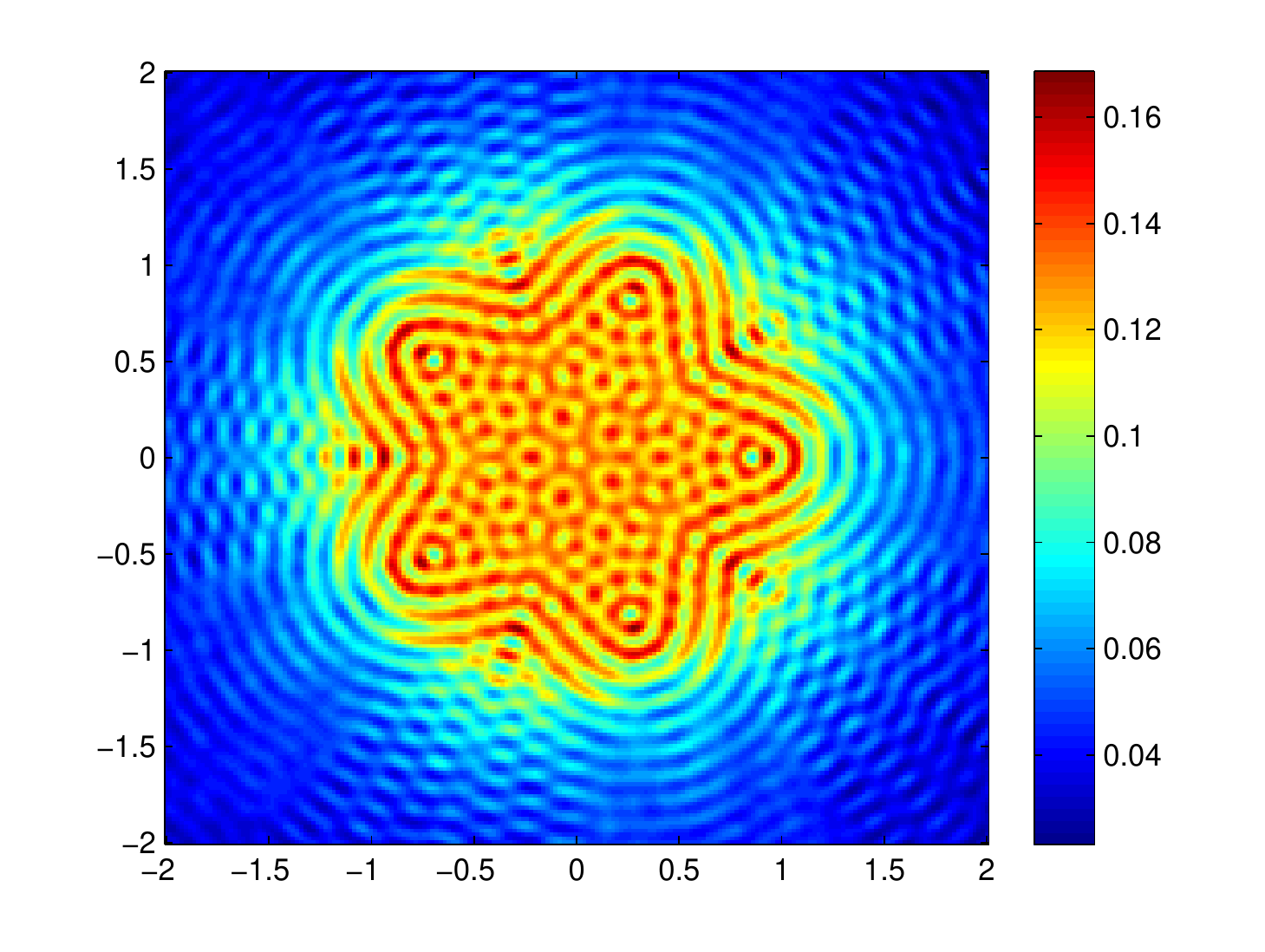}
\includegraphics[width=0.24\textwidth]{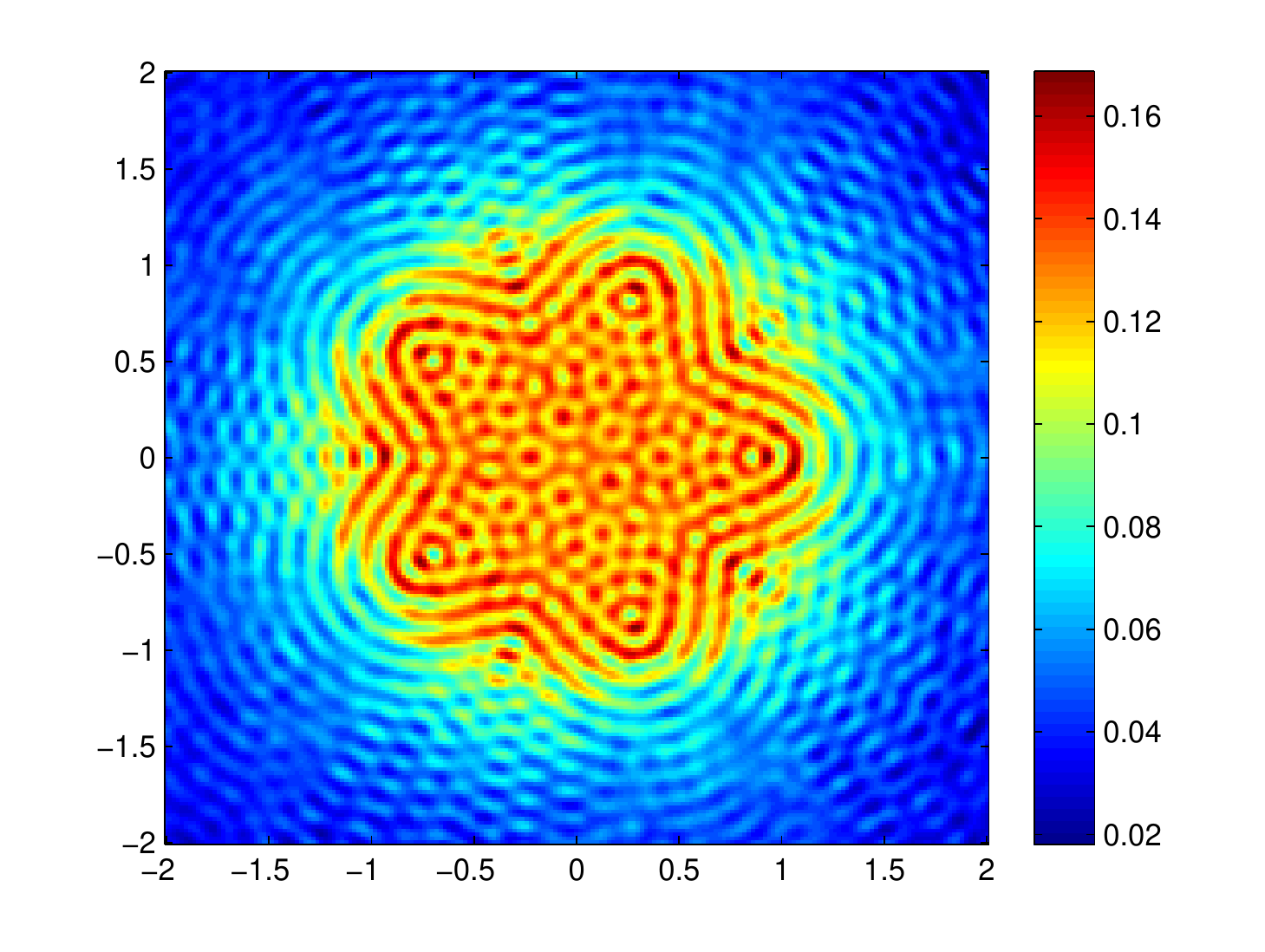}
\includegraphics[width=0.24\textwidth]{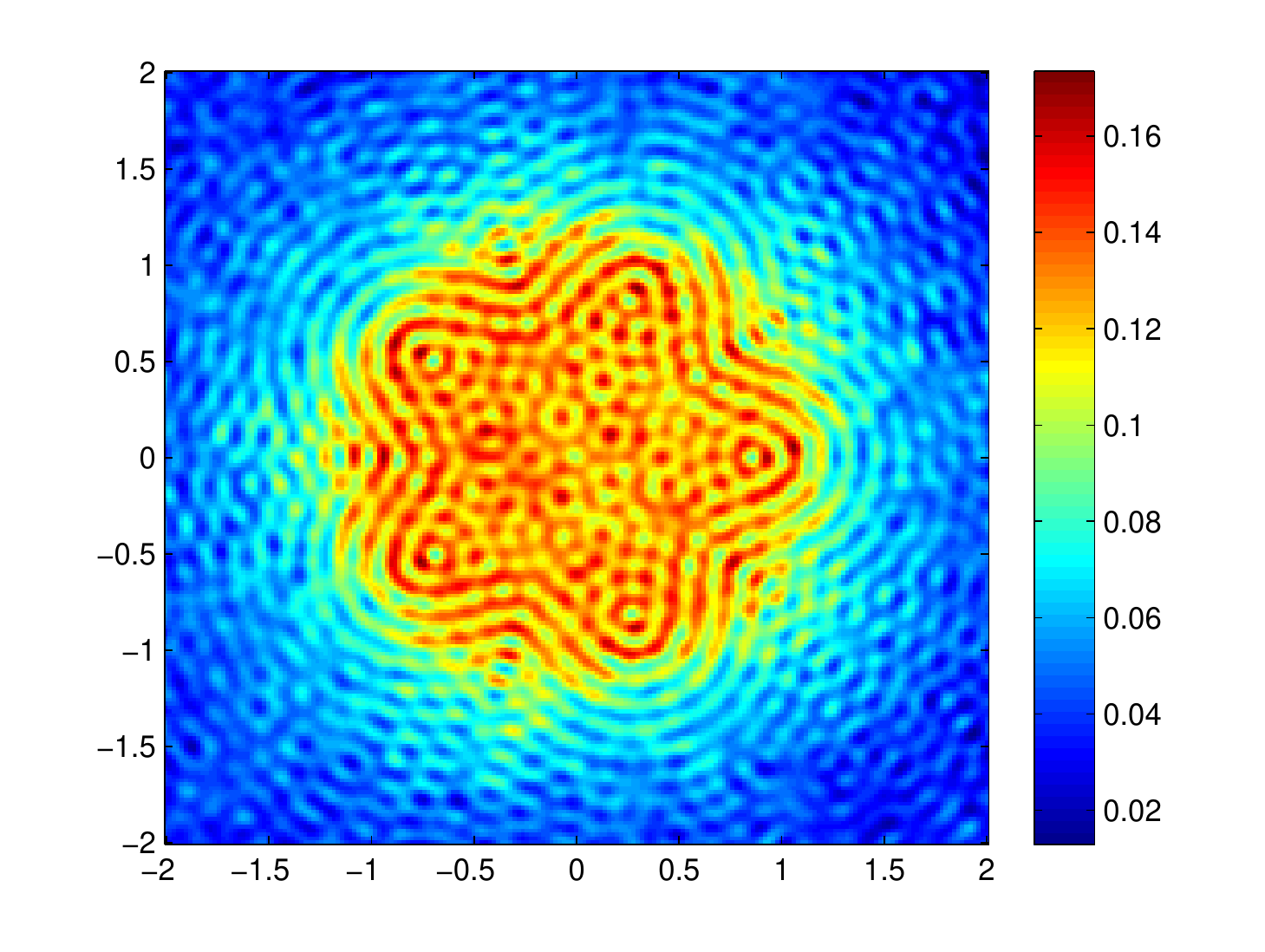}
\includegraphics[width=0.24\textwidth]{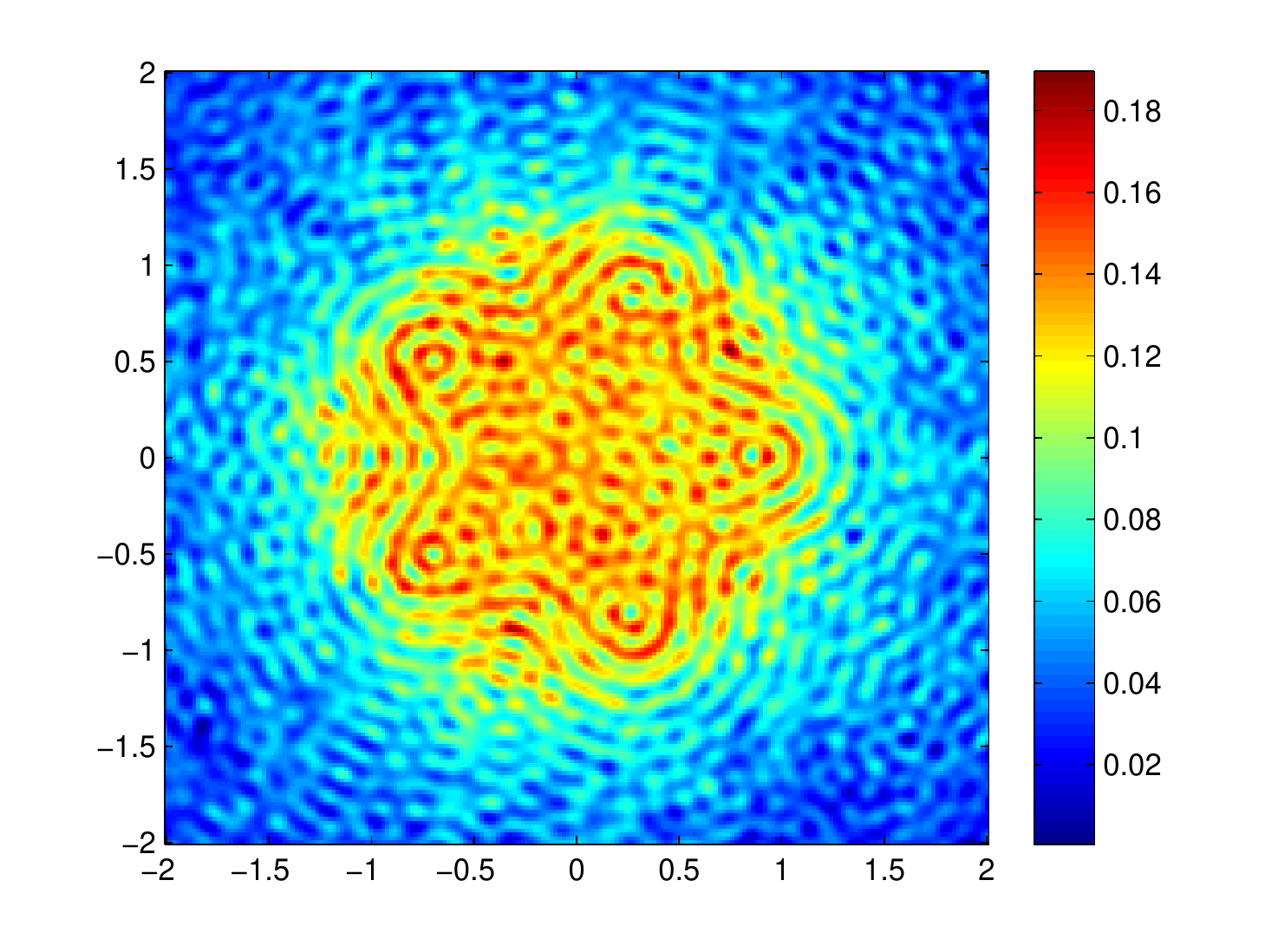}
\caption{The imaging results with respect to multiplicative noise data. The noise levels are $\mu=10\%, 20\%,30\%, 50 \%$ from left to right. The probe wavelength is $\lam=1/4$. The searching domain is $(-2, 2)^2$ and the sampling mesh is $201\times 201$.}\label{fig:31}
\end{figure}

\begin{figure}
\includegraphics[width=0.24\textwidth]{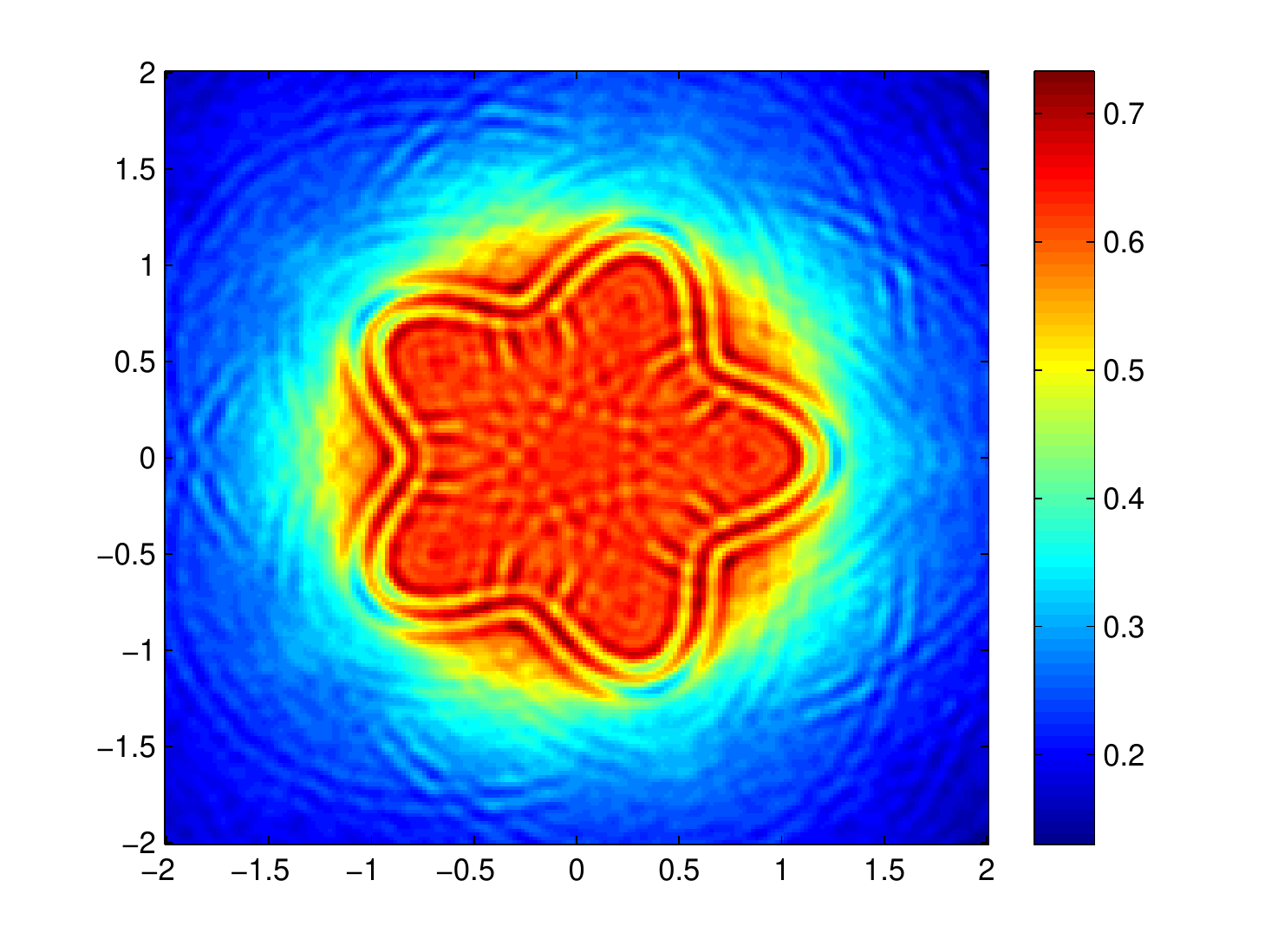}
\includegraphics[width=0.24\textwidth]{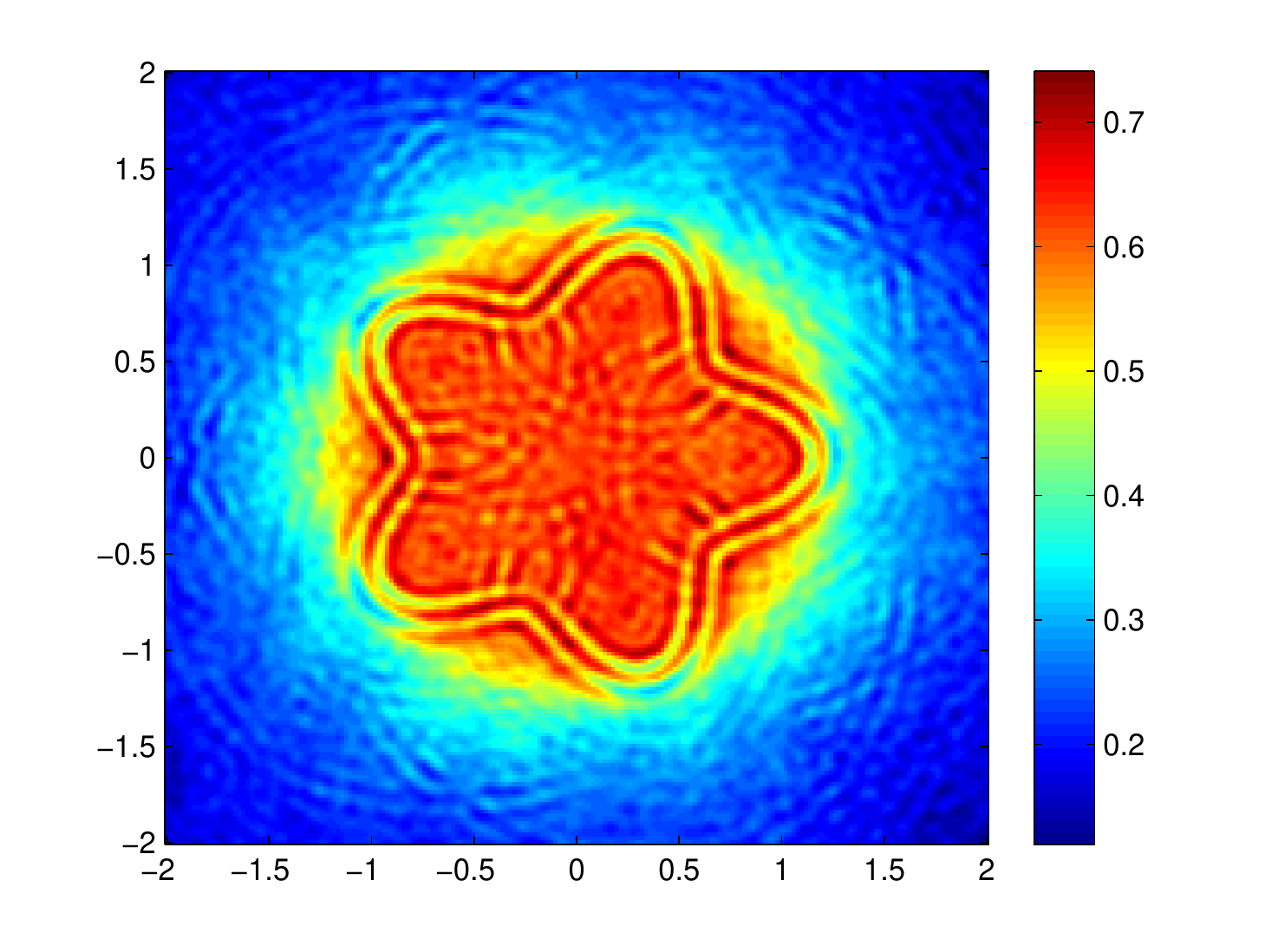}
\includegraphics[width=0.24\textwidth]{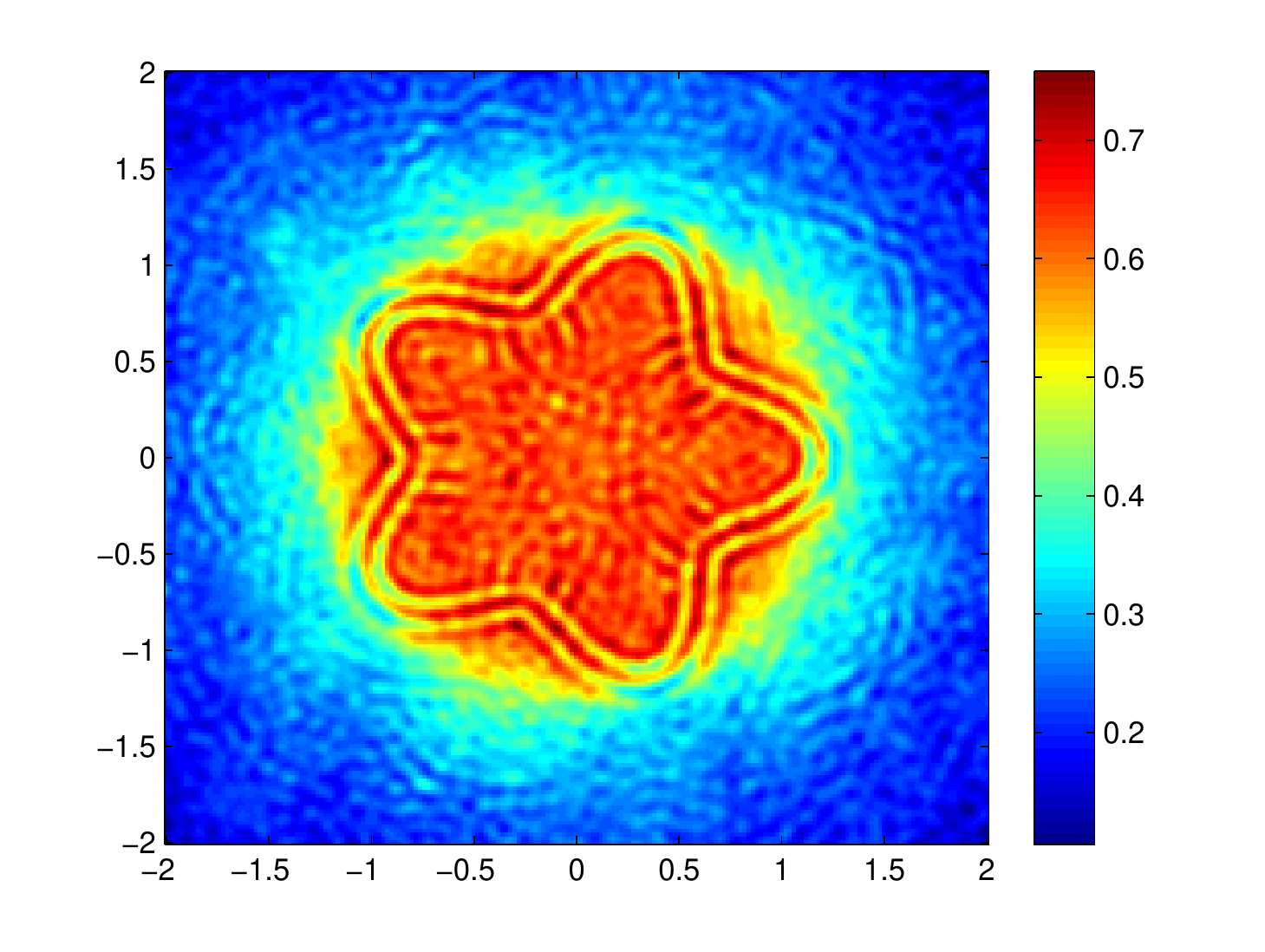}
\includegraphics[width=0.24\textwidth]{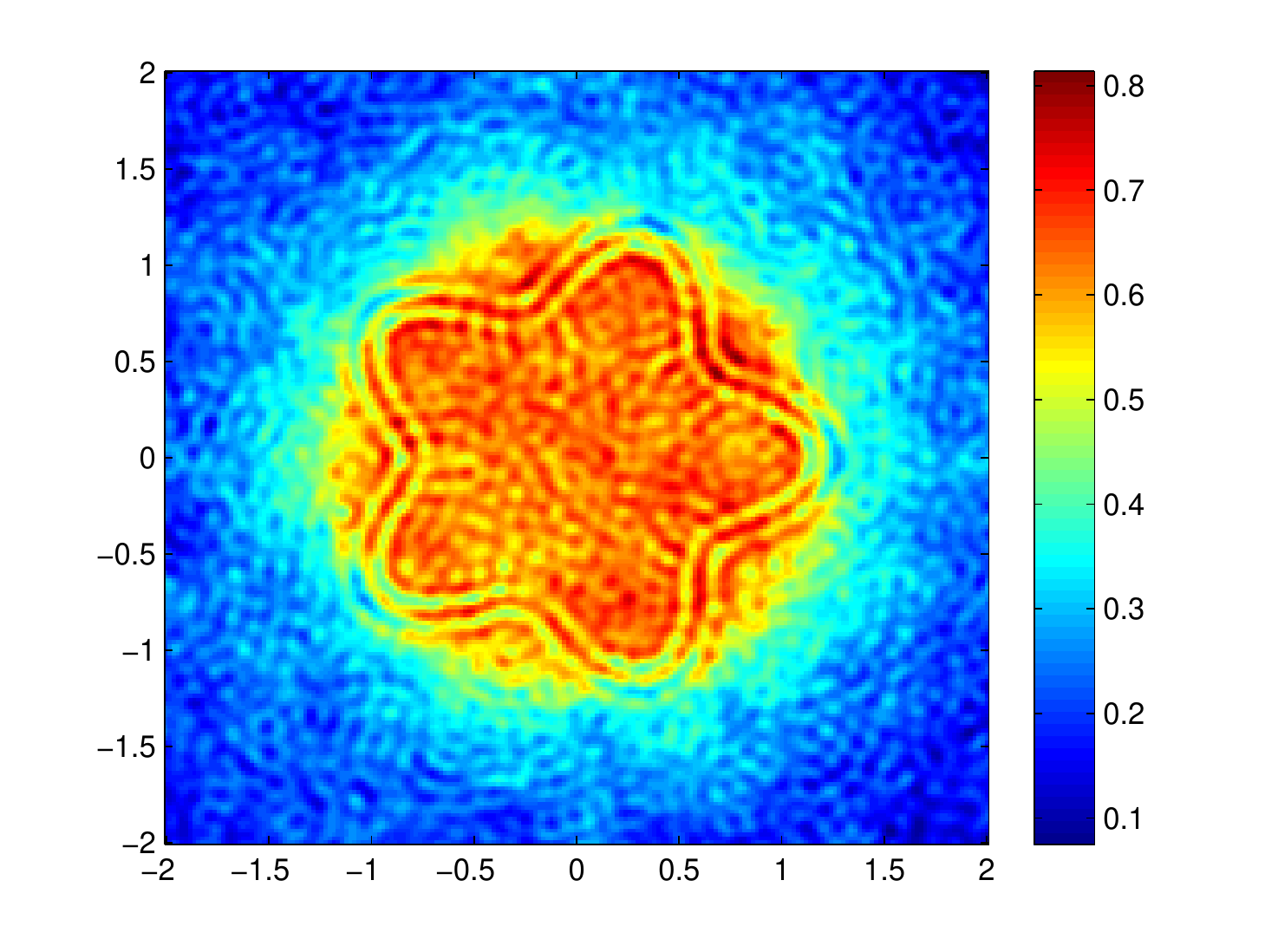}
\caption{The imaging results using multi-frequency data with added Gaussian noise. The noise levels are $\mu=10\%, 20\%,30\%, 50 \%$ from left to right. The multiple wavelength $\lambda=1/3,1/3.5,1/4,1/4.5,1/5$. $N_s=N_r=128$.}\label{fig:32}
\end{figure}

\bigskip
\textbf{Example 3}.
We show the stability of our RTM algorithm in the presence of noise. We introduce the additive Gaussian noise as follows:
 $E^s_{\rm noise}=E^s+\mu \varepsilon$,
 where $\mu$ is the noise level and $\varepsilon$ is normally distributed random variable with mean zero and standard deviation $\max_{x_r,x_s}|E^s(x_r,x_s)|$. The perfectly conducting scatterer is a 5-leaf. Figure \ref{fig:31} shows the numerical results with different noise levels which indicates that our imaging functional is quite stable with respect to the additive Gaussian noise. Figure \ref{fig:32} shows the imaging results can be improved if we sum up the imaging functionals with multi-frequency data with additive Gaussian noise.

\begin{figure}
\includegraphics[width=0.24\textwidth]{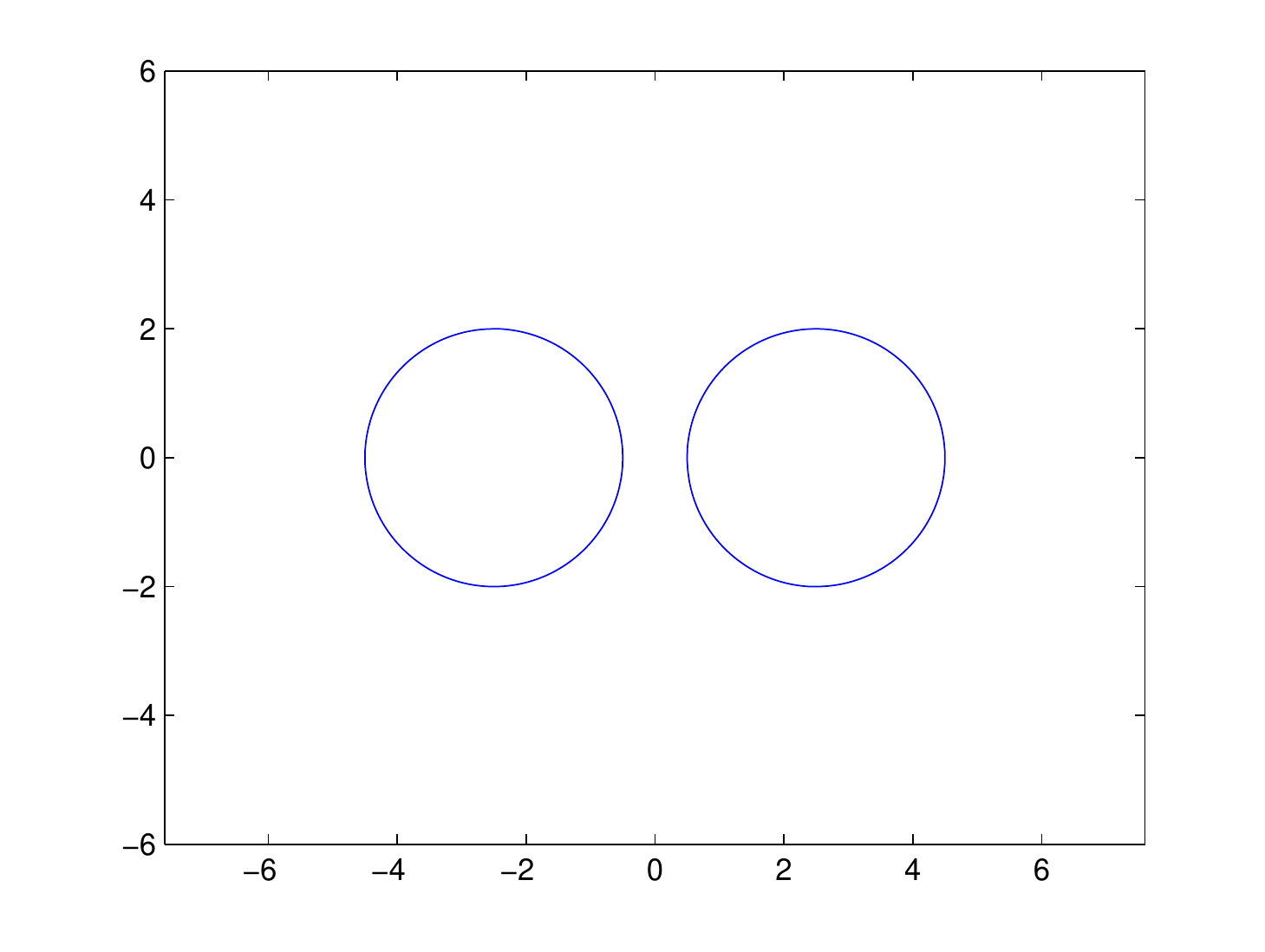}
\includegraphics[width=0.24\textwidth]{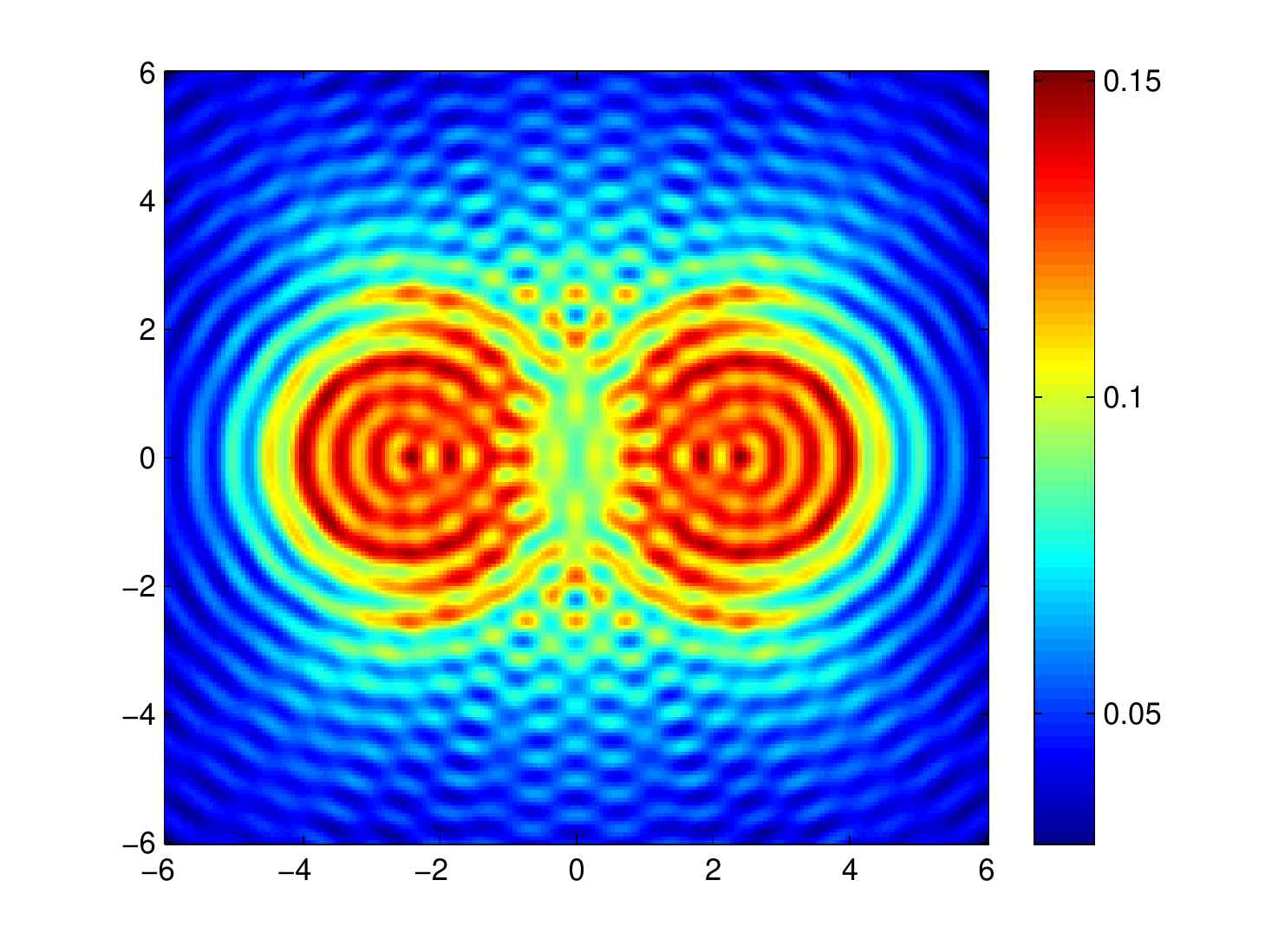}
\includegraphics[width=0.24\textwidth]{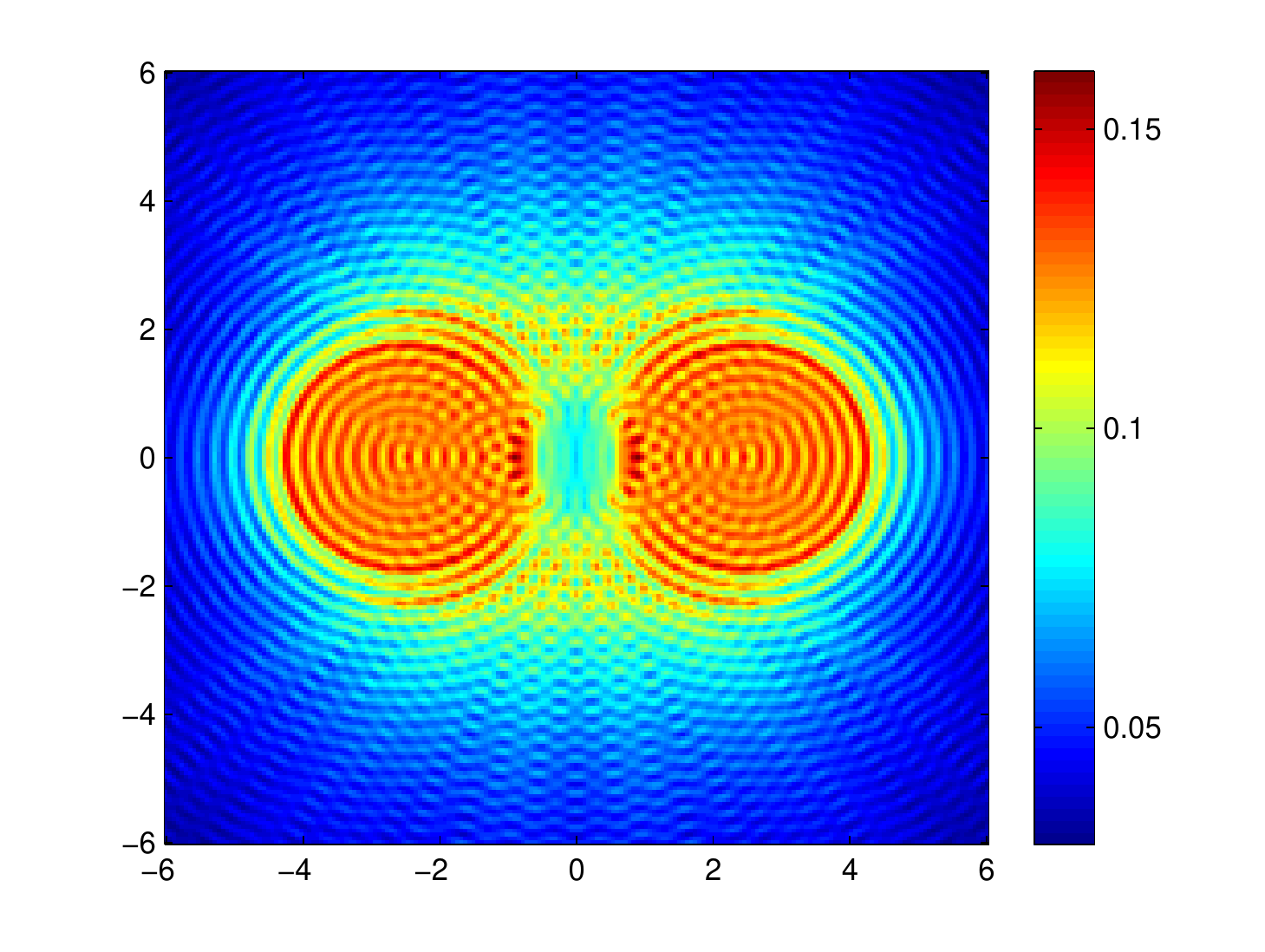}
\includegraphics[width=0.24\textwidth]{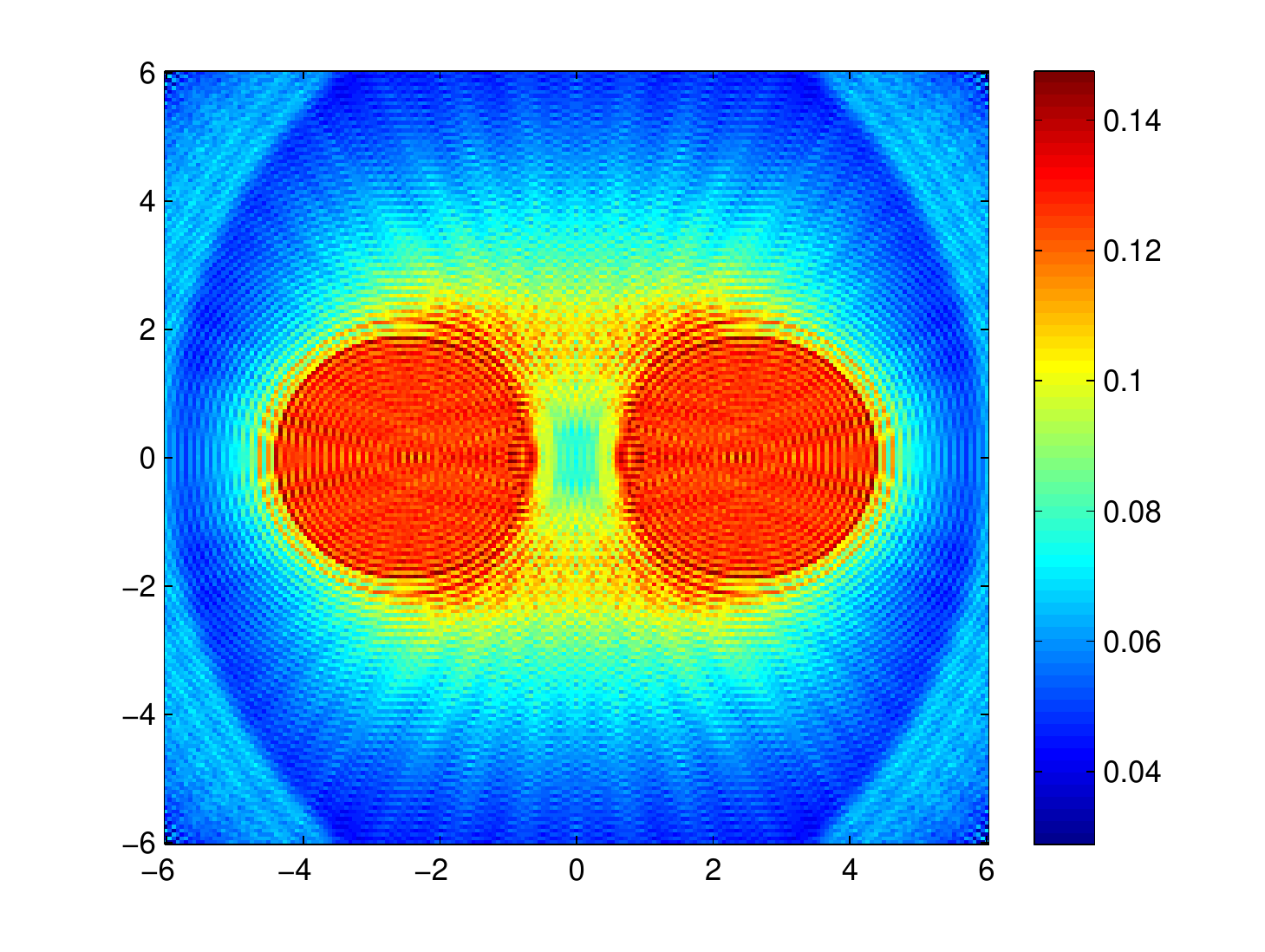}
\caption{Imaging of two perfectly conducting circles.  The first picture is the exact targets, the others are the imaging results with the probe wavelengths $\lam=1,1/2,1/4$ (from left two right). The searching domain is $(-6, 6)^2$ with a $201\times 201$ sampling mesh.}\label{fig:41}
\end{figure}

\begin{figure}
\includegraphics[width=0.24\textwidth]{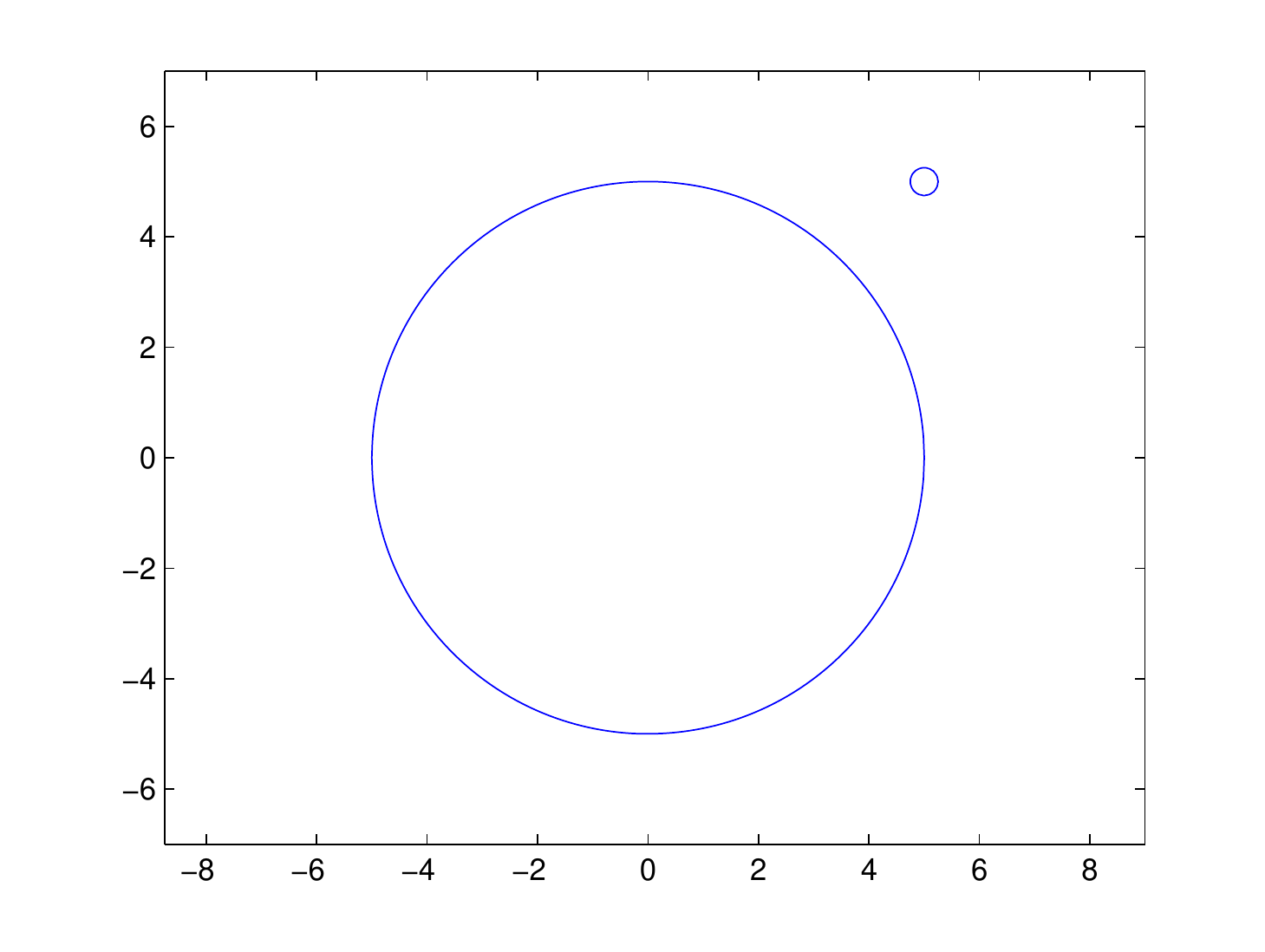}
\includegraphics[width=0.24\textwidth]{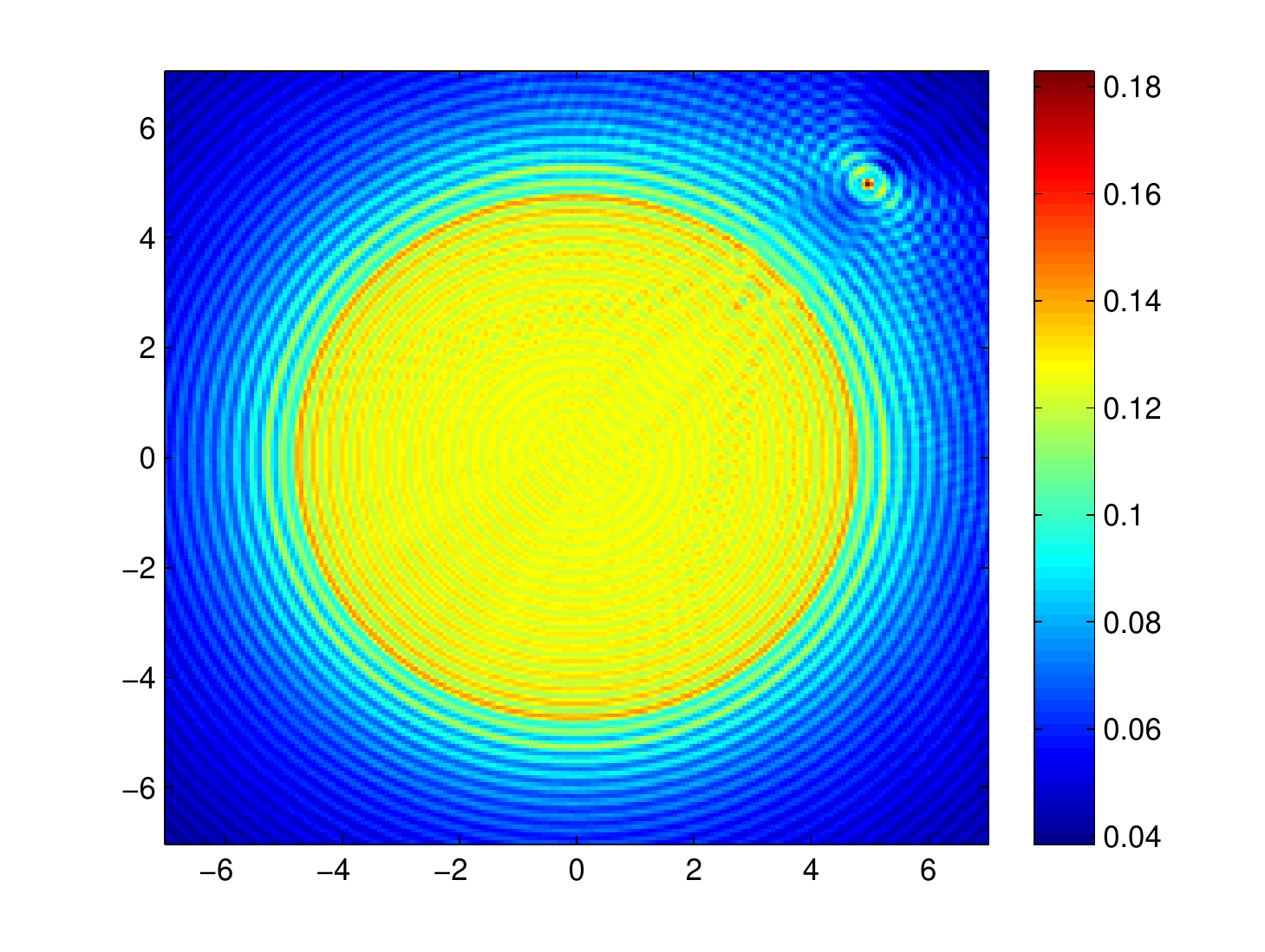}
\includegraphics[width=0.24\textwidth]{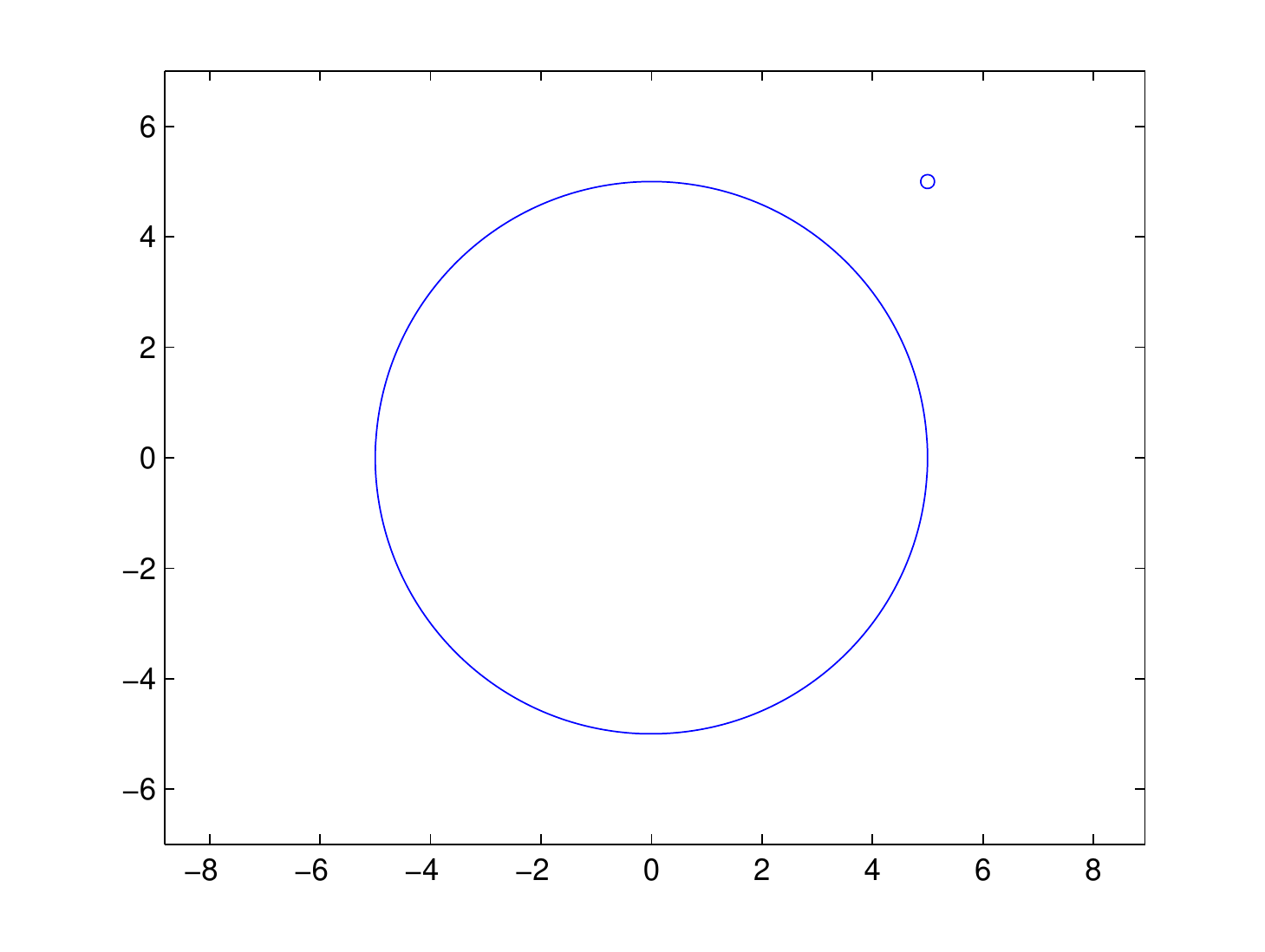}
\includegraphics[width=0.24\textwidth]{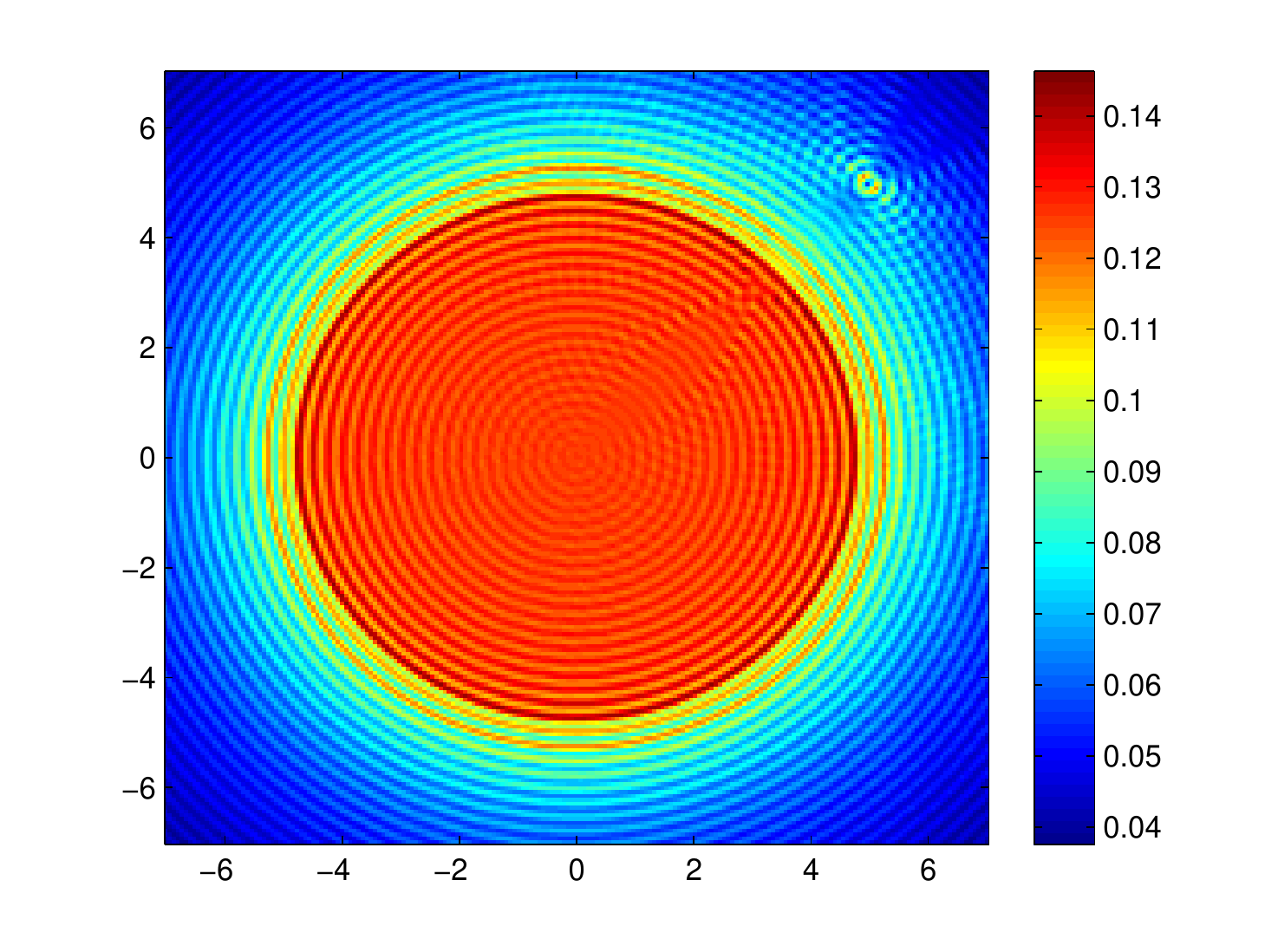}
\caption{The left two pictures give the the imaging results for two circular targets with radius $\rho=5$ and $\rho=0.25$. The other two pictures are the imaging results for two targets with radius $\rho=5$ and $\rho=0.125$. The probe wavelength is $\lam=1/2$. The searching domain is $(-7, 7)^2$ with a $201\times 201$ mesh.}\label{fig:43}
\end{figure}

\begin{figure}
\includegraphics[width=0.24\textwidth]{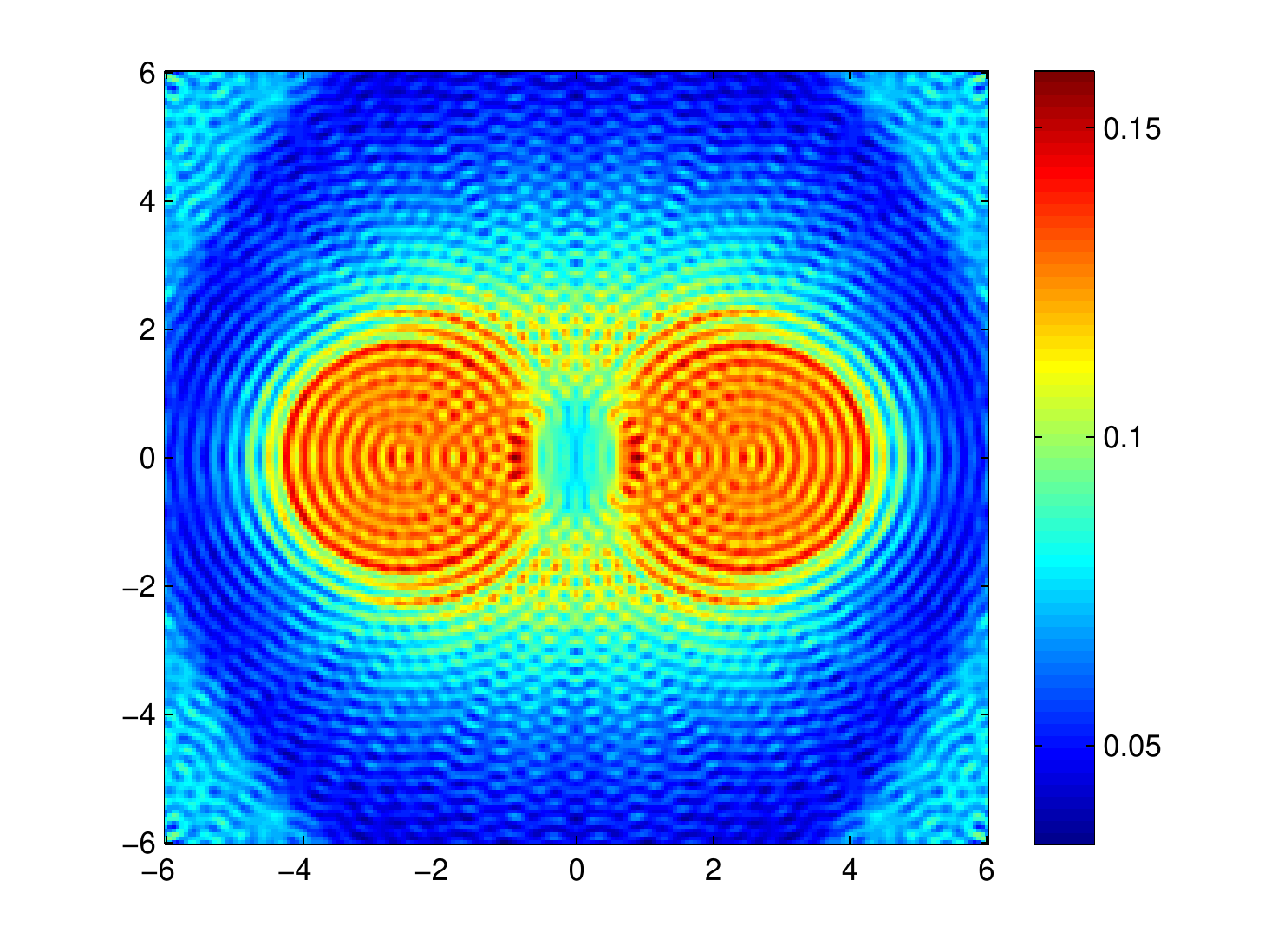}
\includegraphics[width=0.24\textwidth]{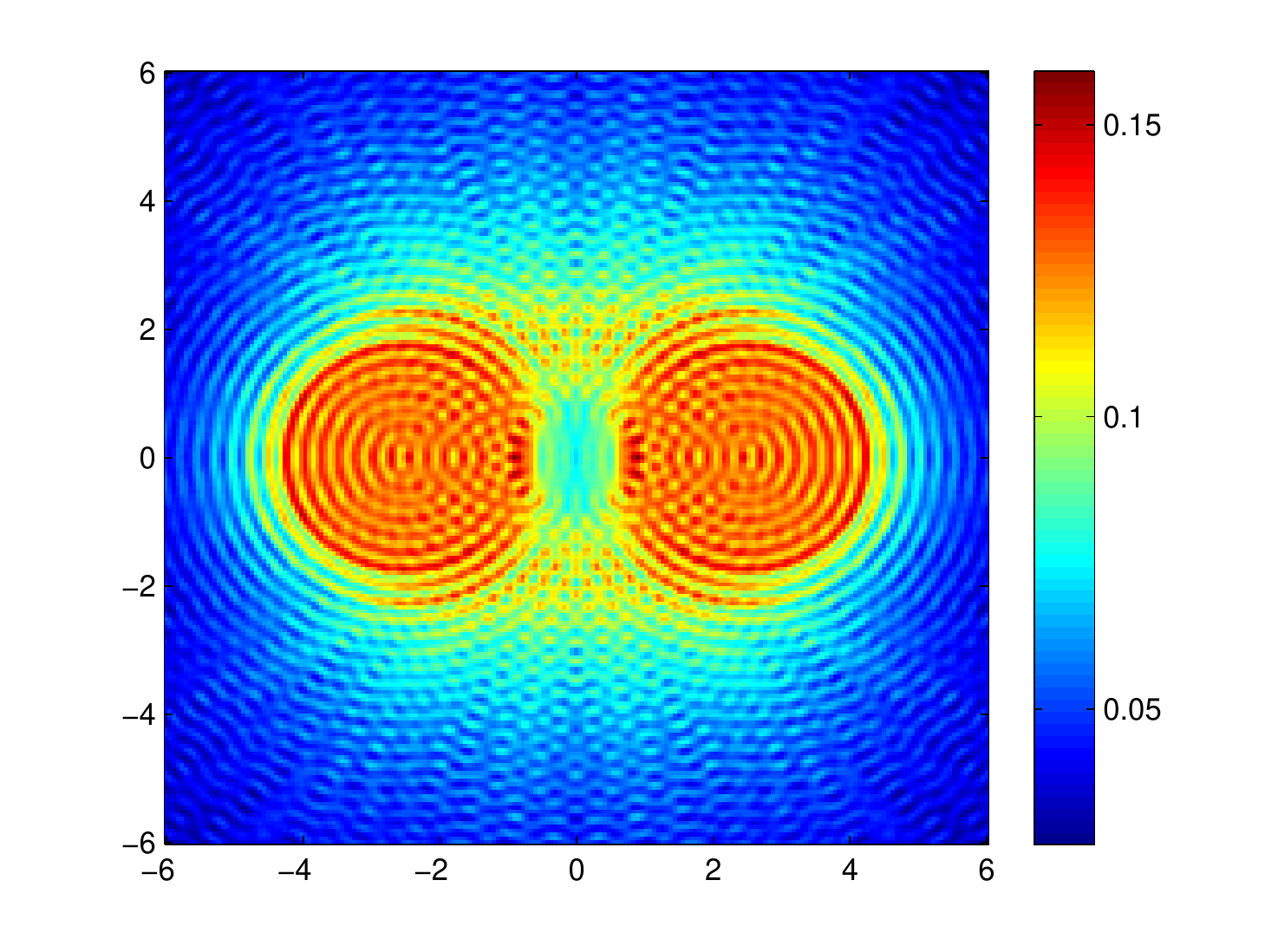}
\includegraphics[width=0.24\textwidth]{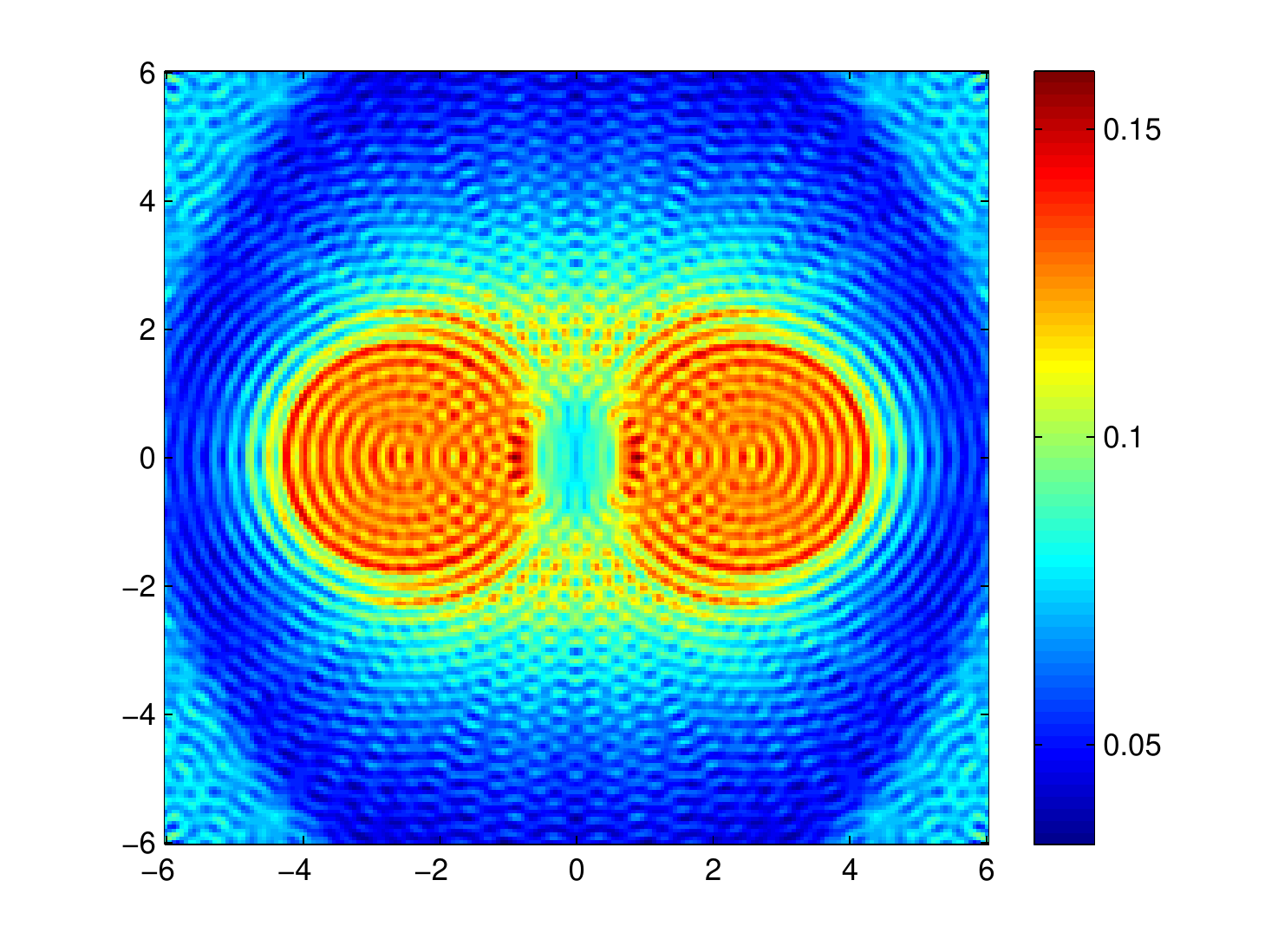}
\includegraphics[width=0.24\textwidth]{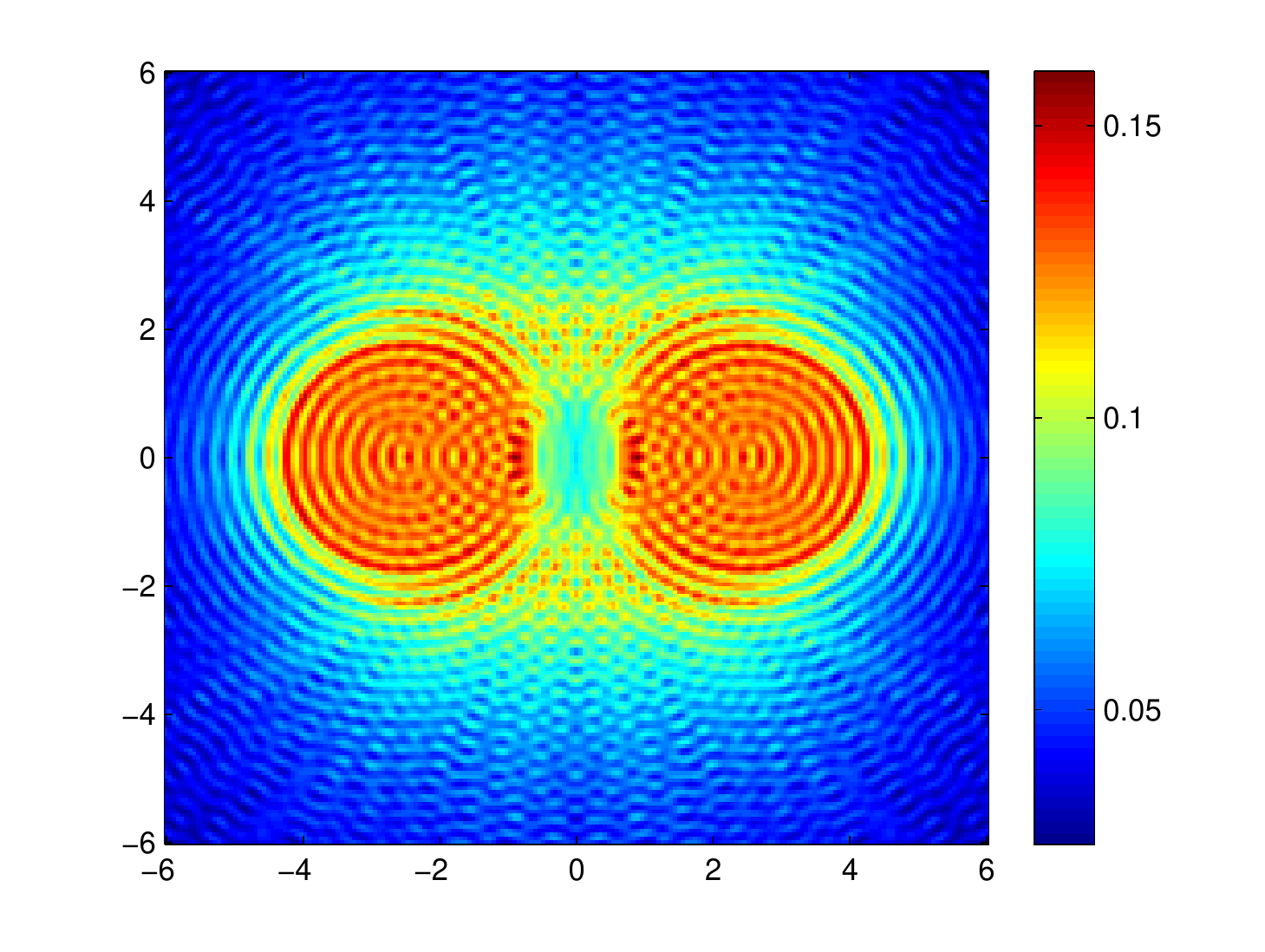}
\caption{Imaging of two perfectly conducting circles for reduced number of sources and receivers: From left to right, ($N_s=64,N_r=128$), ($N_s=64,N_r=256$),($N_s=128,N_r=64$), and ($N_s=256,N_r=64$) respectively. The probe wavelength is $\lam=1/2$. The searching domain is $(-6, 6)^2$ with a $201\times 201$ mesh.}\label{fig:42}
\end{figure}

\bigskip
\textbf{Example 4}.
Figure \ref{fig:41} shows the imaging results of two perfectly conducting circular scatterers of radius $\rho=2$ and centers at $(-2.5,0)$ and $(2.5,0)$.
We see from Figure \ref{fig:41} that with the increase of the wave number, the scatterers become separated. Figure \ref{fig:43} shows the imaging results of two perfectly conducting circular scatterers with different sizes. The bigger scatterer is of radius $\rho=5$ and the smaller ones are of radius $\rho=0.25$ and $\rho=0.125$, respectively. From Figure \ref{fig:43}, we observe that the algorithm can locate the boundary of extended target and the small target simultaneously. By small target we mean that the radius of the scatterer is smaller than the wavelength.
Figure \ref{fig:42} shows the imaging results when the number of sources $N_s$ and receivers $N_r$ is reduced.

\subsection{Numerical examples in 3D}

In this subsection we show the efficiency of our imaging algorithm in the 3D setting. We consider the imaging of perfect conducting objects and use finite element package PHG \cite{phg} to generate the synthetic data. The method of perfectly matched layer is used to truncate the computational domain \cite{pml}. We use the following imaging functional which is the sum of (\ref{cor2}) with three polarization directions $e_1=(1,0,0)^T$, $e_2=(0,1,0)^T$, and $e_3=(0,0,1)^T$:
\ben
\fl I_2(z)=-k^2\sum_{p=e_1,e_2,e_3}\Im\left\{\frac{1}{N_sN_r}\sum^{N_s}_{s=1}\sum^{N_r}_{r=1}|\Delta(x_r)|\,|\Delta(x_s)|\,g(z,x_s)p\cdot\mathbb{G}(z,x_r)^T\overline{E^s(x_r,x_s)}\right\}.
\een
In this subsection we always assume $R_s=R_r=10$.

\begin{figure}
\includegraphics[width=0.3\textwidth]{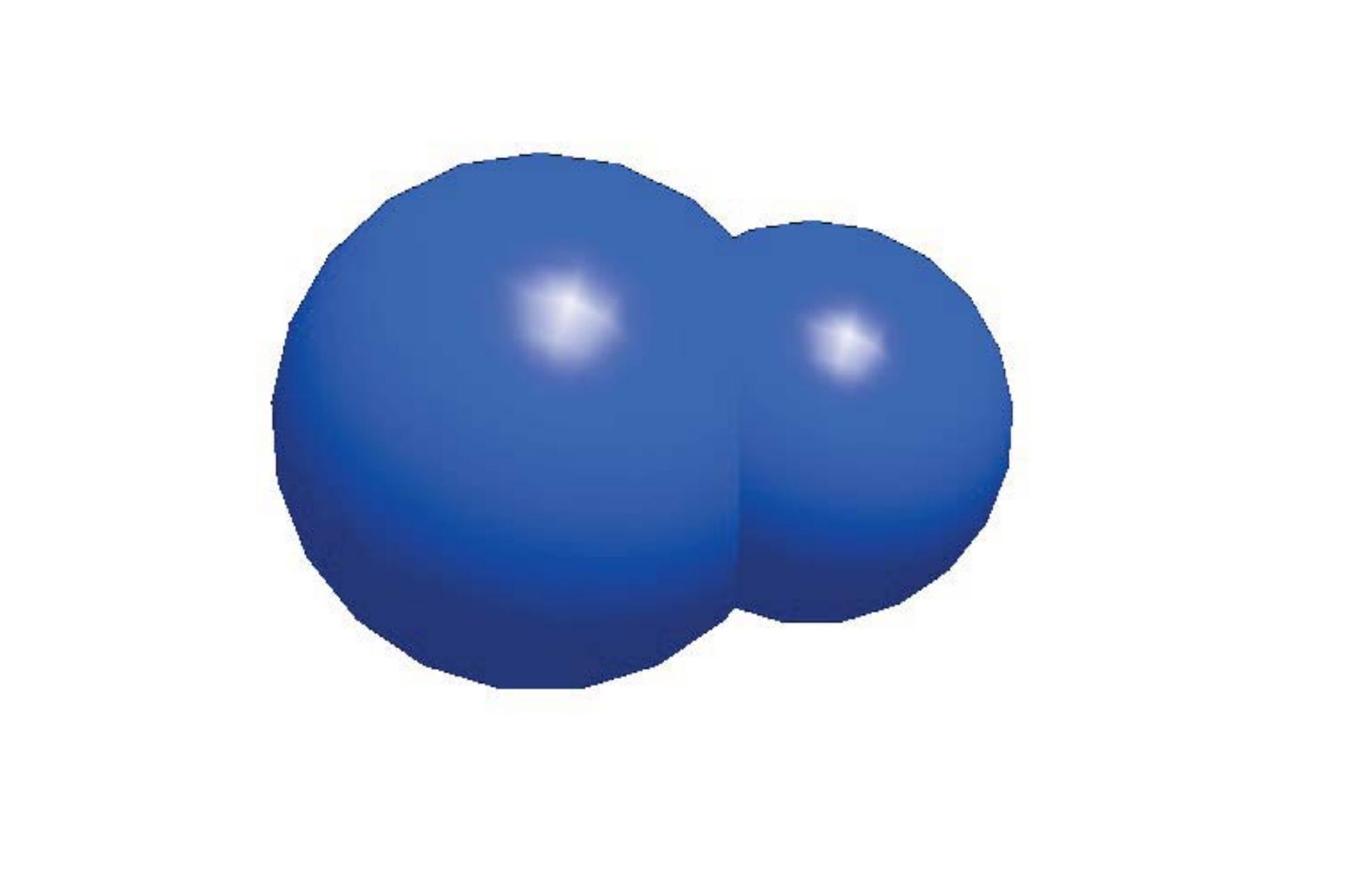}
\includegraphics[width=0.3\textwidth]{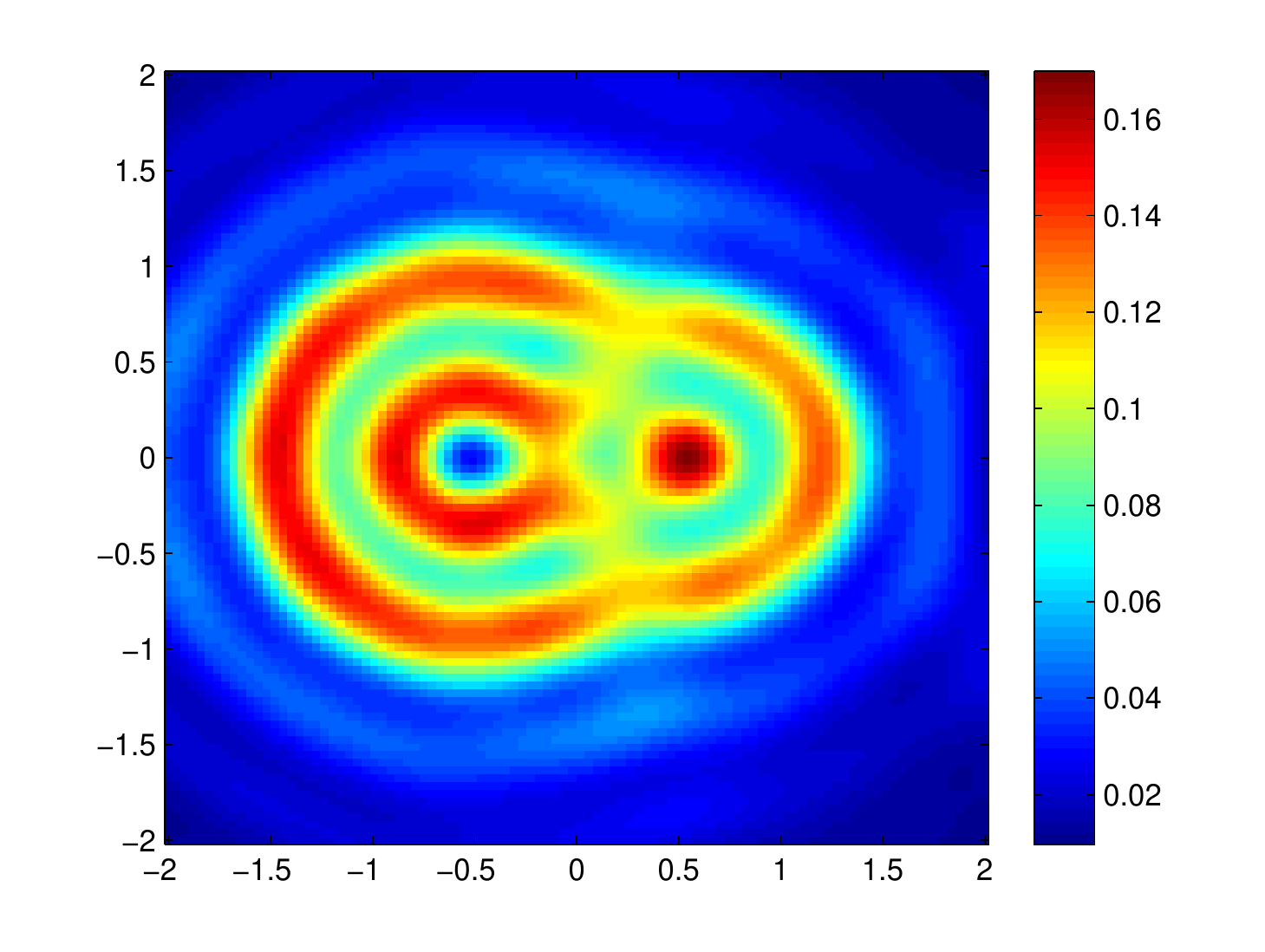}
\includegraphics[width=0.3\textwidth]{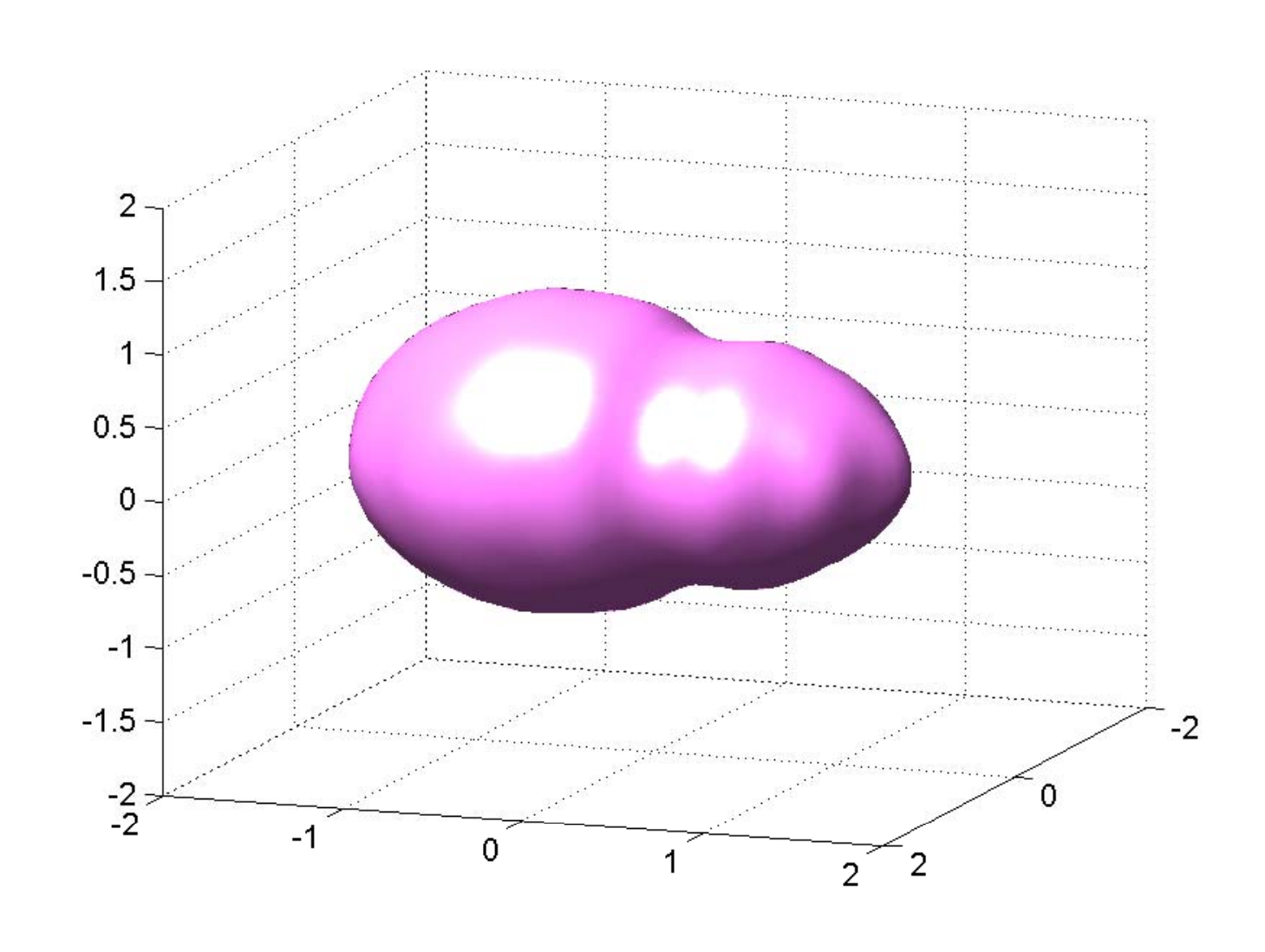}
\caption{Left: The scatterer; Center: Imaging functional at the cross-section $x_3=0$. Right: Imaging result in 3D view. The wavelength $\lambda=1$ and the sampling domain is $(-2, 2)^3$ with the sampling mesh $80\times80\times 80$, $N_s=N_r=256$.}\label{fig:51}
\end{figure}

\bigskip
\textbf{Example 5}.
We consider the scatterer which is like a calabash that includes two balls of radius 1 and 0.75, centered at $(-0.5,0.0,0.0)^T$ and $(0.5,0.0,0.0)^T$, respectively. The imaging results are shown in Figure \ref{fig:51}. In the third picture, we plot the isosurface with isovalue 0.16.  We find
that the imaging functional recovers the scatterer well even with $\lambda=1$. Notice that the radius of bigger ball is 1 and the radius of smaller one is 0.75, the probe wavelength is comparable with the size of the scatterer.

\begin{figure}
\includegraphics[width=0.3\textwidth]{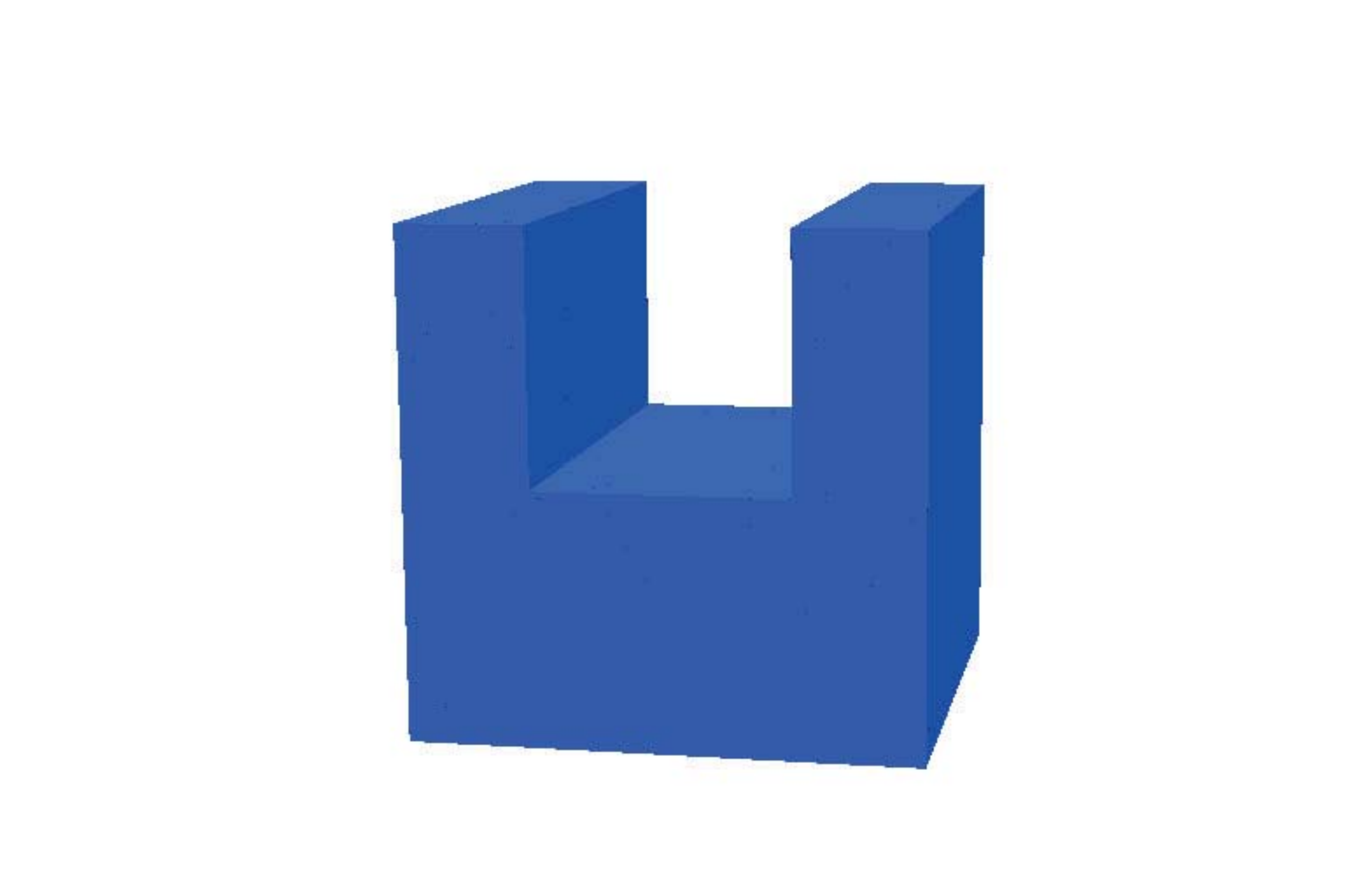}
\includegraphics[width=0.3\textwidth]{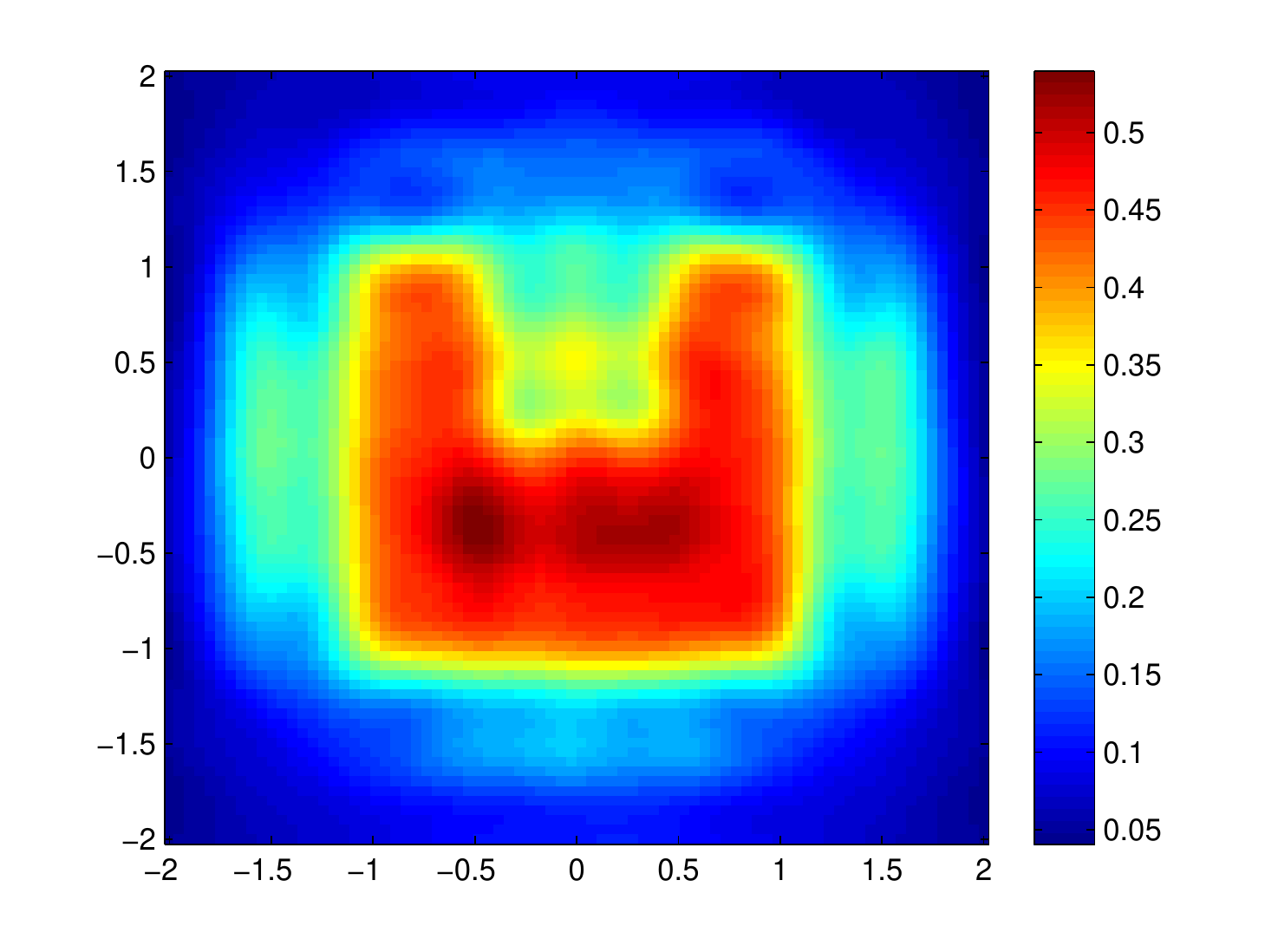}
\includegraphics[width=0.3\textwidth]{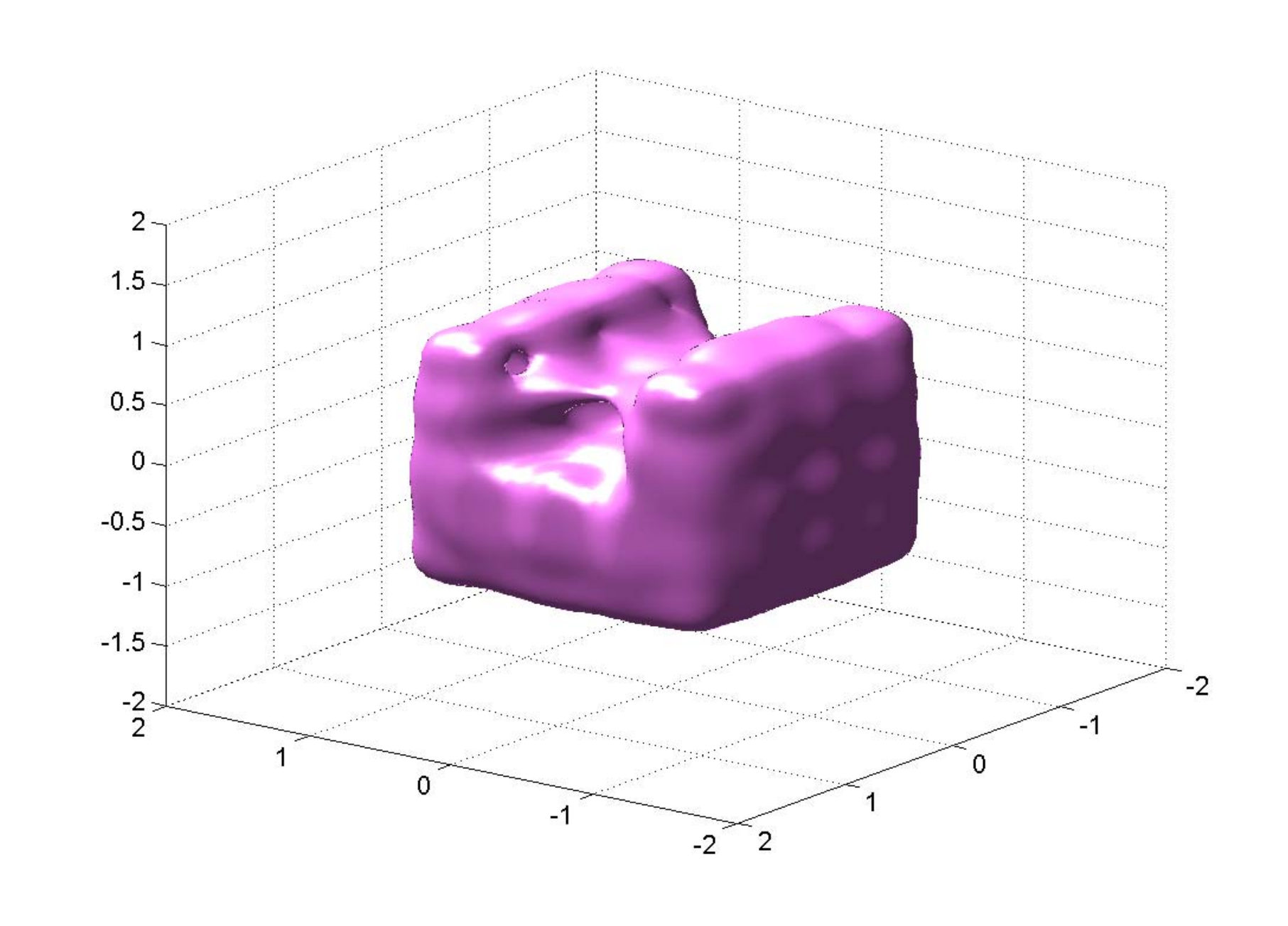}
\caption{Left: the scatterer; Middle: Imaging functional at the cross-section $x_1=1$; Right: Imaging result in 3D view. The wavelength $\lambda=1$ and sampling domain is $(-2,2)^3$ with sampling mesh $80\times80\times 80$,$N_s=N_r=256$.}\label{fig:52}
\end{figure}

\bigskip
\textbf{Example 6}.
We consider in this example the scatterer with polyhedral boundary. The domain of the scatterer is $(-1,1)^3\backslash[-1,1]\times[-0.5,0.5]\times[0,1]$. Figure \ref{fig:52}
shows the exact scatterer, the cross-section of the imaging functional at the plane $x_1=1$, and the isosurface with isovalue 0.42 of the imaging functional. We observe that our imaging method works well for scatterer with non-smooth boundaries.

\section*{Acknowledgments}
The work of J. Chen is supported in part by China NSF under the grant 11001150, 11171040, and that of Z. Chen is supported in part by National Basic Research Project under the grant
2011CB309700 and China NSF under the grant 11021101. The authors would like to thanks the referees for the helpful comments that improved the paper.


\section*{References}
\begin{thebibliography}{99}
\bibitem{anomaly}
Ammari H and Kang H 2004 {\it Reconstruction of Small Inhomogeneities from Boundary Measurements
Lecture Notes in Mathematics} vol~1846 (Berlin: Springer Verlag)

%\bibitem{am10}
%Ammari H 2010 Differential Polarization Imaging  {\it National Institute for Mathematical Sciences NIMS Lecture Note Series TP1003} (South Korea)

\bibitem{ber84}
Berkhout A J 1984  {\it Seismic Migration: Imaging of Acoustic Energy by Wave Field Extrapolation } (New York: Elsevier)

\bibitem{bcs}
Bleistein N Cohen J and Stockwell J 2001 {\it Mathematics of Multidimensional Seismic Imaging, Migration, and Inversion} (New York: Springer)

%\bibitem{boj}
%Bojarski N N 1973 Inverse Scattering {\it Naval Air Systems Command Report N00019-73-C-0312} (Washington D.C.)

\bibitem{boj82}
Bojarski N N 1982 A survey of the near-field far-field inverse scattering inverse source integral equation, {\it IEEE Trans. Antennas Propagation} {\bf AP-30} 975-979

\bibitem{BHV}
Bruhl M, Hanke M and Vogelius M 2003 A direct impedance tomography algorithm for locating small inhomogeneities
{\it Numer. Math.} {\bf 93} 635-654

\bibitem{buffa}
Buffa A, Costabel M and Sheen D 2002 On traces for $H(\curl,\Omega)$ in Lipschitz domains, {\it J. Math. Anal. Appl.} {\bf 276} 845-876

\bibitem{ccm}
Cakoni F, Colton D and Monk P 2004 The electromagnetic inverse scattering problem for partially coated Lipschitz domains
{\it Proc. Royal Soc. Edinburgh} {\bf 134} A 661-682

\bibitem{ccm01}
Cakoni F, Colton D and Monk P 2001 The direct and inverse scattering problems for partially coated obstacles {\it Inverse Problems} {\bf 17} 1997-2015

\bibitem{cla85}
Claerbout J F 1985 {\it Imaging the Earth's Interior} (Oxford: Blackwell Scientific Publication)

\bibitem{pml}
Chen J and Chen Z 2008 An adaptive perfectly matched layer technique for 3-D time-harmonic
electromagnetic scattering problems, {\it Math. Comp.} {\bf 77} 673-698

\bibitem{cch}
Chen J, Chen Z and Huang G 2013 Reverse Time Migration for Extended Obstacles: Acoustic Waves, {\it Inverse Problems}, to appear.

\bibitem{LSM-book}
Cakoni F, Colton D and Monk P 2011 {\it The linear Sampling Method in Inverse Electromagnetic Scattering}, (Philadelphia: SIAM)

\bibitem{LSM}
Colton D and Kirsch A 1996 A simple method for solving inverse scattering problems in the resonance region {\it Inverse Problems} {\bf 12} 383-393

\bibitem{colton-kress}
Colton D and Kress R 1998 {\it Inverse Acoustic and Electromagnetic Scattering Problems} (Heidelberg: Springer)

\bibitem{ck06}
Colton D and Kress R 2006 Using fundamental solutions in inverse scattering {\it Inverse Problems} {\bf 22} R49-R66

\bibitem{fink1}
de Rosny J, Lerosey G, Tourin A, and Fink M 2008  {\it Time reversal of electromagnetic wave, In Modeling and Computations
in Electromagnetics eds  Ammari H} ( Berlin: Springer-Verlag) 187-202

\bibitem{Devaney}
Devaney A J Super-resolution processing of multi-static data using time-reversal and MUSIC,
http://www.ece.neu.edu/faculty/devaney/preprints/paper02n$_{-}$00.pdf

\bibitem{g11}
Garnier J 2010 Sensor array imaging in a noisy environment, in NIMS Lecture Note Series TP1003, National Institute for Mathematical Sciences,
South Korea, 2010.

\bibitem{kirsch_1998}
Kirsch A 1998 Characterization of the shape of a scattering obstacle using the spectral data of the far field operator {\it Inverse Problems} {\bf 14} 1489-1512

\bibitem{fm_book}
Kirsch A and Grinberg N 2008 {\it The Factorization Method for Inverse Problems} (Oxford: Oxford University Press)

\bibitem{kir-mon}
Kirsch A and Monk P 1995 A finite element/spectral method for approximating the time-harmonic Maxwell's system in $\mathcal{R}^3$ {\it SIAM J. APPL. MATH.} {\bf 55} 1324-1344

\bibitem{leis}
Leis R 1986 {\it Initial Boundary Value Problems in Mathematical Physics} (Stuttgart: B.G. Teubner)

\bibitem{GPR}
Leuschen C and Plumb R 2001 A matched-filter-based reverse time migration algorithm for ground-penetrating-radar data
{\it IEEE Trans. Geosci. Remote Sensing} {\bf 39} 929-936

\bibitem{mclean00}
McLean W 2000 {\it Strongly Elliptic Systems and Boundary Integral Equations} (Cambridge: Cambridge University Press)

\bibitem{monk}
Monk P 2003 {\it Finite Element Methods for Maxwell's Equations}(Oxford: Oxford University Press )

\bibitem{nec01}
N\'ed\'elec J C 2001 {\it Acoustic and Electromagnetic Equations: Integral Representations for Harmonic Problems} (Heidelberg: Springer)

\bibitem{peterson}
Peterson A F Ray S L and Mittra R 1997 {\it Computational Methods for Electromagnetics} (Wiley-IEEE Press)

\bibitem{phg}
PHG, Parallel Hierarchical Grid, available online at http://lsec.cc.ac.cn/phg/.

\bibitem{p96}
Potthast R 1996 A fast new method to solve inverse scattering problems {\it Inverse Problems} {\bf 12} 731-742

\bibitem{potthast}
Potthast R 2001 {\it Point Sources and Multipoles in Inverse Scattering Theory} (Boca Raton: Chapman \& Hall/CRC)

\bibitem{music}
Schmidt R 1986 Multiple emitter location and signal parameter estimation {\it IEEE Trans. Antennas. Propag. } {\bf 34} 276-280

\end{thebibliography}
\end{document}